\documentclass[pre,twocolumn,nofootinbib,twoside,showkeys,floatfix,10pt]{revtex4}
\usepackage{dcolumn}%
\usepackage{bm}%
\usepackage{graphicx,textcomp}
\usepackage{array}
\usepackage{bm}
\usepackage{mathtools}
\usepackage{amssymb,amsfonts,amsmath}
\usepackage{color}
\usepackage{nicefrac}
\usepackage{algorithm}  
\usepackage{algorithmicx}  
\usepackage{enumerate}
\usepackage[all]{nowidow}
\usepackage{booktabs}

\usepackage{xcolor}
\newcommand{\nocontentsline}[3]{}
\newcommand{\tocless}[2]
{\bgroup\let\addcontentsline=\nocontentsline#1{#2}\egroup}
\newcommand{\argminE}{\mathop{\mathrm{argmin}}}

\begin{document}

\title{  A mini-review on combinatorial solutions to the Marcus--Lushnikov irreversible aggregation   }

\author{Micha\l{} \L{}epek\footnote{e-mail: michal.lepek@pw.edu.pl}, Agata Fronczak, Piotr Fronczak}

\affiliation{Faculty of Physics, Warsaw University of Technology, Koszykowa 75, Warsaw, Poland, PL-00-662}


\begin{abstract}

Over the last decade, a combinatorial approach to discrete, finite, and irreversibly aggregating systems has been progressively developed. In this work, we review its achievements up to the present moment, focusing on the practical aspects and discussing its limitations. First, we present the assumptions and combinatorial foundations of the approach, which are based on direct counting of the system states, in contrast to the previous approaches of Smoluchowski and Marcus--Lushnikov. A method to obtain combinatorial expressions for the average number of clusters of a given size and, {importantly}, the corresponding standard deviation is described by solving the simplest example of a constant kernel. {An expression for a complete probability distribution for a number of clusters of a given size is also presented.} Then, we extend consideration to a number of kernels (e.g., additive, product, linear-chain, condensation), which were recently solved by explicitly finding the number {of ways to create} a cluster of a given size. {We show that, for a general case, the present framework yields approximate solutions. In this way,} theoretical predictions for any given kernel may be obtained with no need to find an explicit solution but using a recursive expression. We exploit this opportunity to present the use of combinatorial expressions to solve kernels related to the real processes of aerosol growth and planetesimal formation. At this point, a comparison to numerical results appears. Finally, issues related to the {validity} and varying precision of the theoretical predictions are summarized. In the last section, we propose open problems. {Appendix contains partial Bell polynomials, }{generating function method, Lagrange inversion, potential fields of further application, and} considerations on the relation of the presented combinatorial solutions to the scaling solutions of the Smoluchowski equation.
\end{abstract}

\maketitle

\tableofcontents

\begin{figure*}[ht]
 \includegraphics[width=0.65\textwidth]{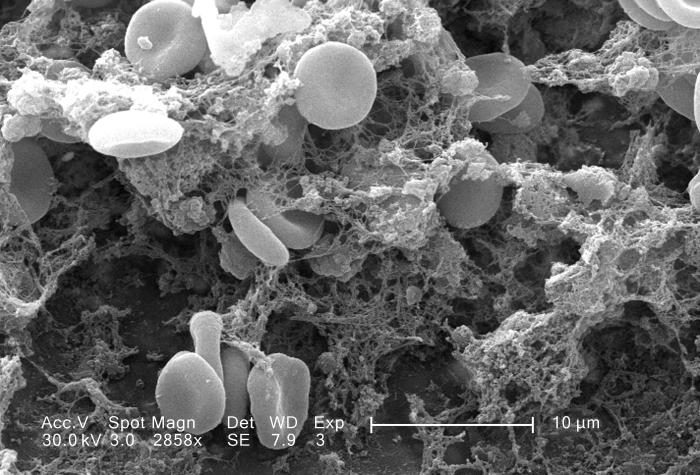}
  \caption{Blood coagulation, or clotting, is the process by which blood changes from a liquid to a gel, forming a blood clot. This image shows a scanning electron micrograph of a number of red blood cells found enmeshed in a fibrinous matrix on the luminal surface of an indwelling vascular catheter. Reproduced from the public domain \cite{wiki_commons_2005}. }
  \label{fig_blood_clot}
\end{figure*}


\section{Introduction}\label{SectIntro}

Although aggregation processes are ubiquitous, we are still far from a complete aggregation theory. The process itself involves merging clusters irreversibly into larger clusters. Despite this straightforward formulation, predicting the system's behavior with high precision is usually an involved task. Typically, binary aggregation acts are considered to prevent the analysis of three-body collisions that correspond to diluted systems. The reaction scheme is
\begin{equation} \label{general_scheme}
   \left(s\right)+\left(g\right){{\stackrel{K\left(s,g\right)}{\longrightarrow}}}\left(s+g\right),
\end{equation}

\noindent where $\left(s\right)$ { and $(g)$ stand for clusters of property $s$ and $g$, respectively,} and $K\left(s,g\right)$ is the coagulation kernel representing the rate of the process. Property $s$ is usually regarded as mass or size but may also be regarded as other additive quantities, such as length or electric charge, or (this is often the case in real processes) as an effective quantity considering all related properties of the particle. Since the system is closed and the clusters grow in time, all particles eventually join to form a single giant cluster.

Aggregation is widely found in nature and technology. In particular fields, it is known as coalescence (e.g., in aerosols, to denote coalescing smaller particles into a larger one), agglomeration (e.g., in polymer or dust aggregation, to denote merging of large aggregates), coagulation (in medicine and food and material processing), or flocculation (water treatment).

\begin{figure*}[ht]
 \includegraphics[width=0.92\textwidth]{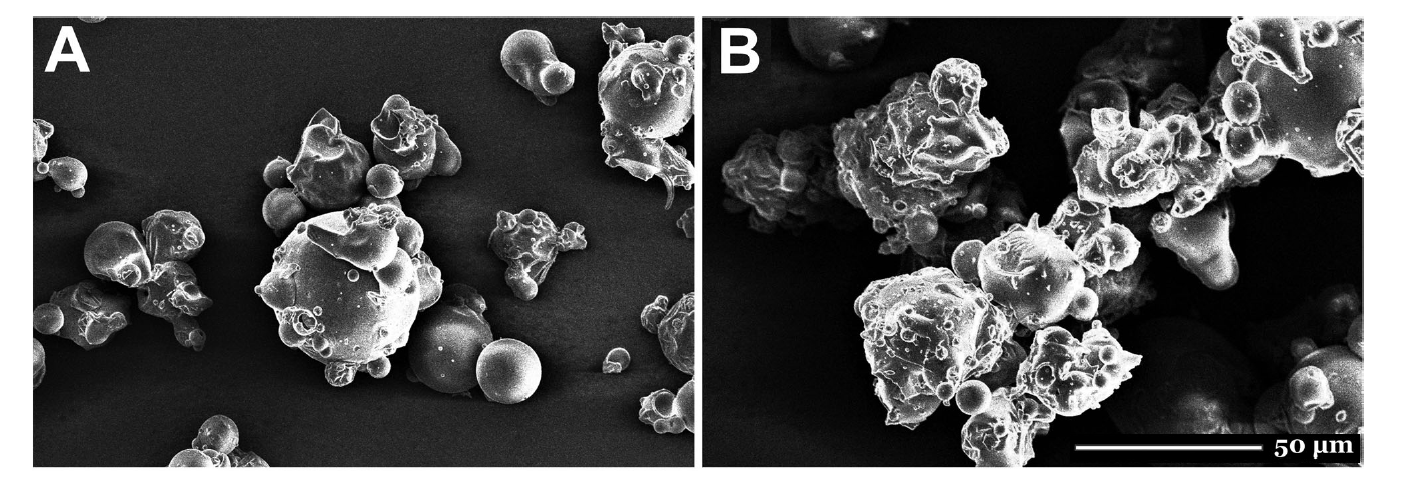}
  \caption{Scanning electron microscope images of the agglomerates with a low number of primary particles per agglomerate (A) and agglomerates with a high number of primary particles per agglomerate (B). These images originate from a recent study on the control of agglomerate formation in spray dryers \cite{Frohlich_2023}. Reprinted by permission of the publisher (Taylor \& Francis Ltd, http://www.tandfonline.com).}
  \label{fig_sem_agglomerates}
\end{figure*}

Figs.~\ref{fig_blood_clot} and \ref{fig_sem_agglomerates} present examples of processes from different fields where particles undergo clustering. In Fig.~\ref{fig_blood_clot}, physiological blood clotting is depicted \cite{wiki_commons_2005}. Fig.~\ref{fig_sem_agglomerates} illustrates two snapshots from a system of drying aerosol \cite{Frohlich_2023}.

The first to study aggregation was Smoluchowski \cite{Smoluchowski_1917}. Kinetic differential equations of his approach require several assumptions characteristic of the mean-field theory, such as the { infinite number of particles, continuous cluster concentrations, and perfect mixing of the system (spatial invariance)}. They were intensively studied for decades and are widely used nowadays. However, a limited number of aggregation processes have explicit solutions to the Smoluchowski equation. It has been used to derive analytical solutions for the three classic kernels, i.e., the constant, $K(s,g)=\mathrm{const}$, additive, $K(s,g) \propto s+g$, and multiplicative (product), $K(s,g) \propto sg$, and for selected initial conditions. { The above solutions were found for both continuous and discrete versions (here, continuous and discrete relate to the allowed sizes of particles). A variety of} other kernels {(see, e.g., Refs.~\cite{daCosta_1998, Aldous_1999})} may be studied {within this approach} with approximations, numerically, or using the so-called scaling solutions. Several extensions are known, e.g., aggregation--fragmentation, aggregation with mass injection \cite{Blum_2006, Leyvraz_2003, Krapivsky_2010, Banasiak_2019}, or {the so-called truncated Smoluchowski equations used in numerical solutions and finite-mass systems \cite{daCosta_1998}}.  {As such, they find their applications (see, e.g., \cite{Rouwhorst_2020_Nature}). However, }{ due to the nature of the Smoluchowski formalism}{, one can ask several fundamental questions. How large does the system have to be for a correct comparison to the results from the Smoluchowski equation? What (obvious) discrepancies are expected to arise between the Smoluchowski theory and the ground truth aggregation process of the finite coagulating system after the initial stage,} { when the Smoluchowski assumptions are not met? The above questions, together with the limited number of explicit solutions known, motivate the search for new approaches.}

The next successful attempt at analytical description {of the irreversible aggregation} was the stochastic approach by Marcus \cite{Marcus_1968}, later studied and developed by Lushnikov \cite{Lushnikov_1978, Lushnikov_2004, Lushnikov_2005, Lushnikov_2011}. {It may be regarded as a microscopic model of the deterministic Smoluchowski model, both describing spatially averaged systems {(spatially invariant)} and being quantitatively similar for sufficiently large systems. However, the Marcus--Lushnikov approach { gives a more detailed description of the process than Smoluchowski. It} allows us to ask questions---such as the probability distribution of particles of given size at a given time or the variability between realizations---that are beyond the scope of the deterministic Smoluchowski rate equations \cite{vanDongen_1987_fluct,vanDongen_1987_fluct_II, Leyvraz_2022}. Another key difference concerns large clusters {(large with respect to the system size)}. While the Smoluchowski equation allows for clusters of unbounded size and represents averages over all possible orders of aggregation events, the Marcus--Lushnikov model represents one possible sequence of coagulation events and restricts clusters to size less than a finite number (finite system size, $N$). This makes it possible to study { the distributions of clusters for any given size, including those similar to $N$ in both non-gelling and gelling systems} \cite{Leyvraz_2022}.} { In general, this \textit{stochastic} Marcus--Lushnikov model provides expected time-dependent size-distribution profiles (either by analytical solutions or averaging over many numerical simulations), which can be compared to solutions of the Smoluchowski model. It also gives information on the distribution of possible resulting states of the system, which has no equivalent in the Smoluchowski approach (cf. Sec.~\ref{subsection_P_of_ns}).}

Within the Marcus-Lushnikov approach, exact solutions of the master equation for the basic kernels (constant, additive, product) were obtained for the simplest case of monodisperse initial conditions, as well as for the mixture of monomers and dimers \cite{Fronczak_2019}. { Mathematical complexity of this approach prevented further extension to other kernels.}

Later, the basic kernels were also revisited in a thermodynamic-focused description using the so-called linear ensembles \cite{2018_Matsoukas}.

In a recent decade, a novel (combinatorial) approach has been proposed { to further overcome the shortcomings of the Smoluchowski approach by providing detailed information on the cluster size distribution in finite systems, even for the late stage of their evolution, when { the continuous cluster concentration conditions are no longer valid.} This discrete model is based on the Marcus--Lushnikov assumptions, but, in this case,} probability distributions were obtained through direct counting of the system states {for the three basic kernels} \cite{grassberger1, grassberger2, Fronczak_2018, Lepek_2019}. Recently, this approach has proved its usefulness in solving more complex forms of kernels with arbitrary parameters under monodisperse initial conditions, i.e., for the linear-chain kernel \cite{Lepek_2021}, $K\propto(1/s+1/g)^\alpha$, condensation kernel, $K\propto(A+s)(A+g)$, and the sum of constant and additive kernels \cite{Lepek_2021_ROMP}, $K\propto A+s+g$. For these kernels, explicit expressions for the average number of clusters of a given size and corresponding standard deviation were obtained.

{ The assumptions behind the work on the combinatorial approach were only achieved to a certain extent: in general, the solutions found were only approximate.} The accuracy of those combinatorial theoretical solutions tested against numerical simulation and experimental data varied from approximate to exact depending on the parameters and coagulation time. By the term ``exact'', we mean solutions to the {finite-system} coagulation process in contrast to the exact solutions of the Smoluchowski coagulation equation, which, in general, is not the same. {As we will see, the relation of the asymptotic behavior of the combinatorial solutions to the known results from the Smoluchowski approach (scaling theory and similarity solutions) presents a complex view \cite{Leyvraz_2022},} { with the asymptotic forms of the combinatorial solutions agreeing closely with the Smoluchowski solutions for the constant and the additive kernels, while differing, especially for small clusters, for the general case.}

{ Despite the limited number of works in the area} { of the combinatorial solutions}{, they address a variety of different aggregation processes and provide a tool to make an attempt to model an arbitrary process, which was not explicitly described before. Therefore, we believe that a concise report of the topic in the form of this (mini-)review article, providing advantages, limitations, and connections to the usual Smoluchowski equations, may be valued in several fields, including finite-size experimental setups.}

{ 
At this point, we will also signal another key difference between the Smoluchowski framework and the combinatorial (discrete Marcus--Lushnikov) framework, which relates to defining the time; continuous in the Smoluchowski equation and discrete in the latter case. Although a large-scale and long-time transformation between these two timescales may be used, comparing combinatorial solutions to the Smoluchowski solutions needs utmost caution. This issue is described in detail in Section~\ref{Remarks_on_time}. Another consequence of the discrete time of the combinatorial approach is its characteristic insensitivity to multiplying kernel forms by a constant factor. This is why we usually use the proportionality sign, ``$\propto$'', when defining kernels in this work. Implications are discussed in Sec.~\ref{Remarks_on_time}.
}

This { mini-}review will expose that the combinatorial approach may be easily used to obtain theoretical predictions of the average number of clusters of a given size for any given kernel, {although yielding, in general, only approximate results}. To do so, we produced auxiliary numerical results of planetesimal formation and aerosol growth to compare them to the theoretical predictions. We show the importance of the ability to predict standard deviation from the mean, together with theoretical solutions (using the example of planetesimals). We propose and briefly outline several practical fields where this method may be helpful: polymer growth, protein aggregation, coagulation of soot, milk, and magnetorheological fluids, cosmic matter aggregation in planetary rings, and dust agglomeration. 

{ Moreover, besides presenting an insight into the first moment (average number of clusters) and the second moment (standard deviation) of the time-dependent probability distribution of finding a number of clusters of a given size, we show a method to find a complete cluster size distribution.}

Throughout this work, we will relate to tests of the theoretical predictions against numerical simulations. A comment on that must be given here. In general, an aggregation process may be studied with several traditional methods: direct numerical simulation of the system, solving the Smoluchowski equation analytically, solving the Smoluchowski equation numerically, or using other (indirect) methods such as scaling solutions. Still, even the numerical solution may be challenging, as the system of equations is usually infinite. Monte Carlo methods have been recently used as a possible way to deal with this problem \cite{Osinsky_2022}. Recently, semi-analytical methods have also been used to study aggregation (still aimed at solving the Smoluchowski equation) \cite{Kaushik_2023}.

Authors sometimes refer to numerical simulations, meaning, for instance, numerical solutions to the Smoluchowski equation. In our work, when we refer to numerical simulation, we always mean direct numerical simulation { of a system with perfect mixing,} where the current state of the system is held in the computer memory and is updated step by step { with binary aggregation acts} strictly due to the kernel rule. {It is done by assigning probabilities to each possible aggregation act and by randomly selecting one.} Such a simulation may be regarded as {reference} for the analytical { Marcus--Lushnikov} approach, as it simulates the aggregation process itself, giving insight into any property of the system without any ``proxy'' tool. Details on how to implement such a {numerical} simulation for any given kernel may be found, for instance, in Refs.~\cite{Kang_1986,Lepek_2021}. {Throughout this work, we use the implementation from Ref.~\cite{Lepek_2021}.} For the analytically-derived expressions (both explicit or recursive), we use the terms \textit{theoretical solution} or \textit{theoretical prediction}. { For the cases for which the precision of the analytical solution is already determined, we clearly specify whether it is an exact solution or an approximation. For the cases that are still challenging in such determination, we give an appropriate reference to the section where the details are considered.}

{
At this moment, to clearly outline the key contributions and methodological framework discussed in this mini-review, we may summarize the main points as follows. An exact combinatorial solution for the constant kernel is presented, expressed in terms of Bell polynomials. Building on this, an \textit{ansatz} is introduced to extend the solution to arbitrary kernels. This extension results in an uncontrolled approximation in all cases other than the constant (and possibly the additive) kernel. Nonetheless, the approximation remains qualitatively correct for large particles. In addition, for a large variety of kernels, this approximate solution can still be evaluated in closed form.
}

The {\color{violet} mini-}review is built as follows. In Section \ref{Approach}, we thoroughly described the initial assumptions and construction of combinatorial expressions. Also, a primary tool to obtain explicit theoretical solutions (a generating function method) has been presented. { Derivations of expressions are shown for an average number of clusters of a given size, corresponding standard deviation, and probability distribution of finding a given number of clusters of a given size.} Section \ref{Solutions} presents a review of known explicit theoretical solutions to several aggregation kernels. Section \ref{NoExplicitSolution} shows that no explicit solution is needed to obtain theoretical predictions {(approximations for a general case)} if a recursive expression may be used. Two examples of use for two different aggregation processes are presented to give a better picture of usability. Readers not keenly interested in the mathematical foundations of the approach may find this section illuminating. Sections \ref{Remarks} and \ref{Summary} summarize the approach and discuss its limitations. {In particular, in Section \ref{Remarks}, we provide a brief discussion on how combinatorial solutions relate to the scaling solutions of the Smoluchowski equation.} Open problems are proposed in the last section. Mathematical tools (partial Bell polynomials and Lagrange inversion) required to understand derivations and transformations are outlined in the corresponding sections in the Appendix. {Also, further fields of potential use are listed.} { In Appendix C, we provide an introduction to the scaling theory of aggregating systems and present related large-size and large-time limit of a general combinatorial solution as compared to the known results from the scaling theory for several kernels.} {The last part of the Appendix contains usual Smoluchowski solutions to the three basic kernels.}

\section{Combinatorial Approach}\label{Approach}

This section describes essential expressions of the combinatorial approach to finite coagulating systems introduced in Refs. \cite{Fronczak_2018, Lepek_2019}. The formalism presented here can be used to analyze the aggregation process with any given kernel.

\subsection{Initial assumptions}

In the combinatorial methodology to analyze the aggregating system, we assume: (i) relatively low particle concentrations, since we do not consider the possibility of having encounters among three or more particles, or several encounters at a given time, (ii) monodisperse initial conditions (in the initial state, all of the clusters are monomers of size 1), (iii) individual coagulation acts are instantaneous and take place in continuous time, and (iv) discrete-time labels which count subsequent coagulation acts, i.e., subsequent states of the system.

The system consists of $N$ monomers (the smallest units, monomer particles), some of which later join into clusters. $N=\text{const}$ is equivalent to the preservation of mass in the system. A single coagulation act occurs in one time step; therefore, the total number of clusters, $k$, at time step $t$ is
\begin{equation} \label{k}  
k = N - t.
\end{equation}

It is worth noting that both $k$ and $t$ define the state of the system in time. This state at time step $t$ is described by the set $\left\{ n_s \right\}$,
\begin{equation} \label{omega}
 \Omega \left(t\right)=\left\{n_1,n_2,\dots ,n_s,\dots ,n_N\right\},
\end{equation}
\noindent where $n_s \ge 0$ stands for the number of clusters of size $s$ (where $s$ is the number of monomeric units included in the cluster) and $n_1$ corresponds to monomers, $n_2$ to dimers, $n_3$ to trimers, { $n_4$ to tetramers, and so on.

During the coagulation process, the sequence $\left\{ n_s \right\}$ is not arbitrary but satisfies the following conditions corresponding to the preservation of the number of monomeric units in the system,
\begin{equation} \label{constraints}
 \sum^N_{s=1}{n_s=k} \;\;\;\; \textrm{and} \;\;\;\; \sum^N_{s=1}{sn_s=N}.
\end{equation}

The above basic assumptions, Eqs.~(\ref{omega}) and (\ref{constraints}), are analogous to the original Marcus--Lushnikov approach.}

{
They result in the following system state in the initial moment of the process:
\begin{equation}
 \Omega (0)=\left\{N,0,0,\dots,0\right\},
\end{equation}

\noindent and after the first coagulation act, we have
\begin{equation}
 \Omega (1)=\left\{N-2,1,0,\dots,0\right\}
\end{equation}

\noindent as two monomers were used to create a dimer.

{ For the next state $\Omega(2)$, there are two possible coagulation steps: two monomers can join to create a second dimer with relative probability $K(1,1)$, or a monomer and a dimer can join to create a trimer with relative probability $K(1,2)$.  In this way, }{ probabilities of the next states, $\Omega(2)$, $\Omega(3)$, and so on, are determined by the kernel $K$ applied.} 



{Finding probabilities $P(\Omega)$ for any given time and system is, in general, a non-trivial task.} In the original Marcus--Lushnikov approach, they were found by solving the master equation {(cf., e.g., Ref.~\cite{Fronczak_2019})}. Instead, in the combinatorial approach presented here, finding these probabilities will include direct counting of available system states.}

\subsection{ Discrete vs. continuous time. Relation to Smoluchowski formalism} \label{Remarks_on_time}

As mentioned before, in the combinatorial approach, individual coagulation acts are instantaneous and take place in continuous time, while the time labels are discrete (i.e., subsequent coagulation acts are marked with time labels).

Time labels mark subsequent coagulation acts (time steps). The first coagulation act occurs at time step $t=1$, the second at time step $t=2$, and so on. These discrete time labels are a counter to subsequent states of the system. { The last aggregation act occurs in $t=N-1$ (thus, for large systems, $t\approx N$). 

This is a key difference of the timescale in this model as compared to the continuous-time models of Smoluchowski and the continuous-time (original) Marcus--Lushnikov model \cite{Marcus_1968, Lushnikov_1978}. A continuous Marcus--Lushnikov time variable equivalent to the discrete combinatorial timescale $t$ may be simply found as $\theta = t/N$. Therefore, at a fixed time point in the continuous-time Marcus--Lushnikov model, the number of particles is a random variable, whereas for the present discrete (combinatorial) model, the number of particles (clusters) is given at each time step \cite{Leyvraz_2022}. }

{
With this formulation, a transformation may be found to connect the time in the combinatorial approach, $t$, to the time in the Smoluchowski approach, $\tilde{t}$. As $t=N-k$ is the discrete time in the combinatorial approach, and if $c_s( \tilde{t})=n_s/N$ corresponds to concentrations  of clusters of size $s$, then
\begin{equation} \label{time_S_time_comb}
\sum_{s=1}^\infty c_s(\tilde{t})=1-\frac{t}{N},
\end{equation}

\noindent provides a correct relation between $\tilde{t}$ of the Smoluchowski equations and $t$ of the discrete Marcus--Lushnikov model (cf. Eqs.~(\ref{AppD_limit_1}) and (\ref{AppD_limit_2}) in Appendix C) \cite{Leyvraz_2022}.

Using the above relation, it can be shown that the distribution of the cluster numbers $z=n_s/N$ as a function of $t$ tends towards $\delta(z-c_s(\tilde{t}))$. As a result, the only difference between the two frameworks is that the Smoluchowski theory provides a less detailed description than the Marcus--Lushnikov theory, as the former does not describe the full distribution of cluster numbers, and always involves a limit $N\to\infty$. In this way, for large values of $N$, the distribution of $n_s$ is sharply peaked around the value given by the Smoluchowski theory. However, for small $N$, for which the distribution does not reduce to a delta function, Marcus--Lushnikov theory provides a fuller view. What also follows is that the variance of the distribution is beyond the reach of the Smoluchowski theory, though it vanishes for the limit of $N \to \infty$ \cite{Leyvraz_2022}.
}



{
For a formally correct comparison between the above two approaches, one shall find infinite-size approximations of the solutions to the discrete combinatorial model as in Ref.~\cite{Leyvraz_2022} (note that for a particular point in the aggregation time, $\theta$, assuming $N \rightarrow \infty $ implies $t \rightarrow \infty$ as well). For the introduction to the similarity solutions and their relation to the large-scale limits of the combinatorial solutions, please see Appendix C.

At this point, an experimental researcher will raise the following concern. Obviously, both combinatorial solutions and Smoluchowski solutions undergo comparison to the experimental data, and comparing predictions from these two frameworks for finite systems is a logical eventuality. Therefore, for practical purposes, in this work, we will plot the solutions to the Smoluchowski equation compared to the combinatorial solutions and to the result from numerical simulations for the constant, additive, and product kernels. For these three basic kernels, the Smoluchowski equation has been solved exactly, yielding explicit (closed form) solutions. We find empirically the time $\tilde{t}$ for the Smoluchowski solutions that correspond to the discrete time $t$ using Eq.~(\ref{time_S_time_comb}). For complete results of this analysis, please see Fig.~\ref{Fig_AppE}.
}

In a real{-time} coagulating system, particles merge (through binary collisions) with frequency determined by the kernel and by the evolution stage of the system {(i.e., concentrations)}. However, the exact time between these acts remains unknown (it is stochastic in nature). {It was modeled previously in Marcus--Lushnikov} framework \cite{Eibeck_2000, Fronczak_2019} and may be modeled also in the combinatorial approach as its further extension. However, {as we know from the above considerations,} the discrete timescale of binary aggregation acts {(as well as $\theta$)} used in the combinatorial formalism is not the same as the physical time measured, e.g., in seconds. Binary aggregation time labels may be regarded as a rescaled version of the physical time and do not provide instantaneous information on the reaction rate expressed in collisions per second. {For the purposes of transformation, Eq.~(\ref{time_S_time_comb}) may be used.}


{ However, at the cost of a somewhat complex relationship with real time, we obtain a different benefit.} Suppose we have information on the initial number of clusters and a ``snapshot'' of the cluster size distribution at a particular point in time (thus, $N$ and $k$). In that case, the translation to the language of discrete time steps is straightforward because a particular number of clusters in the system clearly defines the time step in the combinatorial formalism {($t=N-k$)}. Such an {empirical} procedure of comparing combinatorial predictions to the experimental data {from a small discrete system} has been shown in detail in Ref.~\cite{Lepek_2021}.

{ Another difference between the two frameworks is the definition of the gelation point. In the traditional Smoluchowski approach, gelation is defined by the mass deficiency that appears after a critical time and is attributed to an infinite cluster (a gel). The gel can be assumed to be either passive or active with respect to the sol fraction. The passive gel grows due to a finite mass flux toward infinite particle sizes, while the active gel grows, in addition, by consuming the sol particles. In turn, in the Marcus--Lushnikov approach, where finite coagulating systems are considered, the gel manifests itself as a narrow hump in the distribution of the average particle numbers over their masses (cf. Fig.~1 in Ref.~\cite{Fronczak_2019} or Fig.~\ref{Fig_kernele_zbiorczy}b below). This hump appears after the critical time at macroscopically large mass $s \propto N$ and behaves like the active gel \cite{Lushnikov_2006_review}.} { Nevertheless, the appearance of a hump at large cluster sizes cannot be taken as a unique indicator of gelation, since, due to the finite system size, this feature eventually emerges for all kernels within the Marcus--Lushnikov framework, regardless of whether they exhibit gelation in the Smoluchowski model (cf. Fig.~\ref{Fig_kernele_zbiorczy}b and Fig.~\ref{Fig_kernele_zbiorczy}d). This contrasts with the Smoluchowski formulation, where gelation is a distinctive phenomenon that occurs only for kernels with aggregation rates that grow rapidly with cluster size (e.g., $K=sg$).}

\subsection{Probability distribution over state space} \label{Sec_II.C}

The main aim of the combinatorial analysis is to derive the probability distribution over the state space of the system, i.e., the probability $P(\Omega,t)$ of finding the coagulating system in a given state  { $\mathrm{\Omega}(t)$, cf. Eq. (3)},
\begin{equation} \label{probability_general}  
 P(\Omega,t)=\frac{W(\Omega,t)}{Z(t)},
\end{equation}

\noindent where $W(\Omega,t)$ is the thermodynamic probability~{(a number of microstates
that correspond to a single macrostate)} { of a {(macro)}state $\Omega(t)$} and $Z(t)=\sum_{\mathrm{\Omega}}{W(\Omega,t)}$ stands for the normalizing factor { analogous to partition function in statistical mechanics} \cite{Fronczak_2018}. Allowed states of the system are not equiprobable and $P(\mathrm{\Omega }) \neq \mathrm{const}$, which is a result of non-equilibrium character of the process. 

{ An important comment must be given here. Eq.~(\ref{probability_general}), written analogously as for the microcanonical distribution, is only as assumption that will later allow do derive further quantities: an average number of clusters of given size $s$, $\langle n_s \rangle = \sum_{\Omega}{n_s(\Omega)P(\Omega)}$, and its standard deviation, and probability distribution $P(\Omega,t)$ itself. Eq.~(\ref{probability_general}) assumes that the probability of a given state is proportional to the number of microstates realizing such a given state. Thus, Eq.~(\ref{probability_general}) describes a system where all the ways to create a state $\Omega$ (possible realizations of the system over time), their number being $W(\Omega,t)$, are equally probable. { In other words, by the equiprobability of those ways, we mean that if there are multiple ways {(microstates)} of ending up at the same state $\Omega$ {(macrostate)}, then all the routes have equal probability of occurring.}

The assumption of equiprobable ways of obtaining the state $\Omega$ is entirely valid for the constant kernel, for which it was initially used in Ref.~\cite{Fronczak_2018}. However, it does not hold for arbitrary kernels, being only an approximation for them. Further comments on that issue may be found in Sec.~\ref{Exactness} and in Appendix C.
}

{ In Eq.~(\ref{probability_general}), we formally shall use notation $P(\Omega(t))$, $W(\Omega(t))$, but as it would be obfuscating in reading, we used notation with a comma. The sense of the notation is that, at time $t$, there is a set of possible states $\Omega$ that could exist in the system. In further writing, we will usually omit explicit dependence on $t$ as the state $\Omega$, although formally dependent on $t$, constitutes a more fundamental and unambiguous quantity, being a base for further analysis. We believe it will also shorten the expressions.}

Throughout the following sections, we will derive expressions to calculate the thermodynamic probabilities $W(\Omega,t)$ and the corresponding $Z(t)$ factor. We will take advantage of the three observations: (i) the set of monomers can be divided into subsets in a particular number of ways, (ii) a cluster of a given size $g$ can be created in a particular number of ways, $x_g$ (the number of possible histories of a cluster), and (iii) coagulation acts for a particular cluster can be distributed in different time steps.

\begin{figure*}[ht]
 \includegraphics[width=1.0\textwidth]{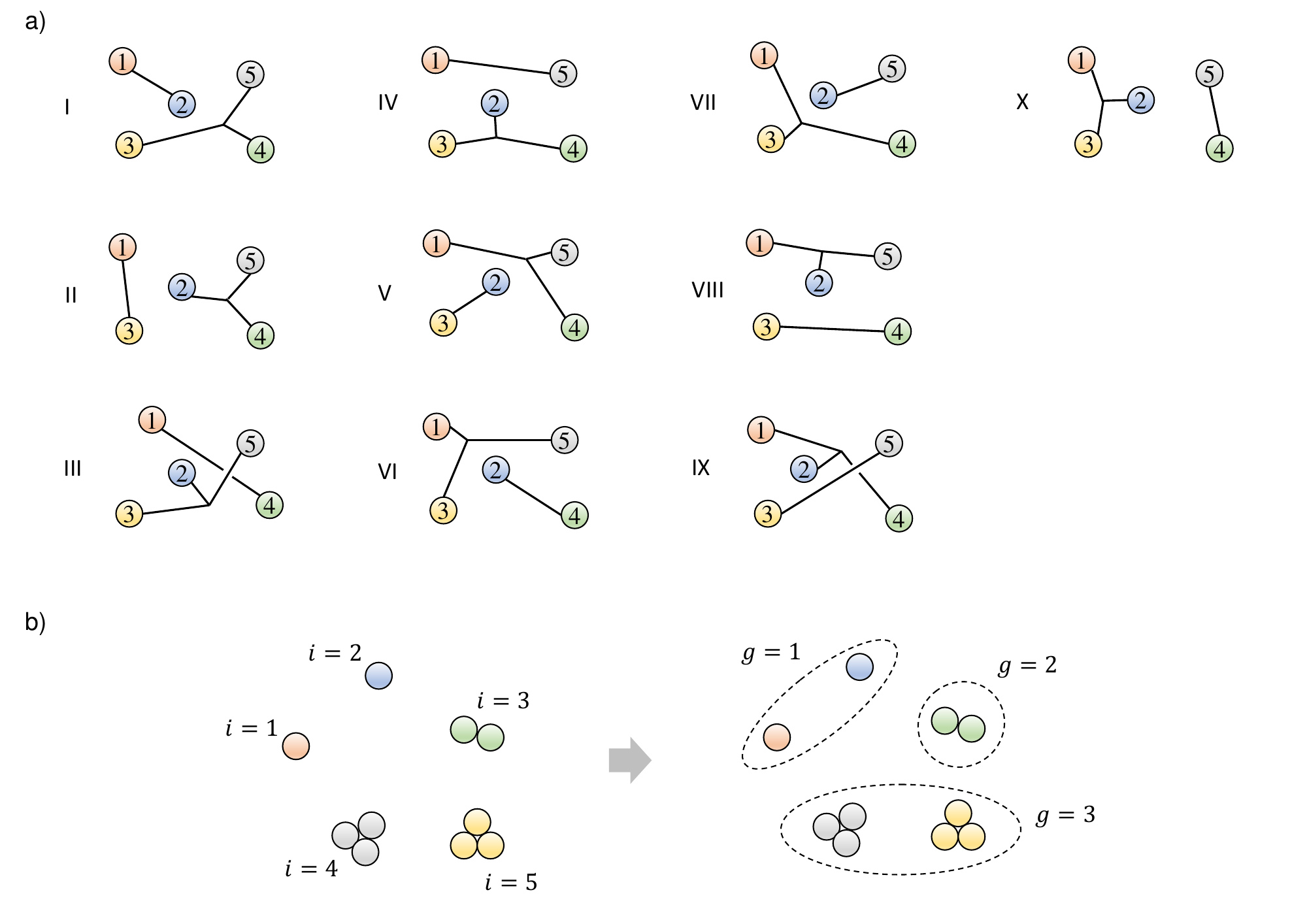}
  \caption{a) Possible ways to group a set of $N=5$ elements into two clusters of sizes 2 and 3. All particles are distinguishable. There are ten ways to do so (cf. Eq.~(\ref{partitioning_long})). {b) Illustration of the change of the multiplication index from $i$ to $g$, cf. Eq.~(\ref{partitioning_long}), meaning that we no longer enumerate over the subsets (i.e., over clusters {which are assigned with their respective numbers $i=1,2,3,4,5$}) but over the sets of subsets (i.e., the sets of clusters) of a given size determined by $g$ {(for this example, $g=1,2,3$, {$\Omega=(2,1,2,0,0,0,0,0,0,0)$})}.} Illustration by the authors.}
  \label{rys_przyklad_grupowanie}
\end{figure*}



\begin{figure*}[ht!]
 \includegraphics[width=0.98\textwidth]{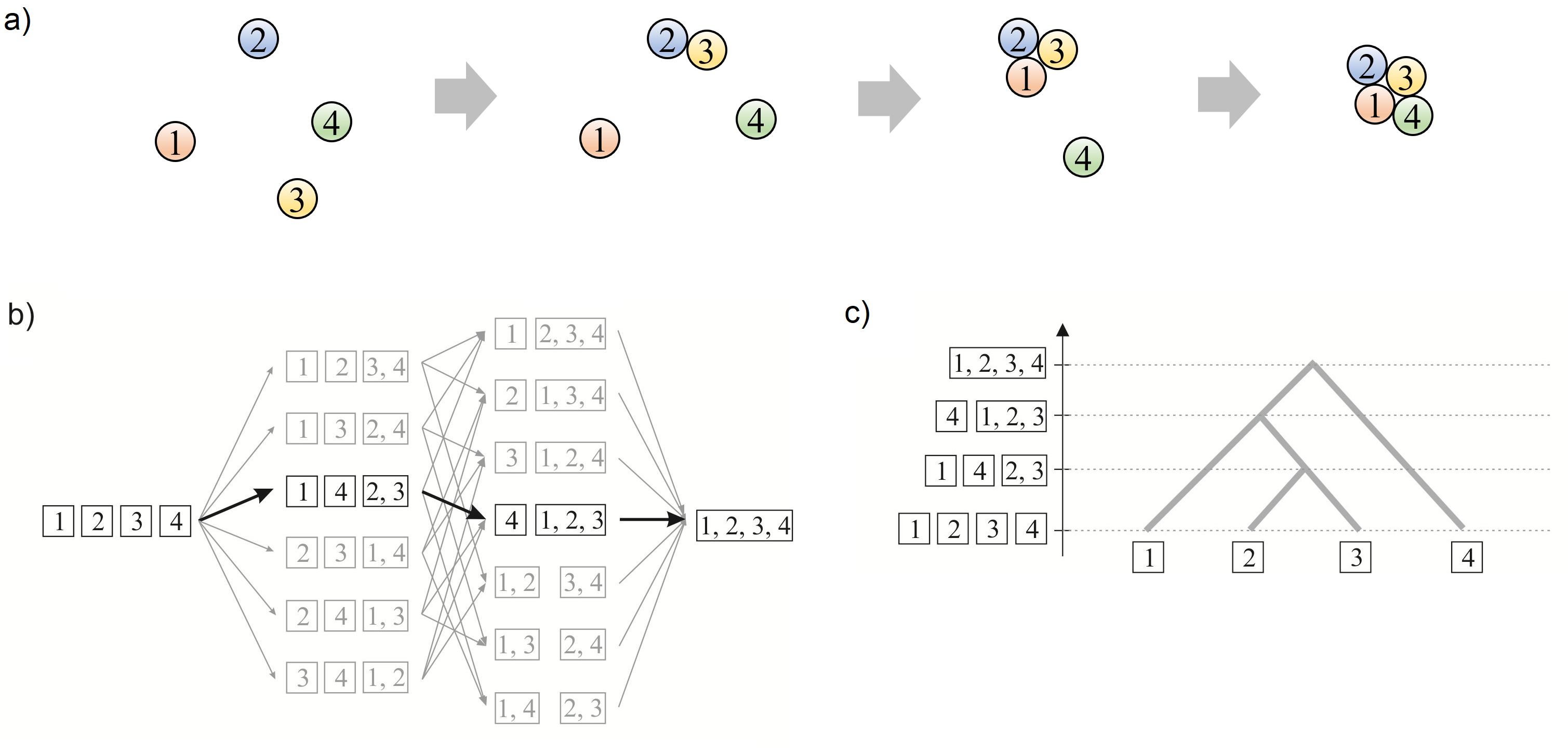}
  \caption{a) A graphical representation of the cluster selection process in the case of coagulation, in which the probability of two particles merging does not depend on their sizes. Our task is to create a particle of size $g$, i.e., consisting of $g$ monomers. The monomers are distinguishable (numbered for clarity). In the first step, we combine two out of $g$ available particles (here $g=4$). In the next step, we combine two further freely selected clusters. We merge the particles until we obtain a single cluster. In this way, we implement one of $x_4= 4!3! / 2^3 = 18$ ways to create a cluster of size $4$. b) Diagram illustrating all possible growth histories of such a particle of size $4$. The number of different growth histories, $x_g$, equals the number of different paths drawn by arrows and leading through different states of the diagram. c) A tree corresponding to the bold path in the diagram in b). Single particle growth history may be illustrated as a rooted tree, with leaf nodes standing for monomeric units, internal nodes representing the history-dependent transition states of the cluster, and the root node being the final step in the growth process. Each vertical line represents a state of the system at a particular time step. Illustration by the authors.}
  \label{rys_przyklad_xg}
\end{figure*}

\subsection{Distributing monomers into clusters}

In this section, we will calculate the number of ways to distribute monomers into the given number of clusters (i.e., partitioning the system of $N$ monomers into $k$ subsets) \cite{Fronczak_2018}. { As both $s$ and $g$ denote cluster size, at this point, we will use $g$, which will prove convenient in subsequent derivations, cf. Eqs.~(\ref{eq_33})-(\ref{eq_37}).} Let $N$ distinguishable objects be divided into $k$ non-empty and disjoint subsets of $c_i>0$ elements each, where $\sum^k_{i=1}{c_i=N}$. As we can choose $c_1$ in  $\binom{N}{c_1}$ ways, $c_2$ in $\binom{N-c_1}{c_2}$ ways, there are
\begin{widetext}
\begin{equation} \label{partitioning_long} 
\binom{N}{c_1,c_2,\dots,c_k} = \binom{N}{c_1} \binom{N-c_1}{c_2} \dots \binom{N-\left(c_1+c_2+\dots +c_{k-1}\right)}{c_k} = N!\prod^k_{i=1}{\frac{1}{c_i!}} = N!\prod^{N-k+1}_{g=1}{{\left(\frac{1}{g!}\right)}^{n_g}}
\end{equation}
\end{widetext}

\noindent of such partitions; $n_g\ge 0$ stands for the number of subsets of size $g$, with the largest {(possible)} subset size equal to $N-k+1$ {(some of the subsets may stay empty)}. The change of the multiplication index $i$ to $g$ means that we no longer multiply over the subsets ({i.e., over} clusters) but over the sets of subsets ({i.e., the sets of} clusters) of a given size determined by $g$. { For a better illustration of changing the index, see Fig.~\ref{rys_przyklad_grupowanie}b}.

For example, let us consider a system of $N=5$ monomers grouped into $k=2$ clusters of sizes 2 and 3. Remember that all monomers are distinguishable (i.e., we can assign numbers from $1$ to $N=5$ to these particles). Possible grouping methods are presented in Fig.~\ref{rys_przyklad_grupowanie}a. There are ten ways of doing so.

Let us now try to determine the number of such combinations using the formula (\ref{partitioning_long}), remembering that the state of the system is described by $\Omega = \left\{0,1,1,0,0\right\}$ (cf. Eq.~(\ref{omega})). We obtain, as expected,
\begin{equation}
N!\prod^{N-k+1}_{g=1}{{\left(\frac{1}{g!}\right)}^{n_g}} = \frac{5!}{(1!)^0 (2!)^1 (3!)^1 (4!)^0 } = 10.
\end{equation}

If in such a composition, defined by Eq.~(\ref{partitioning_long}), each of $n_g$ subsets (clusters) of size $g$ can be in any of $x_g\ge 0$ internal states and the order of clusters does not matter (division by $n_g!$) then the number of partitions becomes
\begin{equation} \label{partitioning_short}
{ W_p (\mathrm{\Omega}) =} N!\prod^{N-k+1}_{g=1}{\frac{1}{n_g!}{\left(\frac{x_g}{g!}\right)}^{n_g}}.
\end{equation}

Summing Eq.~(\ref{partitioning_short}) over all integers $\left\{ n_g \right\}$ specified by Eq.~(\ref{constraints}) one obtains the partial Bell polynomial $B_{N,k} \left( \left\{ x_g \right\} \right)$, an important fact that will be used in Sec.~\ref{Expressions_for_ns}. Readers are encouraged to consult Appendix A for the definition and interpretation of partial Bell polynomials.

Internal state $x_g$ of the cluster, { introduced in Eq.~(\ref{partitioning_short}), in this context,} simply corresponds to the number of ways in which the cluster of size $g$ could be created in the system (in other words, it is the number of possible histories of { a given} cluster). In general, $x_g$ varies for different kernels. The following subsection shows how to obtain $x_g$ for any given kernel.

To sum up this part, Eq.~(\ref{partitioning_short}) gives the number of possible ways of partitioning $N$ monomers into $k$ clusters for a given state $\mathrm{\Omega}$ of the system. However, it implicitly assumes that all the clusters in the system arose at once, which is not valid, as only one merging act is allowed per time step. Coagulation acts corresponding to different clusters may alternate, increasing the number of ways in which a given (final) state can be created.  This issue leads to the need to calculate the number of possible distributions of merging acts in time (see Sec.~\ref{Sec_Ditr_time}).


\subsection{Calculating number $x_g$ of { ways to create} cluster of given size for constant kernel} \label{Calculating_xg_const}

In Ref.~\cite{Fronczak_2018}, it has been shown that in the case of the constant kernel, the number of all possible internal states of a cluster of size $g$ (i.e., all possible histories of its growth) can be simply calculated using binomial coefficients. Since the reaction rate is constant, the probability of two clusters coalescing into a larger one is also constant throughout the evolution.

We will now focus on $g$ monomers contained in the cluster of size $g$ and the time steps when the merging acts for this particular cluster occurred. In the first time step for this particular cluster, two monomers can be chosen out of $g$ available monomers and merged. In the next time step, two clusters out of $g-1$ available are chosen and merged. In the third time step, the next two clusters available out of $g-2$ are merged, and so on. The above results in
\begin{equation} \label{xg_binomials}
 x_g = \binom{g}{2} \binom{g-1}{2} \dots \binom{2}{2} =\frac{g!\left(g-1\right)!}{2^{g-1}}. 
\end{equation}

The process described by Eq.~(\ref{xg_binomials}) is illustrated in Fig.~\ref{rys_przyklad_xg} for a cluster of size 4.

Unfortunately, such a simple calculation cannot be extended to cover other kernels for which the reaction rate depends on the size of the merging clusters. To solve this problem, a recursive equation for $x_g$ has been proposed \cite{Lepek_2019}. By this method, for the constant kernel, we can write 
\begin{equation} \label{xg_recursive_definition_const}
 x_g = \frac{1}{2}\sum^{g-1}_{h=1}\binom{g}{h}\binom{g-2}{h-1}x_hx_{g-h},
\end{equation}

\noindent where $x_h$ and $x_{g-h}$ represent the numbers of ways to create two clusters of size $h$ and $(g-h)$ that joined and became a cluster of size $g$, { and $x_1=1$}.

At this point, we must describe the formulation of Eq.~(\ref{xg_recursive_definition_const}). We can divide the cluster of size $g$ into sub-clusters of size $h$ and size $(g-h)$ in exactly $\binom{g}{h}$ ways. Thus, the first binomial factor denotes the number of ways of choosing a cluster of size $h$ out of $g$ monomers.  

The second binomial factor, $\binom{g-2}{h-1}$, covers the fact that coagulation acts, resulting in clusters $h$ and $(g-h)$, could occupy different possible time steps (cf. Fig.~\ref{Fig_sciezki}).

The sum in Eq.~(\ref{xg_recursive_definition_const}) is taken over all possible pairs of clusters that can join to become a cluster of size $g$. The factor of 1/2 is used to prevent double-counting of the coagulation acts. This factor is arbitrary in the sense of the final result, $\langle n_s \rangle$, which would not change if the factor of 1/2 were removed. More comments on this factor may be found in Ref.~\cite{Lepek_2019}.

{ In the Appendix \ref{Gen_Func_Meth}, we show how to obtain Eq.~(\ref{xg_binomials}) from Eq.~(\ref{xg_recursive_definition_const}) via the generating function method. Thus, we show that they define the same sequence. The recursive form will be later used to obtain $x_g$ for \textit{any} given kernel.}

\subsection{Distributing joining acts in time} \label{Sec_Ditr_time}

\setlength{\fboxsep}{2pt} 

\begin{figure*}[ht]
 \includegraphics[scale=0.75]{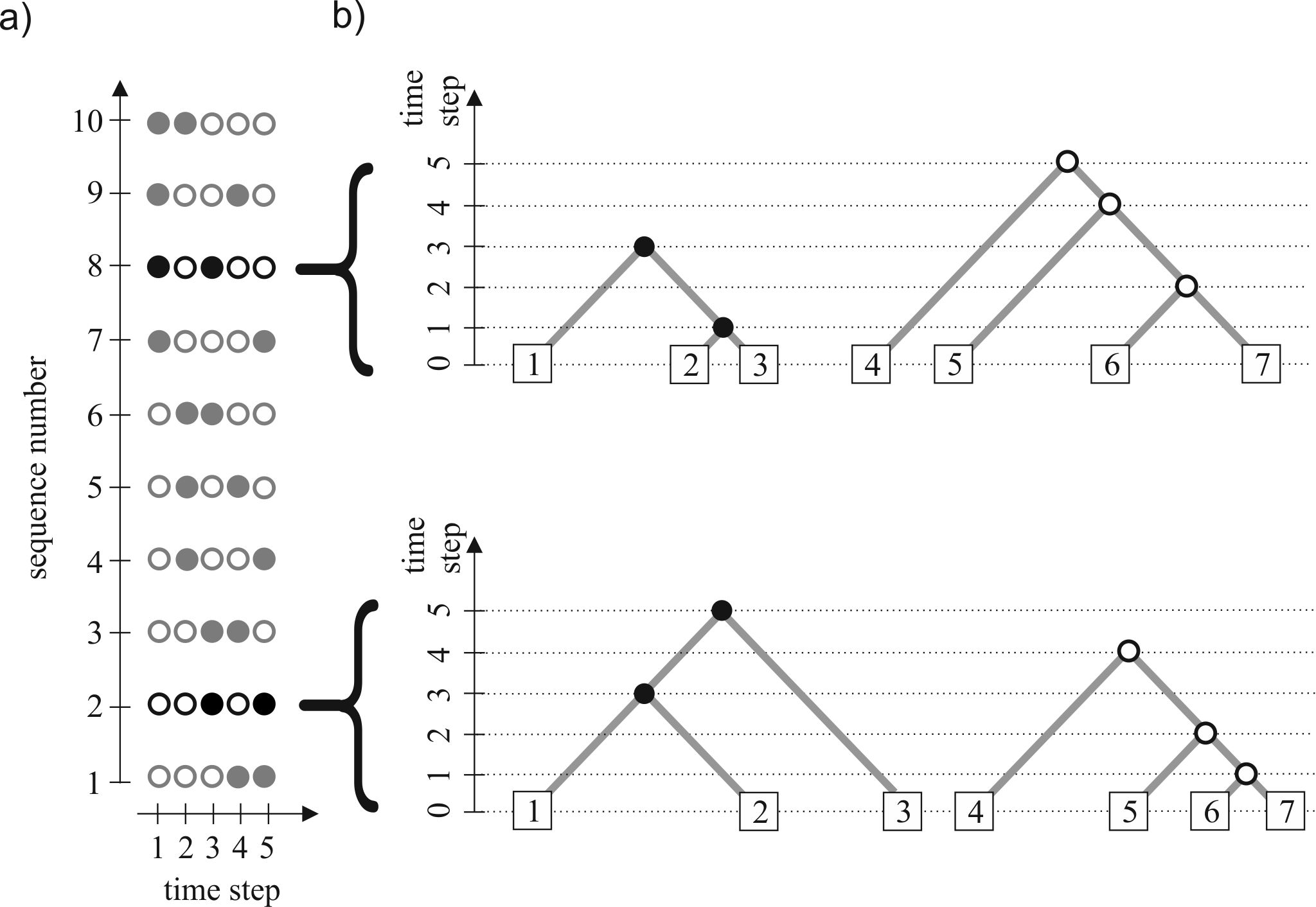}
  \caption{A graphical representation of {the time evolution of a coalescing system (with a constant kernel)} consisting of $N=7$ monomers, in which two particles with sizes 3 and 4 were formed. {In the above figure, for illustration, we consider a state where the particle of size 3 was formed of monomers numbered as $\boxed{1}$, $\boxed{2}$, and $\boxed{3}$, and the particle of size 4 was formed of monomers $\boxed{4}$, $\boxed{5}$, $\boxed{6}$, and $\boxed{7}$ (monomers are distinguishable). For this selected case, which is one of the many possible microscopic realizations of the system,} there are ten possible sequences of coagulation acts leading to the formation of these two clusters. To create them, we need five acts of coagulation, i.e., five time steps. The smaller cluster, size 3, is created in two time steps, and the larger cluster, size 4, is created in three time steps. In part a), we can see that the black dots, representing the coagulation acts that lead to the formation of a cluster of size 3, and the white dots, representing the coagulation acts that lead to the formation of a cluster of size 4, can be ordered in ten different ways in time. Part b) of the figure shows two of the ten possible system evolution scenarios. Coagulation acts associated with forming a cluster of size $ h $ can occur in $ h-1 $ time steps out of the total number of $ N-2 = 5 $ steps needed to form clusters of sizes 3 and 4. {For an arbitrary state $\Omega$, the total number of such sequences is given by Eq.~(\ref{distribution_time}), cf. Eq.~(\ref{distribution_time_example})}. Reprinted from \cite{Lepek_2021}, with permission from Elsevier. }
  \label{Fig_sciezki}
\end{figure*}


As signaled in the previous subsections, a given state can arise as a result of different sequences of intermixed coagulation acts corresponding to different clusters. In other words, if we study a given cluster, its related coagulation acts can occur in different steps in time. So far, we have not considered that fact. Therefore, the distribution of coagulation acts in time is the second origin of combinatorial equations.

Let us resume the basic properties of the dynamics of the system: (i) each cluster of size $g$ requires $g-1$ acts of coagulation, (ii) the process starts from the monodisperse initial state, and (iii) the total number of coagulation acts is equal to $t$. We will now analyze dimers, trimers, and so on. Each dimer requires one step in time to be created. Thus, we can choose this time step in $\binom{t}{1}$ ways as we currently have $t$ available time steps. Then we can choose the time step for the second dimer in $\binom{t-1}{1}$ ways, for the third dimer in $\binom{t-2}{1}$ ways, and so on until for the last dimer we have $\binom{t-n_2+1}{1}$ ways as $n_2$ is the number of dimers. Performing analogous reasoning for trimers (trimers require two time steps to be created, and available space of time is decreased) as well as for larger particles, we obtain the number of possible sequences in time corresponding to each microscopic realization of the system (Eq.~(\ref{partitioning_short})) as
\begin{widetext}
\begin{equation} \label{distribution_time}
\begin{split}
\left[\binom{t}{1} \binom{t-1}{1} \dots \binom{t-n_2+1}{1} \right]  \left[\binom{t-n_2}{2} \binom{t-n_2-2}{2} \dots \binom{t-n_2-2(n_3-1)}{2} \right] \dots \\ = \frac{t!}{{\left(1!\right)}^{n_2}{\left(2!\right)}^{n_3}\dots {\left(\left(g-1\right)!\right)}^{n_g}\dots }=t!\prod^{N-k+1}_{g=1}{\frac{1}{{\left(\left(g-1\right)!\right)}^{n_g}}} { = W_t (\mathrm{\Omega}) }.
\end{split}
\end{equation}
\end{widetext}

For a better understanding of Eq.~(\ref{distribution_time}), consider, as before, the system of $N=5$ monomers after $t=3$ time steps. { Thus, $k=2$ clusters exist in that system. Let us assume those are:} one cluster of size 2 (dimer) and one cluster of size 3 (trimer). From Eq.~(\ref{distribution_time}) we have
\begin{equation}
{ W_t(\mathrm{\Omega}) = W_t(\{0,1,1,0,0\}) = } \frac{3!}{ {\left(0!\right)}^{0} {\left(1!\right)}^{1}{\left(2!\right)}^{1}{\left(3!\right)}^{0}} = 3,
\end{equation}

\noindent meaning that there are three sequences of coagulation acts over time that could have led to the formation of these two clusters.

In turn, for the system of $N=7$ monomers after time $t=5$, in which we have $k=2$ clusters, now, with sizes 3 and 4, the number of possible sequences is
\begin{equation} \label{distribution_time_example}
\begin{split}
{ W_t (\mathrm{\Omega}) } &= { W_t(\{0,0,1,1,0,0,0\}) } \\ &=  \frac{7!}{ {\left(0!\right)}^{0} {\left(1!\right)}^{0}{\left(2!\right)}^{1}{\left(3!\right)}^{1}{\left(4!\right)}^{0}{\left(5!\right)}^{0}} = {420}.
\end{split}
\end{equation}

Sequences that can be observed for {a selected microscopic realization of} that system are illustrated in Fig.~\ref{Fig_sciezki}.

{
It is worth noting that the form of Eq.~(\ref{distribution_time}) assumes that coagulation acts corresponding to different clusters may occupy unrestrictedly any of the available time steps; thus, it clearly results from the assumption of equiprobable ways of creation of a given state $\Omega$, Eq.~(\ref{probability_general}).} 

By multiplying Eqs.~(\ref{partitioning_short}) and (\ref{distribution_time}), one calculates the number of ways in which the system state could be created, $W(\Omega)$. Assuming equiprobability of these ways {(it does not hold for an arbitrary case, see Sec.~\ref{Sec_II.C} and Sec.~\ref{Exactness})}, $W(\mathrm{\Omega})$ defines the thermodynamic probability to find the system in the state $\mathrm{\Omega}$.


\subsection{Expressions for average number of clusters of given size and its standard deviation} \label{Expressions_for_ns}

Armed with the methodology to calculate $x_g$, we can proceed to find an average number of clusters of a given size {(average over the realizations of the process)}, taking advantage of Eq.~(\ref{probability_general}) \cite{Fronczak_2018, Lepek_2019}. 

First, we will write down the expression for the thermodynamic probabilities $W(\mathrm{\Omega})$. { Thermodynamic probability $W(\mathrm{\Omega})$ defines the relative probability of realization of a given state $\Omega$ (the number of ways in which this state could be created). To calculate $W(\mathrm{\Omega})$, we multiply Eqs.~(\ref{partitioning_short}) and (\ref{distribution_time}) to have} 
\begin{equation} \label{W_omega} 
\begin{split}
 W(\mathrm{\Omega}) &= \left[t!\prod^{N-k+1}_{g=1}{\frac{1}{{\left(\left(g-1\right)!\right)}^{n_g}}}\right]\left[N!\prod^{N-k+1}_{g=1}{\frac{1}{n_g!}{\left(\frac{x_g}{g!}\right)}^{n_g}}\right] \\ &= t!N!\prod^{N-k+1}_{g=1}{\frac{1}{n_g!}{\left(\frac{x_g}{(g-1)!g!}\right)}^{n_g}}.
\end{split}
\end{equation}

{ We find the normalizing factor $Z$, playing an analogous role as the partition function in physics of statistical ensembles, by calculating a sum} over all states of the system,
\begin{equation} \label{Z_calculations}
\begin{split}
 Z &= \sum_{\mathrm{\Omega }}{W(\mathrm{\Omega})} \\ &= t!\left[N! \sum_{\left\{n_g\right\}}{ \prod^{N-k+1}_{g=1}{\frac{1}{n_g!}{\left(\frac{x_g}{\left(g-1\right)!g!}\right)}^{n_g}} } \right].
\end{split}
\end{equation}

Here, we will take a key step as we may observe that formulation of Eq.~(\ref{Z_calculations}) naturally takes the form of a partial Bell polynomial (see Appendix A for definition). Thus, using Bell polynomials to simplify the expression for $Z$, we have
\begin{equation} 
 Z = t!B_{N,k}\left(\left\{\frac{x_g}{\left(g-1\right)!}\right\}\right) = t!B_{N,k}\left(\left\{{\omega }_g\right\}\right),
\end{equation}

\noindent where, for further simplification, we used
{
\begin{equation} \label{omega_s}  
\omega_g = \frac{x_g}{(g-1)!} \;\;\;\; \textrm{and} \;\;\;\; \left\{\omega_g\right\} =  \left\{ \frac{x_g}{(g-1)!} \right\}.
\end{equation}
}

Considering the above, the probability distribution is specified as 
\begin{equation} \label{P_omega} 
 P(\mathrm{\Omega}) = \frac{W(\mathrm{\Omega })}{Z} = \frac{N!}{B_{N,k}\left(\left\{{\omega }_g\right\}\right)}\prod^{N-k+1}_{g=1}{\frac{1}{n_g!}{\left(\frac{x_g}{g!}\right)}^{n_g}}.
\end{equation}

Eq.~(\ref{P_omega}) { is} the main contribution of Ref.~\cite{Fronczak_2018} as it provides the most detailed information about the finite-size coalescing system. Let us recall that $t=N-k$; thus, Eq.~(\ref{P_omega}) holds for { any $k$ (any $t$)}.

In Ref.~\cite{Fronczak_2018}, it has been shown that having found $P(\Omega)$ the average number of clusters of given size $s$, $\langle n_s \rangle$, may be derived as
\begin{align}
  \phantom{\left\langle n_s\right\rangle}
  &\begin{aligned} \label{eq_33}
    \mathllap{\left\langle n_s\right\rangle} &= \sum_{\Omega }{n_s(\Omega) P(\Omega)}
  \end{aligned}\\
  &\begin{aligned}
    \mathllap{} &= \frac{N!}{B_{N,k}\left(\left\{{\omega }_g\right\}\right)} \sum_{\left\{n_g\right\}} {n_s \prod^{N-k+1}_{g=1}{\frac{1}{n_g!}{\left(\frac{\omega_g}{g!}\right)}^{n_g}} }
  \end{aligned}\\
  &\begin{aligned}
    \mathllap{} &= \frac{N!}{B_{N,k}\left(\left\{{\omega }_g\right\}\right)} \sum_{\left\{n_g\right\}} { \left( \omega_s \frac{\partial}{\partial \omega_s} \right) \prod^{N-k+1}_{g=1}{\frac{1}{n_g!}{\left(\frac{\omega_g}{g!}\right)}^{n_g}} }
  \end{aligned}\\
  &\begin{aligned}
    \mathllap{} &= \frac{1}{B_{N,k}\left(\left\{{\omega }_g\right\}\right)}  \left( \omega_s \frac{\partial}{\partial \omega_s} \right) B_{N,k}\left(\left\{{\omega }_g\right\}\right)
  \end{aligned}\\
  &\begin{aligned} \label{eq_37}
   \mathllap{} &= \binom{N}{s}{\omega }_s\frac{B_{N-s,k-1}\left(\left\{{\omega }_g\right\}\right)}{B_{N,k}\left(\left\{{\omega }_g\right\}\right)},
   \end{aligned}
\end{align}

\noindent where, in the last transformation, we use the expression for the derivative of the Bell polynomial (described in Appendix A, Eq.~(\ref{Bell_pochodne_definicja})).

Finally, we obtain (cf. Figs.~\ref{Fig_kernel_const} and \ref{Fig_kernele_zbiorczy})
\begin{equation} \label{ns_general}  
 \left\langle n_s\right\rangle = \binom{N}{s}{\omega }_s\frac{B_{N-s,k-1}\left(\left\{{\omega }_g\right\}\right)}{B_{N,k}\left(\left\{{\omega }_g\right\}\right)}.
\end{equation}

Several remarks must be made here. Eq.~(\ref{ns_general}) describes the average number of clusters of size $s$ after $t$ time steps of aggregation. Although $t$ is not present in the expression, $k$ plays its role as $k$ is the total number of clusters in the system, and it decreases linearly in time (counted by subsequent merging acts), $t=N-k$. Of course, for clarity, one may rewrite the expression for explicit dependence of $t$,
\begin{equation} \label{ns_general_t}  
 \left\langle n_s\right\rangle (t) =\binom{N}{s}{\omega }_s\frac{B_{N-s,N-t-1}\left(\left\{\omega_g\right\}\right)}{B_{N,N-t}\left(\left\{{\omega }_g\right\}\right)},
\end{equation}

\noindent but we will use the form with $k$ for shorter notation.

It should also be noted that $\omega_s$ is a single value depending on the cluster size $s$ while $\omega_g$ is a sequence which is independent of $s$, where $g$ varies from $1$ to $N-k+1$ (i.e., to $t+1$). Clearly, for cluster sizes $s \ge t+2$ we have $\left\langle n_s\right\rangle = 0$ as clusters of such sizes had no chance to arise.

A significant element that makes the combinatorial approach different from the previous approaches to coagulation (Smoluchowski, Marcus--Lushnikov) is the fact that using Eq.~(\ref{P_omega}), one can calculate the standard deviation corresponding to the average number of clusters of a given size as \cite{Fronczak_2018}
\begin{equation} \label{std_dev_general}  
 {\sigma }_s = \sqrt{\left\langle n_s^2 \right\rangle -{\left\langle n_s\right\rangle }^2} = \sqrt{\left\langle n_s\left(n_s-1\right)\right\rangle +\left\langle n_s\right\rangle -{\left\langle n_s\right\rangle }^2},
\end{equation}

\noindent where {(see Appendix \ref{trans_for_std_dev} for the transformations)}
\begin{equation} \label{std_dev_general_addition}  
 \left\langle n_s\left(n_s-1\right)\right\rangle =\binom{N}{s,s}{{\omega }_s}^2\frac{B_{N-2s,k-2}\left(\left\{{\omega }_g\right\}\right)}{B_{N,k}\left(\left\{{\omega }_g\right\}\right)}\
\end{equation}

\noindent for $2s \leqslant N$, and $\left\langle n_s\left(n_s-1\right)\right\rangle = 0$ for other cases. For short, $\binom{N}{s,s}=\binom{N}{s}\binom{N-s}{s}$.

Note that Eqs.~(\ref{P_omega}), (\ref{ns_general}), and (\ref{std_dev_general_addition}) were defined for any $x_g$ (or $x_s$). The sequence $x_g$ (or value $x_s$) is the only point in the framework where the considered kernel actually impacts the expressions. Up to this point, we have calculated $x_g$ for the constant kernel only. In the next subsection, we show how to modify the expression for $x_g$ to cover any given kernel.

\subsection{Expression for $x_g$ generalized for arbitrary kernel}

The number $x_g$ (or $x_s$; we usually use the subscript $g$ to emphasize that we study a single cluster of size $g$ that \textit{could} grow in the system) is the only place in the framework where kernel $K$ is used. Therefore, it requires special attention. It is the number of internal states of a cluster of a given size. It can also be interpreted as proportional to the number of possible ways to create such a cluster. We used the phrase ``proportional'' as we will soon find that, generally, the expression for $x_g$ may output real numbers. It has been shown \cite{Lepek_2019,Lepek_2021_ROMP} that this number can be calculated for any given kernel as (cf. Eq.~(\ref{xg_recursive_definition_const}))
\begin{equation} \label{xg_recursive_definition_general}
 x_g = \frac{1}{2}\sum^{g-1}_{h=1}\binom{g}{h}\binom{g-2}{h-1}x_hx_{g-h}K(g-h,h),
\end{equation}

\noindent where, again, $x_h$ and $x_{g-h}$ represent the numbers of ways to create two clusters of size $h$ and $(g-h)$ that joined and became a cluster of size $g$. The kernel $K(g-h,h)$ is an arbitrary kernel translated into variables $g$ and $h$. Obviously, the kernel of the form $K(\Diamond,\square)$ needs to be translated into the form of $K(g-h,h)$, but it is straightforward as $g=\Diamond+\square$ and $h$ is one of the merging sub-clusters, e.g., $h=\Diamond$. For instance, for the additive kernel, $K(\Diamond,\square)=\Diamond+\square$, we obtain $K(g-h,h)=g$, and for the product kernel, $K(\Diamond,\square)=\Diamond \square$, we obtain $K(g-h,h)=(g-h)h$.

The correctness of the above modification may be checked by simply drawing and counting possible ways to create a cluster of size $g$ for the additive and product kernels for $g=1, 2, 3, \dots$. By mathematical induction, one may find that such a modification guarantees the correct value of $x_g$ for any $g$. { However, as we will soon observe, the results for arbitrary kernels obtained by using Eq.~(\ref{xg_recursive_definition_general}) are only approximate. Presumably, this is due to the use of a limiting assumption of equiprobable ways of creating a given state, Eq.~(\ref{probability_general}), which is valid for the constant kernel {(see Sec.~\ref{Exactness} for further details)}. Eq.~(\ref{xg_recursive_definition_general}), being a successor of the assumption~(\ref{probability_general}), inherits its scope of validity. Exhaustive considerations on the relation to the scaling theory are given in Appendix C.}

{
For the combinatorial approach, modification in Eq.~(\ref{xg_recursive_definition_general}) was originally guessed \cite{Lepek_2019}, but it may also be derived, using involved analytics, from the master equation (work in progress). Moreover, expressions of that kind were previously studied in relation to the very short time limit of the Smoluchowski equation, and to the scaling limit for large sizes \cite{Ernst_1984, vanDongen_1987, Leyvraz_2022}.}

It is worth emphasizing that the kernel $K$ present in Eq.~(\ref{xg_recursive_definition_general}) is, in general, defined up to a multiplicative constant,
\begin{equation}
K \propto f(s,g),
\end{equation}

\noindent which is characteristic of the combinatorial approach and a consequence of the discrete time. This is why we will use the proportionality sign, ``$\propto$'', when defining kernels later in this work. Such a formulation has several implications, which were discussed in Sec.~\ref{Remarks_on_time} (also, cf. comments on the factor of $1/2$ in Sec.~\ref{Calculating_xg_const}).

{
\subsection{Probability distribution for number
of clusters of given size} \label{subsection_P_of_ns}

Up to this moment, we have been focused on finding the average number $\langle n_s \rangle$ of clusters of a given size $s$, Eq.~(\ref{ns_general}), and its corresponding standard deviation $\sigma_s$, Eq.~(\ref{std_dev_general}). In fact, these two quantities constitute the first and the second moment of the time-dependent probability distribution for the number of clusters of a given size, $P(n_s, t)$. It is the probability that there are exactly $n_s$ clusters of size $s$ in the system consisting of $N$ monomers, in which there are $k=N-t$ clusters in total.

This probability is \cite{Fronczak_2018}
\begin{equation} \label{P_ns}
    P(n_s, t) = \sum_{\Omega^{\ast}{(s,t)}}{P(\Omega^{\ast}{(s,t)})},
\end{equation}

\noindent where the summation runs over all states $\Omega^{\ast}{(s,t)}$ of the system, in which $n_s$ is fixed. Such states can be defined as (cf. Eqs.~(\ref{omega}) and (\ref{constraints}))
\begin{equation}
    \Omega^{\ast}{(s,t)} = \{ n_g: n_s=\mathrm{const} ~\wedge~ \sum_{g\neq s}{n_g=k^{\ast}} ~\wedge~ \sum_{g\neq s}{gn_g=N^\ast} \},
\end{equation}

\noindent where $k^\ast = k-n_s$ and $N^\ast = N-sn_s$. { In the above equation, naturally, we consider only those cluster numbers $n_s$ and those cluster sizes $s$ which may be observed in the system at time $t$, i.e. $k^*, N^* \ge 0$, and $n_s = 0$ for $s \ge N-k+2$.}

By inserting Eq.~(\ref{probability_general}) into Eq.~(\ref{P_ns}) and by using the definition of Bell polynomials, Eq.~(\ref{Bell_polynom_def}), one finds the general expression for the probability distribution of the number $n_s$ of clusters of size $s$ as \cite{Fronczak_2018}
\begin{equation} \label{P_ns_full}
\begin{split}
 P(n_s, t) &= \frac{N!}{B_{N,k}\left(\left\{{\omega }_g\right\}\right)} \frac{1}{n_s!} \left( \frac{\omega_s}{s!} \right)^{n_s} \sum_{\Omega^\ast} {\prod_{g\neq s}{\frac{1}{n_g!}{\left(\frac{\omega_g}{g!}\right)}^{n_g}} } \\ &= \frac{1}{n_s!} \left( \frac{\omega_s}{s!} \right)^{n_s} \frac{N!}{N^{\ast}!}  \frac{B_{N^\ast,k^\ast}\left(\left\{ (1-\delta_{gs}) \omega_g \right\}\right)}{B_{N,k}\left(\left\{{\omega }_g\right\}\right)}.
\end{split}
\end{equation}

\noindent where $\omega_s$ and $\omega_g$ are defined as before (Eq.~(\ref{omega_s})), $\delta_{gs}$ is the Kronecker delta, and the corresponding sequence of parameters $\{(1-\delta_{gs}) \omega_g \}$ stands for $\{\omega_1, \dots, \omega_{s-1}, 0, \omega_{s+1}, \dots, \omega_N \}$.

Below, we present a simplification of Eq.~(\ref{P_ns_full}) for the case of the constant kernel, which will be particularly illustrative.

\begin{figure}[t] 
 \includegraphics[scale=0.55]{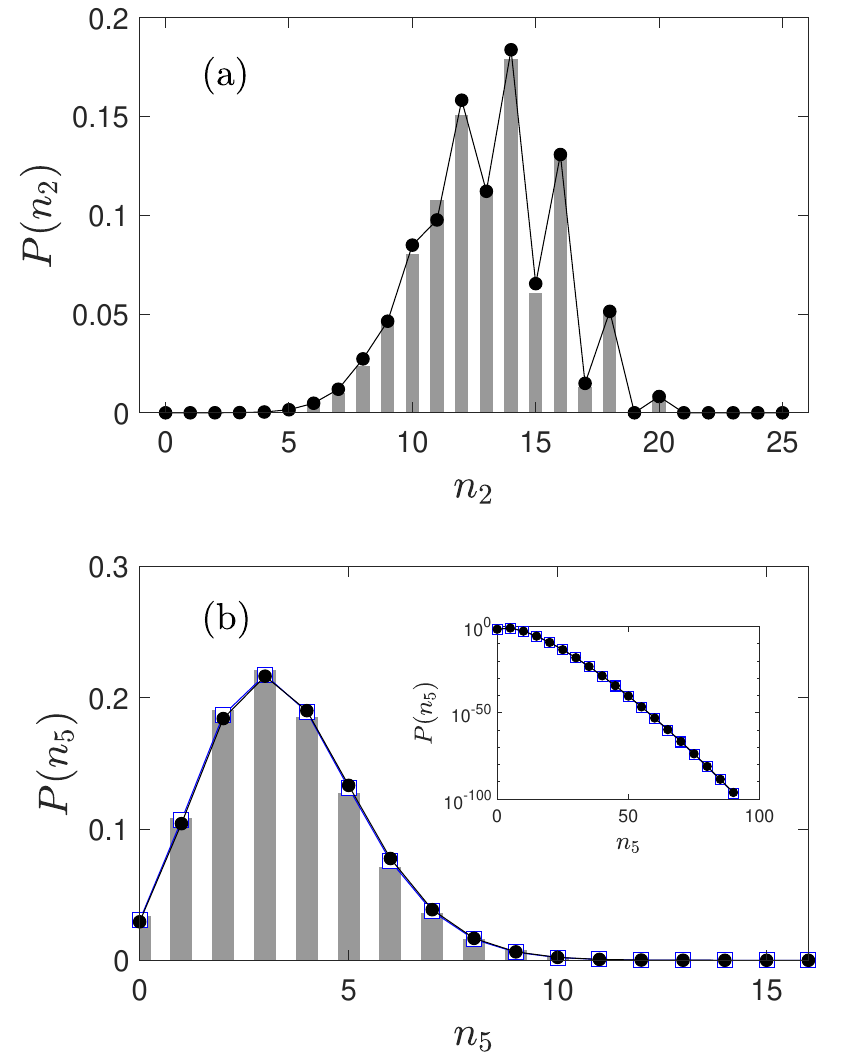}
 \caption{ { Probability distributions for the number of clusters of a given size, $P(n_s)$, for (a) $N = 100$, $k = 80$, and $s = 2$ (dimers), and (b) $N = 10^5$, $k = 600$, and $s = 5$ (clusters of size 5). Solid black circles stand for the exact expression, Eq.~(\ref{P_ns_t_exact}). Open squares represent the approximated formula, Eq.~(\ref{P_ns_simplified}). Bars represent results of the numerical simulation averaged over $10^4$ independent runs. Inset: $P(n_s)$ vs. $n_s$ in semi-logarithmic scale. $P(n_s)$ in the plots is normalized, $\sum_{i} P(i) = 1$. Lines connecting data points are only guidelines for the eyes.} }
 \label{Figure_Pns}
\end{figure}

Having previously found that for the constant kernel $\omega_s = s!/2^{s-1}$ (Eqs.~(\ref{xg_binomials}) and (\ref{omega_s})) and by using several Bell polynomial relations (for the detailed transformations, see Ref.~\cite{Fronczak_2018}), we obtain the following exact expression for the probability distribution of the number of clusters $n_s$ of a given size $s$ in the coagulating system with the constant kernel for a given time $t=N-k$,
\begin{equation} \label{P_ns_t_exact}
 P(n_s,t) = \frac{\binom{k}{n_s}}{\binom{N-1}{k-1}} \sum_{\kappa=0}^{\kappa_{max}} { \binom{k-n_s}{\kappa} \binom{N-sn_s-s\kappa-1}{k-n_s-\kappa-1} (-1)^\kappa }.
\end{equation}

In the above expression, for the limit of large $N \gg 1$ and $kN^{-1} \ll 1$, the fraction of successive sum components behaves as $N^{-1}$. Effectively, by neglecting all terms in the sum except the first one, for $\kappa = 0$, one obtains an approximate expression for the distribution $P(n_s,t)$ for a coagulating system with the constant kernel,
\begin{equation} \label{P_ns_simplified}
 P(n_s,t) \simeq \frac{\binom{k}{n_s}\binom{N-sn_s}{k-n_s}}{\binom{N}{k}}.
\end{equation}

It is worth observing that Eq.~(\ref{P_ns_simplified}) defines the hypergeometric distribution and its expected value, for $sn_s \ll N$, coincides with Eq.~(\ref{EQ42}),
\begin{equation} \label{ns_from_P_ns}
    \langle n_s \rangle \simeq k \frac{\binom{N}{k-2}}{\binom{N}{k-1}} \simeq \frac{k^2}{N}.
\end{equation}

Fig.~\ref{Figure_Pns} shows a perfect agreement of the exact expression, Eq.~(\ref{P_ns_t_exact}), with numerical simulations of the aggregating system with the constant kernel. As in Fig.~\ref{Figure_Pns}b, the difference between the exact and the approximate formula for $P(n_s)$, Eq.~(\ref{P_ns_simplified}), for $sn_s \ll N$, stays slight.

The correctness of the presented probability distribution was directly evaluated for the three basic kernels, coinciding exactly with the true values for the constant and additive kernels. For further comments, see Sec.~\ref{Exactness}.

} 

\section{Finding explicit solutions for average number of clusters of given size} \label{Solutions}

In this section, we will review solutions to kernels where explicit final expressions for an average number of clusters of a given size, $\langle n_s \rangle$, were found. We will see that by using known relations for Bell polynomials { (see Appendix~\ref{AppA_1} for details)}, these final expressions can be simplified, for some cases even to forms without Bell polynomials. For some other kernels, attempts have been made but with no success---these will be listed at the end of the section.

As the only point in the framework involving $K$ is the recursive expression for $x_g$, finding a solution to kernel $K$ comes down to transforming $x_g$ to explicit form (using generating function method and, when needed, Lagrange inversion) and later---if possible---simplifying the final expression for $\langle n_s \rangle$ using Bell polynomial relations.

\subsection{Constant kernel, $K=\mathrm{const}$}

In Sec.~\ref{Approach}, we have gone through the scheme of solving a kernel for the constant rate process, i.e., we have found corresponding $x_g$, Eq.~(\ref{xg_binomials}), cf. Eq.~(\ref{xg_const_appendix}). Combining that result with the expressions for $\langle n_s \rangle$, Eqs.~(\ref{omega_s}) and (\ref{ns_general}), we have
\begin{equation} \label{ns_const_kernel}
 \left\langle n_s\right\rangle = \binom{N}{s} \frac{s!}{2^{s-1}} \frac{B_{N-s,k-1}\left(\left\{ \frac{g!}{2^{g-1}}  \right\}\right)}{B_{N,k}\left(\left\{  \frac{g!}{2^{g-1}} \right\}\right)}.
\end{equation}

The above expression (\ref{ns_const_kernel}) may be significantly simplified \cite{Fronczak_2018} to the form with no Bell polynomials. For that purpose, one can use the relation~(\ref{relacja_abz}) from Appendix A. Bell polynomial $B_{N,k}(\left\{  g!/2^{g-1} \right\})$ has been transformed { to the following form (see Eqs.~(\ref{A8})--(\ref{uproszczenie_const_1})):
\begin{equation}
  B_{N,k}\left(\left\{ \frac{g!}{2^{g-1}} \right\}\right) = 2^k \left( \frac{1}{2} \right)^{N} \binom{N-1}{k-1} \frac{N!}{k!}.
\end{equation} 
}

Using the above result, we can use $N-s$ instead of $N$ and $k-1$ instead of $k$  to obtain
\begin{equation} \label{uproszczenie_const_2}
\begin{split}
 B_{N-s, k-1} & \left(\left\{  \frac{g!}{2^{g-1}} \right\}\right) \\ &=   2^{k-1} \left( \frac{1}{2} \right)^{N-s} \binom{N-s-1}{k-2} \frac{(N-s)!}{(k-1)!}.
 \end{split}
\end{equation}

After applying Eq.~(\ref{uproszczenie_const_2}) to Eq.~(\ref{ns_const_kernel}) the expression for $\langle n_s\rangle$ is simplified to
\begin{equation} \label{EQ42} 
 { \langle n_s \rangle } = k\frac{\binom{N-s-1}{k-2}}{\binom{N-1}{k-1}},
\end{equation}

{
\noindent which coincides with previously found Eq.~(\ref{ns_from_P_ns}).
}

Analogously, the expression~(\ref{std_dev_general_addition}) to calculate standard deviation may be simplified for the constant kernel to the form of
\begin{equation} \label{EQ43}  
 \left\langle n_s\left(n_s-1\right)\right\rangle = k(k-1) \frac{\binom{N-2s-1}{k-3}}{\binom{N-1}{k-1}}.
\end{equation}

\begin{figure}[ht]
\includegraphics[scale=0.64]{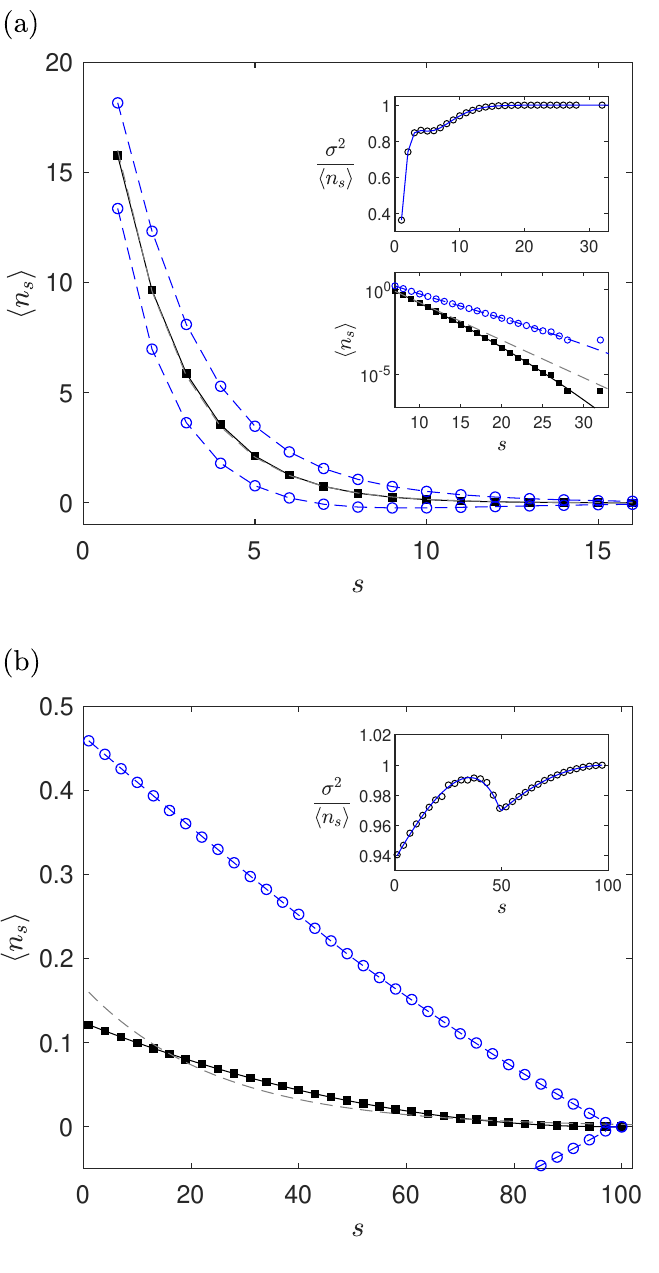}
\caption{Constant kernel aggregation of a system consisting of $N=100$ monomeric units. An average number of clusters of a given size $s$ and its standard deviation for (a) $t=60$ when $k=40$ clusters are left, and (b) $t=96$ when only $k=4$ clusters are left in the system. Lines (guidelines for eyes) correspond to the { exact theoretical solutions}: black solid ones for $\langle n_s \rangle$, Eq.~(\ref{EQ42}), and blue dashed ones for $\langle n_s \rangle \pm \sigma_s$, Eqs.~(\ref{std_dev_general}), (\ref{EQ42}), and (\ref{EQ43}).  The scattered points represent the results of numerical simulations, averaged over $10^6$ independent runs. Upper inset figures in (a) and (b): Variance divided by the mean. Lower inset in (a): Semi-logarithmic plot of $\langle n_s \rangle$ for higher $s$. { Low $\langle n_s \rangle$ and a skewed distribution as in (b) mean that, in a single realization of the process, many of the sizes $s$ stay unoccupied, i.e., no clusters are observed for those $s$. Additionally, the solutions from the Smoluchowski theory are presented (gray dashed lines). For details of the Smoluchowski solutions, see Appendix D.}}
\label{Fig_kernel_const}
\end{figure}

A comparison of the statistics obtained by the above equations to the results of numerical simulations of a system of $N=100$ monomers aggregating at a constant rate is presented in Fig.~\ref{Fig_kernel_const}. { What deserves emphasis, variance increases at long times as the set of possible system states widens. Later, when the system approaches its final state ($t = N-1$) with only a single cluster, variance quickly decreases as the number of possible states decreases. Also, observing values of $\sigma_s^2 / \langle n_s \rangle$ close to $1$ tells us that the actual fluctuations are of the same intensity as the mean; thus, we can expect that for particular sizes $g$, we may not find such clusters in the system at all.}

\subsection{Product kernel, $K \propto sg$}

To calculate $x_g$ for the product kernel, we use Eq.~(\ref{xg_recursive_definition_general}) to obtain the following form \cite{Lepek_2019},
\begin{equation} \label{EQ23}  
 x_g=\frac{1}{2}\sum^{g-1}_{h=1}{\binom{g}{h}\binom{g-2}{h-1}}x_hx_{g-h} h(g-h).
\end{equation}

It can be seen that the coefficient $h(g-h)$ modifies $x_g$ in accordance with the sizes of two contributing sub-clusters. For instance, if one of the sub-clusters is a monomer ($h=1$), then only the larger sub-cluster of size $(g-h)$ alternates $x_g$.

Expanding binomial signs and, this time, using another substitution,
\begin{equation}
y_g = \frac{x_g}{\left(g-1\right)!\left(g-1\right)!}, 
\end{equation}

\noindent we have
\begin{equation} \label{EQ24} 
 {\left(g-1\right)y}_g=\frac{1}{2}g\sum^{g-1}_{h=1}{y_hy_{g-h}}.
\end{equation}

Applying analogous steps and transformations as for the constant kernel, we obtain the equation for the generating function $G(z)$ (see Eq.~(\ref{EQ17})) for the product kernel as
\begin{equation} \label{EQ25} 
 G(z) e^{-G(z)}=Cz.
\end{equation}

The Eq.~(\ref{EQ25}) has the form of $f(G)=F$ where $f(G)$ stands for the left--hand side of Eq.~(\ref{EQ25}) and $F$ stands for its right--hand side. {This implicit equation is a version of the Lambert's $W$-function (see Ref.~\cite{NIST} for details). Thus, a series representation of the inverse function $G(F)$ reads}
\begin{equation} \label{EQ26}  
 G(F)=\sum_{n\ge 1}{\frac{n^{n-1}}{n!}F^n} 
\end{equation}

\noindent which, in this case, reads
\begin{equation} \label{EQ27} 
 G(z) = \sum^{\infty }_{g=1}{y_gz^g} = \sum^{\infty }_{g=1}{\frac{g^{g-1}}{g!}{\left(Cz\right)}^g}.
\end{equation}

From the initial condition, $y_1=1$, we may find a constant $C=1$. Considering the substitution $y_g$ and the expressions under the summation, we obtain
\begin{equation} \label{EQ28}  
 y_g=\frac{g^{g-1}}{g!}\ =\frac{x_g}{\left(g-1\right)!\left(g-1\right)!}.
\end{equation}

Thus, we find the number of ways to create a cluster of size $g$ for the product kernel as
\begin{equation} \label{EQ29}  
 x_g = (g-1)! g^{g-2}. 
\end{equation}

Finally, considering Eqs.~(\ref{omega_s}) and (\ref{ns_general}), we find the average number of clusters of given size for the product kernel,
\begin{equation} \label{EQ44_product} 
 {\langle n_s\rangle } = \binom{N}{s} s^{s-2}\frac{B_{N-s,k-1}\left(\left\{g^{g-2}\right\}\right)}{B_{N,k}\left(\left\{g^{g-2}\right\}\right)}.
\end{equation}

\begin{figure*} [ht]
\includegraphics[width=0.93\textwidth]{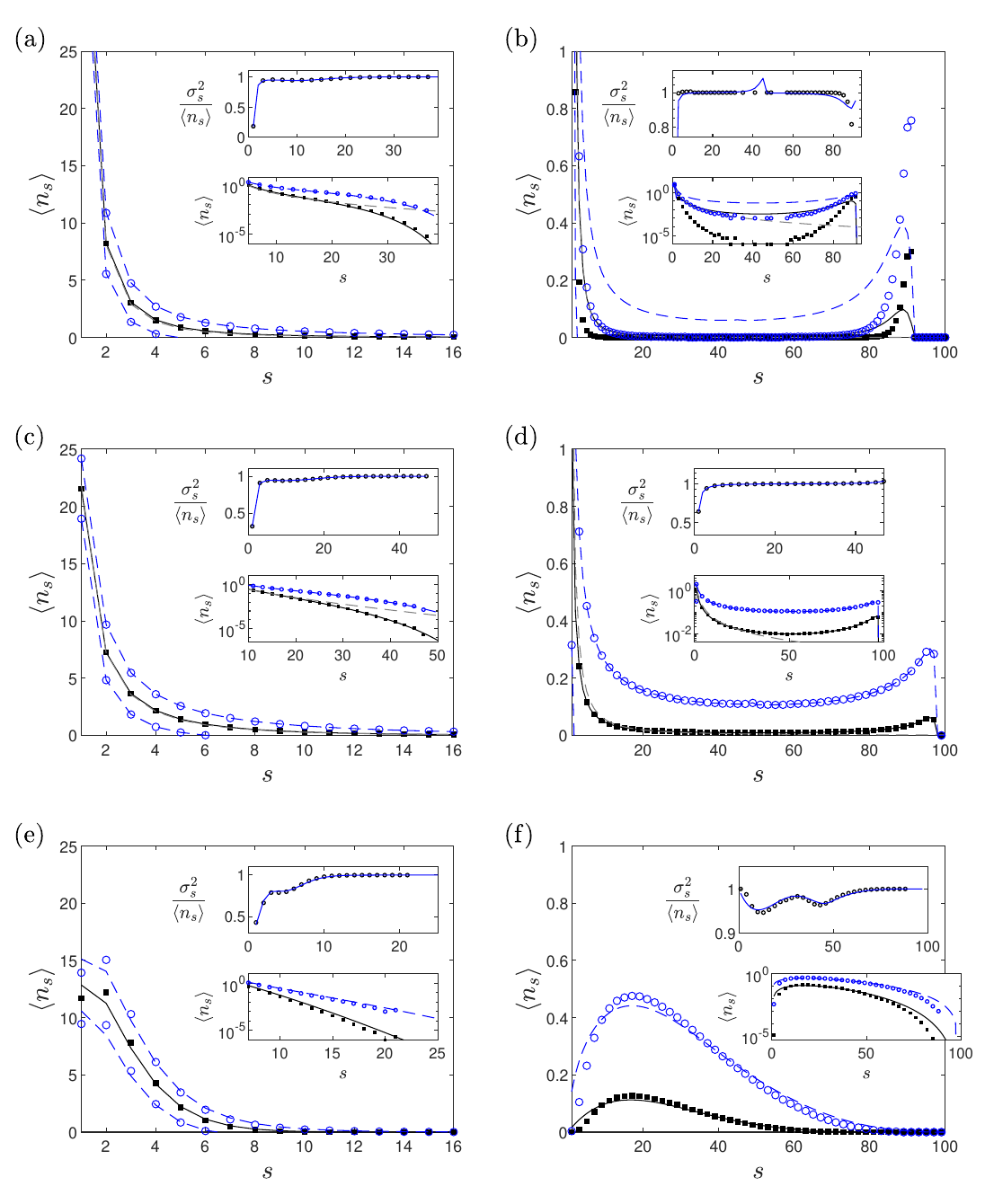}
\caption{Statistics of the coagulating systems arose from the monodisperse initial conditions with particular aggregation kernels: product kernel, $K \propto sg$, for $t=40$ (a), and $t=90$ (b); additive kernel, $K \propto s+g$, for $t=60$ (c), and $t=96$ (d); and linear--chain (LC) kernel, $K \propto s^{-1} + g^{-1}$, for $t=60$ (e), and $t=96$ (f). The systems initially consisted of 100 monomers. Lines represent average numbers of clusters of a given size, $\langle n_{s} \rangle$, and its corresponding standard deviation obtained by the combinatorial expressions: solid lines for theoretical $\langle n_{s} \rangle$, Eq.~(\ref{ns_general}), and dashed lines for its standard deviation, Eq.~(\ref{std_dev_general}). Symbols represent data obtained by numerical simulations: black squares stand for $\langle n_{s} \rangle$, and open circles stand for {$\langle n_s \rangle \pm \sigma_s$}. The precision of the theoretical predictions significantly decreases after the transition point in gelling systems, as in the one with the product kernel, while it may be regarded as { excellent (though, formally approximate)} in the case of the additive kernel (see Tab.~\ref{tab_solved_kernels_precision} and Sec.~\ref{Exactness} for further comments). Numerical simulation results were averaged over $10^6$ independent runs. Inset figures: Variance divided by $\langle n_{s} \rangle$ (upper inset figures), $\langle n_{s} \rangle$ for higher $s$ (lower inset figures, left column), and $\langle n_{s} \rangle$ in semi-logarithmic scale (lower inset figures, right column). { Gray dashed lines in (a)--(d) correspond to the usual solutions from the Smoluchowski theory, see Appendix~\ref{SmolSols}. No explicit Smoluchowski solution for $\langle n_{s} \rangle$ is known for the LC kernel; for detailed considerations on the asymptotic scaling solutions for this case, see Refs.~\cite{Wattis_2009, Leyvraz_2021}.}}
\label{Fig_kernele_zbiorczy}
\end{figure*}

\begin{figure*}[h]
\includegraphics[width=0.98\textwidth]{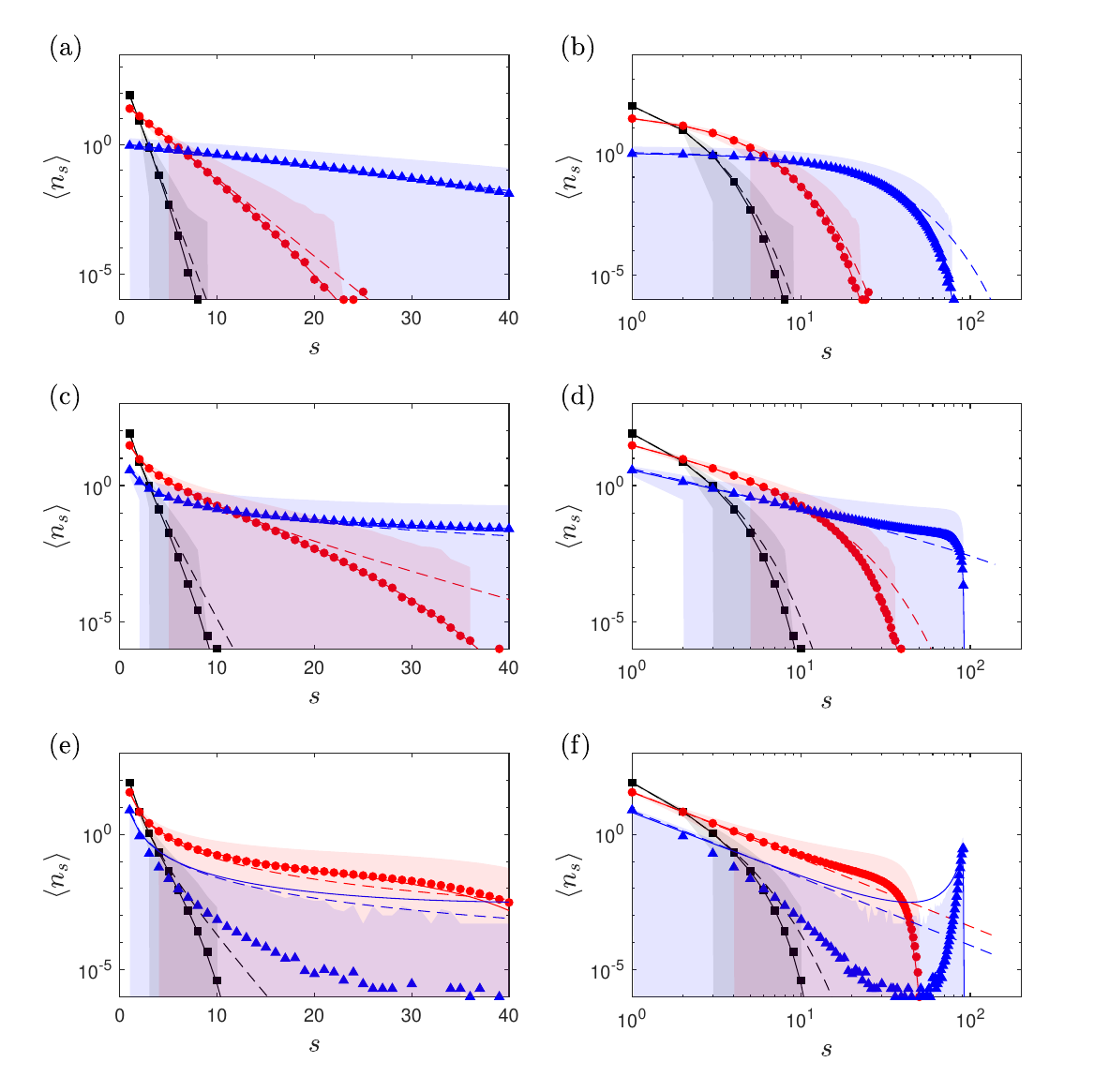}
\caption{  Theoretical predictions of average numbers of clusters of a given size, $\langle n_{s} \rangle$, obtained from the combinatorial approach and from the Smoluchowski theory, plotted against numerical simulations of a finite system of $N=100$ particles for the three basic kernels. (a), (b): The constant kernel. (c), (d): The additive kernel. (e), (f): The product kernel. Three data series are presented: an early stage of the evolution, $t=10$ ($\theta=t/N=0.1$, black); a half-time, $t=50$ ($\theta=0.5$, red); and a late stage of the evolution, $t=90$ ($\theta=0.9$, blue). Solid lines represent combinatorial solutions, dashed lines represent Smoluchowski solutions, and symbols represent data from numerical simulations. The left column presents log-linear plots, while the right column presents log-log plots. { Color shadows represent $\langle n_s \rangle \pm \sigma_s$ obtained from numerical simulation. Sharp drops in $\langle n_s \rangle - \sigma_s$ occur due to the logarithmic scale when $\langle n_s \rangle - \sigma_s \le 0$, while sharp drops in $\langle n_s \rangle + \sigma_s$ occur when $\sigma_s=0$ due to the limited resolution of the numerical results} (averaged over $10^6$ independent runs). { Blue series in (f) clearly show decreased precision of the approximate combinatorial solution for the product kernel in the post-gel phase, where the existence of the giant cluster is reproduced but with significant differences in the distribution of average cluster numbers with respect to the simulation. For a detailed discussion on the precision of the combinatorial results, see Sec.~\ref{Exactness}.} Respective times $\tilde{t}$ for the Smoluchowski theory were found empirically via Eq.~(\ref{time_S_time_comb}). For the details of the Smoluchowski solutions, see Appendix~\ref{SmolSols}.}
\label{Fig_AppE}
\end{figure*}

\subsection{Additive kernel, $K \propto s+g$}

For the additive kernel, we use Eq.~(\ref{xg_recursive_definition_general}) accordingly \cite{Lepek_2019},
\begin{equation} \label{EQ30} 
\begin{split}
 x_g=\frac{1}{2}\sum^{g-1}_{h=1}{\binom{g}{h}\binom{g-2}{h-1}}x_hx_{g-h} ( h+(g-h) ) \\ =\frac{1}{2} g \sum^{g-1}_{h=1}{\binom{g}{h}\binom{g-2}{h-1}}  x_hx_{g-h}.
\end{split}
\end{equation}

If we unfold binomial signs and use substitution, this time
\begin{equation} \label{EQ31}
y_g = \frac{x_g}{g! \left(g-1 \right)!},
\end{equation}

\noindent we obtain
\begin{equation} \label{EQ32} 
 {\left(g-1\right)y}_g=\frac{1}{2}g\sum^{g-1}_{h=1}{y_hy_{g-h}}.
\end{equation}

Eq.~(\ref{EQ32}) has the same form as in the case of the product kernel, though it differs in the substitution $y_g$. Thus, we use the previously found equation for the generating function (Eq.~(\ref{EQ27})) and compare it to the present substitution (Eq.~(\ref{EQ31})) to obtain
\begin{equation} \label{EQ32_2} 
 y_g=\frac{g^{g-1}}{g!}\ =\frac{x_g}{g!\left(g-1\right)!}.
\end{equation}

Finally, for the additive kernel, the number of ways to create a cluster of size $g$ reads
\begin{equation} \label{EQ33}  
 x_g=\left(g-1\right)! g^{g-1}.
\end{equation}


\begin{figure*} 
\includegraphics[width=0.8\textwidth]{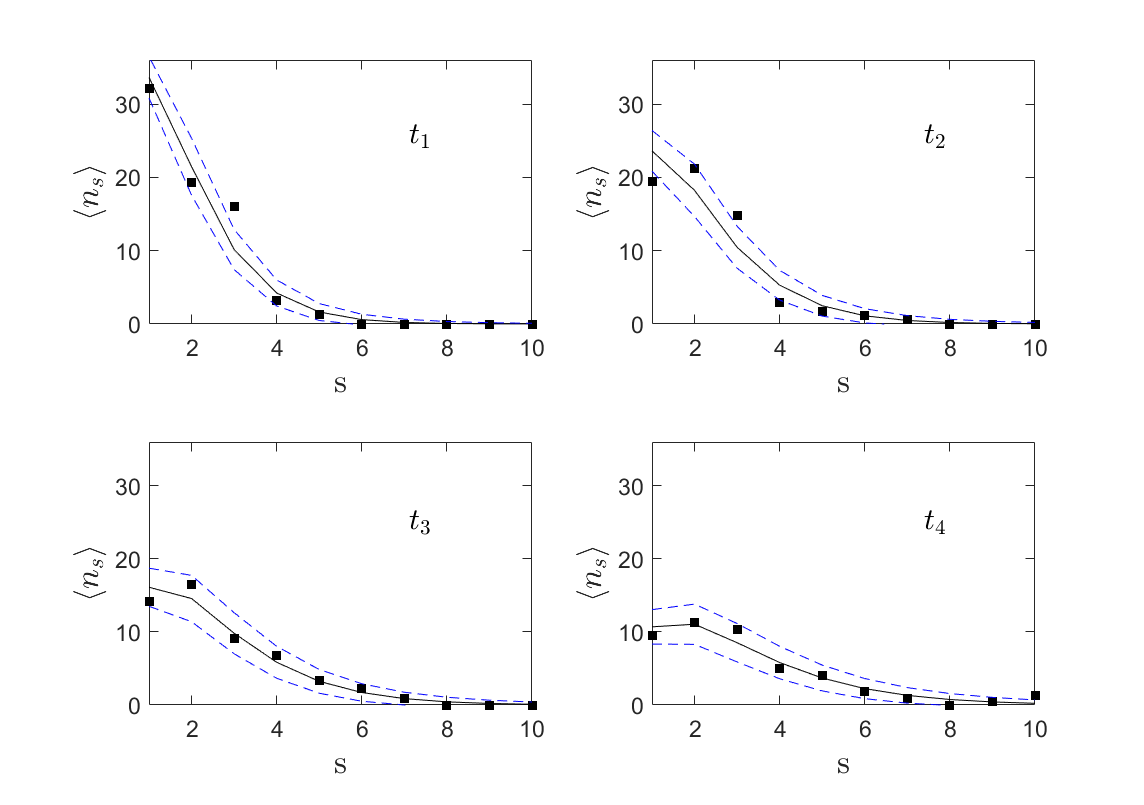}
  \caption{Electrorheological aggregation of particles into linear chains. Theoretical calculations of the average number of clusters of a given size $s$, $\left\langle n_s\right\rangle$, for the ER coagulation with $\alpha = 1$, Eq.~(\ref{ns_solved_final_simplified_alpha_1}), compared to the experimental data of coagulating polystyrene particles (see Refs.~\cite{Mimouni_2007,Wattis_2009,Lepek_2021} for details). Solid and dashed lines represent theoretical { approximations} of $\left\langle n_s\right\rangle$ and its standard deviation, respectively. Squares represent experimental data points. In this case, translation from the physical time to the time of aggregation acts has been performed by counting the number of clusters in the system in subsequent time points (cf. Sec.~\ref{Remarks_on_time}). The number of monomers in the system was estimated as $N = 138$. Four time steps of the process are presented: $t_1 = 1$ min ($t = 83$ in the time counted as binary aggregation acts), $t_2 = 3$ min ($t = 95$), $t_3 = 5$ min ($t = 107$), and $t_4 = 7$ min ($t = 117$). Reprinted from \cite{Lepek_2021}, with permission from Elsevier.}
  \label{LC_data_2021}
\end{figure*}

Combining $x_g$ with the expressions for $\langle n_s \rangle$, Eqs.~(\ref{omega_s}) and (\ref{ns_general}), we have
\begin{equation} \label{EQ44_additive} 
 {\langle n_s\rangle } =\binom{N}{s} s^{s-1}\frac{B_{N-s,k-1}\left(\left\{g^{g-1}\right\}\right)}{B_{N,k}\left(\left\{g^{g-1}\right\}\right)}.
\end{equation}

Analogously as for the constant kernel, using relation~(\ref{relacja_gg-1}) from Appendix A,  Eq.~(\ref{EQ44_additive}) can be simplified to the form of
\begin{equation} \label{EQ45} 
 {\left\langle n_s\right\rangle } = \binom{N}{s} s^{s-1}\frac{\binom{N-1-s}{k-2} (N-s)^{N-s-k+1}  }{ \binom{N-1}{k-1}  N^{N-k}},
\end{equation}

\noindent which does not contain Bell polynomials.

{ The above results for the product and additive kernels are evaluated against the numerical simulation in Fig.~\ref{Fig_kernele_zbiorczy}. Fig.~\ref{Fig_AppE} provides a more detailed comparison to their usual Smoluchowski counterparts.}

\subsection{Linear--chain kernel, $K \propto (s^{-1}+g^{-1})^{\alpha}$}




In this subsection, we will recall a kernel describing systems of particles that coagulate into linear chains (LC). A widely known example of such a process is an aggregation of the electrorheological (ER) fluid. This case may be of particular interest as the combinatorial predictions for this kernel were the only ones tested against experimental data \cite{Lepek_2021}. Most data points fell within the range defined by the theoretical standard deviation (cf. Fig.~\ref{LC_data_2021}).

ER fluids are colloidal suspensions of electrically active particles in insulating fluid, e.g., polyimide particles suspended in silica oil. Applying an external electric field to such a system forces the particles to behave as electric dipoles and, in effect, they aggregate (typically in milliseconds) to form linear chains oriented in the direction of the field (see, for instance, Ref.~\cite{film_semenova}). Although discovered in the 1940s, ER fluids are still researched \cite{Semenov_2021}. Analogous chains are observed in coagulation processes consisting of colloidal particles in the presence of an external magnetic field (magnetorheological materials) \cite{Dominguez_2007, Han_2010, Bossis_2013, Reynolds_2016, Kumar_2022}.

Kinetics of irreversible aggregation of systems with linear chains was the subject of research in several works \cite{Miyazima_1987, Fraden_1989, Melle_2001, Mimouni_2007, Wattis_2009}. In Ref.~\cite{Wattis_2009}, it has been shown that the process is governed by the kernel of the form
\begin{equation} \label{electrorheological_kernel}
K \propto \frac{1}{s}+\frac{1}{g},
\end{equation}

\noindent which is sometimes referred to as an inverse kernel.

The above kernel form found its solution in the combinatorial approach, surprisingly, even generalized with a real power \cite{Lepek_2021},
\begin{equation} \label{generalized_electrorheological_kernel}
K \propto \left( \frac{1}{s}+\frac{1}{g} \right)^\alpha = \left( \frac{s+g}{sg} \right)^\alpha.
\end{equation}

Modifying Eq.~(\ref{xg_recursive_definition_general}) for $x_g$ accordingly, we can write
\begin{equation} \label{Kgk_electro}
 K\left(g-h,h\right) = \left( \frac{h+(g-h)}{h(g-h)} \right)^\alpha = 
 \left( \frac{g}{h(g-h)} \right)^\alpha,
\end{equation}
\begin{equation} \label{xg_recursive_electro}
 x_g = \frac{1}{2}\sum^{g-1}_{h=1}\binom{g}{h}\binom{g-2}{h-1}x_hx_{g-h} \left( \frac{g}{h(g-h)} \right)^{\alpha}.
\end{equation}

Expanding binomial signs and substituting
\begin{equation} \label{yg_electro}
y_g=\frac{x_g}{g!(g-1)!g^\alpha}
\end{equation}

\noindent gives
\begin{equation} \label{electro_2} 
 (g-1)y_g=\frac{1}{2}\sum^{g-1}_{h=1}{y_hy_{g-h}}.
\end{equation}

Eq.~(\ref{electro_2}) took the same form as the analogous expression for the constant kernel (cf. Eq.~(\ref{EQ14})), with the only difference in defining $ y_g $. Hence, the previous solution may be used (cf. Eq.~(\ref{EQ21})),
\begin{equation} \label{yg_general_solution}  
 y_g=\frac{1}{2^{g-1}},
\end{equation}

\noindent which, after returning from the substitution, Eq.~(\ref{yg_electro}), results in
\begin{equation} \label{electro_3}  
 \frac{1}{2^{g-1}} = \frac{x_g}{g!g^\alpha\left(g-1\right)!}.
\end{equation}

Finally, we obtain explicit $ x_g $ for the generalized LC kernel as
\begin{equation} \label{electro_xg}  
 x_g = \frac{g!g!g^{\alpha-1}}{2^{g-1}}.
\end{equation}

What is worth noting is that for $ \alpha = 0 $, the above solution entirely agrees with the solution for the constant kernel, Eq.~(\ref{xg_binomials}). A typical ER aggregation process is described by $ x_g $ with $ \alpha = 1 $.

The final expression for the average number of clusters of a given size for the LC kernel takes the form
\begin{equation} \label{ns_LC_solved_final}  
 \left\langle n_s\right\rangle =\binom{N}{s} \frac{ s!s^{\alpha} }{ 2^{s-1} } \frac{ B_{N-s,k-1}\left(  \left\{ \frac{ g!g^{\alpha} }{ 2^{g-1} } \right\}   \right) }{ B_{N,k}\left(\left\{  \frac{ g!g^{\alpha} }{ 2^{g-1} }  \right\}\right)},
\end{equation}

\noindent where, of course, $g$ varies from $1$ to $N-k+1$.

The above expression is another example of a solution that may be simplified through relation~(\ref{relacja_abz}) for Bell polynomials from Appendix A to the form of
\begin{equation} \label{ns_solved_final_simplified}  
 \left\langle n_s\right\rangle =\binom{N}{s}  s!s^{\alpha}  \frac{ B_{N-s,k-1}\left(  \left\{  g!g^{\alpha} \right\}   \right) }{ B_{N,k}\left(\left\{  g!g^{\alpha}   \right\}\right)}.
\end{equation}

Moreover, further simplification to the form with no Bell polynomials may be found for the typical ER process ($ \alpha = 1 $) by using the relation with the so-called falling factorial, see Eq.~(\ref{relacja_falling}), as \cite{Lepek_2021}
{
\begin{equation} \label{ns_solved_final_simplified_alpha_1} 
 \left\langle n_s  \right\rangle _{\alpha=1} =
\frac{  2sk(k-1)(2k-1)(N-k)!(N-s+k-2)!  }{  (N-s-k+1)!(N+k-1)!  }.
\end{equation}
}

{ The above result is evaluated against the numerical simulation in Figs.~\ref{Fig_kernele_zbiorczy}e and \ref{Fig_kernele_zbiorczy}f.}

As mentioned before, in Ref.~\cite{Lepek_2021}, a comparison between theoretical { approximations}, Eq.~(\ref{ns_solved_final_simplified_alpha_1}), and experimental data from Ref.~\cite{Wattis_2009} has been performed, and a method to translate points in physical timeline to the points in coagulation acts timeline has been presented, a method which was mentioned in Sec.~\ref{Remarks_on_time}, i.e., by counting the number of clusters in particular moments (as $k$ clearly defines the moment of aggregation). The experiment was well-suited for such analysis as it started from monodisperse initial conditions. It was observed that the vast majority of the points fell inside the range defined by the combinatorial estimate of the standard deviation given by Eq.~(\ref{std_dev_general}) as presented in reproduced Fig.~\ref{LC_data_2021}.


\subsection{Condensation kernel, $K \propto (A+s)(A+g)$}

In the case of the condensation (or branched-chain polymerization) process, a kernel may be described as \cite{Aldous_1999}
\begin{equation} \label{condensation_kernel}
K \propto (s+A)(g+A),
\end{equation}

\noindent where $A \ge 0$ stands for an arbitrary process-dependent parameter. This kernel may be regarded as a generalization of the product kernel with $A$ being a factor that relatively decreases the rate of the process for larger particles, thus preventing the creation of a giant cluster (gel).

Condensation was traditionally researched in aerosols growth \cite{Walter_1973, Wagner_1982} and water vapor condensation \cite{Long_1974, Gillespie_1975, Dziekan_2017, Liu_2022} but also in, e.g., supersonic flows \cite{Wegener_1954,Yang_2017} or biology \cite{Spouge_1983, Baskakov_2007, Fang_2011}. Several works were devoted to the physical properties of branched-chain polymers \cite{Gray_2009, Vilaplana_2010, Marshall_2013}.

An explicit solution to the condensation kernel, Eq.~(\ref{condensation_kernel}), was found in Ref.~\cite{Lepek_2021_ROMP}. Analogously as for previous kernels, Eq.~(\ref{xg_recursive_definition_general}) was rewritten as
\begin{equation} \label{xg_recursive_condensation}
 x_g=\frac{1}{2}\sum^{g-1}_{h=1}\binom{g}{h}\binom{g-2}{h-1}x_hx_{g-h} \left( h+A \right) \left( g-h+A \right).
\end{equation}

The strategy of solving Eq.~(\ref{xg_recursive_condensation}) to obtain an explicit expression for $x_g$ is based, again, on the substitution, generating function method, and {(this time)} Lagrange inversion {(see Appendix A)}. Detailed derivations may be found in Ref.~\cite{Lepek_2021_ROMP}. Here, we will only present the final result of that work, being (a little more involved) explicit expression for $x_g$,
\begin{equation} \label{xg_final_cond}
x_g = \frac{(A+1)^g (g-1)!}{2^{g-1}(A+g)} [2g+gA]^{*},
\end{equation}

\noindent where asterisk denotes a product of the form
\begin{equation} \label{asterisk_function_cond}
[2n+nA]^{*}=\left\{
\begin{array}{ccc}
1 & \mbox{for} & n=1, \\
\prod^{n}_{m=2}{(2n+mA)} & \mbox{for} & n \ge 2,
\end{array}
\right.
\end{equation}

{
\noindent which can be expressed in terms of the Gamma function,
\begin{equation} \label{asterisk_function_gamma}
[2n+nA]^{*} = \frac{A^n}{2n+A} \frac{\Gamma(n+1+2n/A)}{\Gamma(1+2n/A)}.
\end{equation}
}

The final expression for the average number of clusters of a given size for the condensation kernel is defined, of course, by Eqs.~(\ref{omega_s}), (\ref{ns_general}) and (\ref{xg_final_cond}). Taking advantage, again, of a relation for Bell polynomials (see Eq.~(\ref{relacja_abz}) in Appendix A), we can simplify this set of equations to the form of
\begin{equation} \label{ns_general_final_simplified}  
 \left\langle n_s\right\rangle =\binom{N}{s} \frac{[2s+sA]^{*}}{(A+s)} \frac{B_{N-s,k-1}\left(\left\{ \frac{[2g+gA]^{*}}{A+g} \right\}\right)}{B_{N,k}\left(\left\{\frac{[2g+gA]^{*}}{A+g} \right\}\right)}.
\end{equation}

\subsection{Combination of constant and additive kernels, $K=A+s+g$}

A sum of the constant and additive kernels, $K=A+s+g$, may be regarded as a special case of Spouge's universal kernel, $K=A+B(s+g)+Csg$ \cite{Spouge_1983b}. In this case, Eq.~(\ref{xg_recursive_definition_general}) was modified as
\begin{equation}
\begin{split}
 x_g = \frac{1}{2}\sum^{g-1}_{h=1}\binom{g}{h}\binom{g-2}{h-1}x_hx_{g-h} \left( A + g \right) \\ =\frac{1}{2} (A+g) \sum^{g-1}_{h=1}{\binom{g}{h}\binom{g-2}{h-1}}  x_hx_{g-h}.
\end{split}
\end{equation}

Luckily, the above expression leads to the same equations as for the condensation kernel (cf. Ref.~\cite{Lepek_2021_ROMP}) with the only difference in the substitution. Thus, it has been instantaneously found that 
\begin{equation} \label{xg_final_comb}
x_g = \frac{(g-1)!}{2^{g-1}} [2g+gA]^{*},
\end{equation}

\noindent where the asterisk denotes, the same as previously, the product given by Eq.~(\ref{asterisk_function_cond}).

Again, using the relation~(\ref{relacja_abz}), the set of Eqs.~(\ref{omega_s}), (\ref{ns_general}), and (\ref{xg_final_comb}) has been simplified to the form of
\begin{equation} \label{ns_general_final_comb_simplified}  
 \left\langle n_s\right\rangle =\binom{N}{s} [2s+sA]^{*} \frac{B_{N-s,k-1}\left(\left\{ [2g+gA]^{*} \right\}\right)} {B_{N,k}\left(\left\{[2g+gA]^{*} \right\}\right)}.
\end{equation}

Interestingly, the above expression seems to give exact results when compared to the numerical simulation \cite{Lepek_2021_ROMP}. For further comment, see Table~\ref{tab_solved_kernels_precision} and Sec.~\ref{Exactness}.

\begin{table*}[t]
\setlength\extrarowheight{2.5pt}
\setlength{\tabcolsep}{10pt}
\begin{tabular}{  l  l  l  l l }
\hline \hline
\rule{0pt}{15pt}
$K\propto$ & $y_g$ & $\omega_g=\frac{x_g}{(g-1)!}$ & Simplified $\langle n_s \rangle$ & Ref. \\[7pt]
\hline \hline
\rule{0pt}{15pt}
$1$ & $\frac{x_g}{g!(g-1)!} $ & $\frac{g!}{2^{g-1}}$ & $k\frac{\binom{N-s-1}{k-2}}{\binom{N-1}{k-1}}$  & \cite{Fronczak_2018} \\[7pt]
\hline
\rule{0pt}{15pt}
$sg$ & $\frac{x_g}{(g-1)! (g-1)!} $ & $g^{g-2}$ & --- & \cite{Lepek_2019} \\[7pt]
\hline
\rule{0pt}{15pt}
$s+g$ & $\frac{x_g}{g!(g-1)!}$ & $g^{g-1}$ & $\binom{N}{s} s^{s-1}\frac{\binom{N-1-s}{k-2} (N-s)^{N-s-k+1}  }{ \binom{N-1}{k-1}  N^{N-k}}$ & \cite{Lepek_2019} \\[7pt]
\hline
\rule{0pt}{15pt}
$\left( s^{-1}+g^{-1} \right)^\alpha$ & $\frac{x_g}{g!(g-1)!g^\alpha}$ & $\frac{g!g^{\alpha}}{2^{g-1}}$ & $\binom{N}{s}  s!sk  \frac{  \sum^{k-1}_{j=0} (-1)^{k-1-j} \binom{k-1}{j} (j+k+N-2)_N  }{  \sum^{k}_{j=0} (-1)^{k-j} \binom{k}{j} (j+k+N-1)_N }$ & \cite{Lepek_2021} \\[7pt]
\hline
\rule{0pt}{15pt}
$(s+A)(g+A)$ & $\frac{x_g(g+A)}{g!(g-1)!}$ & $\frac{(A+1)^g}{2^{g-1}(A+g)} [2g+gA]^{*}$ & --- & \cite{Lepek_2021_ROMP} \\[7pt]
\hline
\rule{0pt}{15pt}
$A+s+g$ & $\frac{x_g}{g!(g-1)!}$ & $\frac{1}{2^{g-1}} [2g+gA]^{*}$ & --- & \cite{Lepek_2021_ROMP}
\\[7pt]
\hline
\rule{0pt}{15pt}
$(s+g)^2(sg)^{-1}$ & $\frac{x_g}{g!(g-1)!g}$ & $g^g $ & --- & \cite{Lepek_thesis_2021}  \\[7pt]
\hline
\rule{0pt}{15pt}
$sg\left(1+(s+g)^{-1}\right)$ & $\frac{x_g}{\left(g-1 \right)! \left(g-1 \right)!} $ & $\frac{(3g)!}{2^{g-1}g(2g+1)!}$ & --- & \cite{Lepek_thesis_2021} \\[7pt]
\hline 
\end{tabular}
\caption{Kernels solved in the combinatorial approach, i.e., where an explicit expression for $x_g$ was found. In the second column, substitutions $y_g$ are presented. The third column shows $\omega_g = x_g/(g-1)!$, the only element characteristic of the given kernel in the expression for $\langle n_s \rangle$, Eq.~(\ref{ns_general}). A simplified form (with no Bell polynomials) of $\langle n_s \rangle$ is given in the fourth column if only such a simplification was obtained. Symbol $[.]^*$ stands for a product given by Eq.~(\ref{asterisk_function_cond}).}
\label{tab_solved_kernels_forms}
\end{table*}

\subsection{Other kernels solved}

Besides the kernels presented in the previous subsections, solutions to two more processes have been found in the combinatorial approach \cite{Lepek_thesis_2021}.

The first one is
\begin{equation} \label{K_antysoc}
    K \propto \frac{(s+g)^2}{sg}.
\end{equation}

Its formulation is similar to the linear-chain kernel (which may be presented in the form of $K=(s+g)/(sg)$), but, in this case, the contribution of the additive part in the numerator is squared. In this kernel, when $i$ and $j$ are close, its value $K$ is relatively low, while significantly different values of $i$ and $j$ result in higher $K$ (disassortativity).

Applying analogous steps as for previous kernels, we obtain $\omega_g = g^g$, and the expression for $\langle n_s \rangle$ takes the form
\begin{equation}
 \left\langle n_s\right\rangle =\binom{N}{s} s^s \frac{B_{N-s,k-1}(\{ g^g \})} {B_{N,k}(\{g^g \})}.
\end{equation}

A family of electrorheological kernels may be generally defined as $(i+j)^\alpha(ij)^{-\beta}$. Unfortunately, for such a form, though a substitution may be found,
\begin{equation}
y_g=\frac{x_g}{g^\beta g!(g-1)!},
\end{equation}

\noindent a problem with obtaining a differential equation for the generating function $G(z)$ immediately occurs.

The last kernel solved in the combinatorial approach is
\begin{equation} \label{K_dziwny}
    K \propto sg \left( 1 + \frac{1}{s+g} \right).
\end{equation}

In the case of  Eq.~(\ref{K_dziwny}), after applying analogous steps as in previous cases and using relation~(\ref{relacja_abz}) for Bell polynomials, we obtain
\begin{equation} \label{ns_final_dziwny}  
 \left\langle n_s\right\rangle =\binom{N}{s} \frac{(3s)!}{s(2s+1)!} \frac{B_{N-s,k-1}\left(\left\{ \frac{(3g)!}{g(2g+1)!} \right\}\right)} {B_{N,k}\left(\left\{ \frac{(3g)!}{g(2g+1)!}  \right\}\right)}.
\end{equation}

\begin{figure*}[]
\includegraphics[width=1\textwidth]{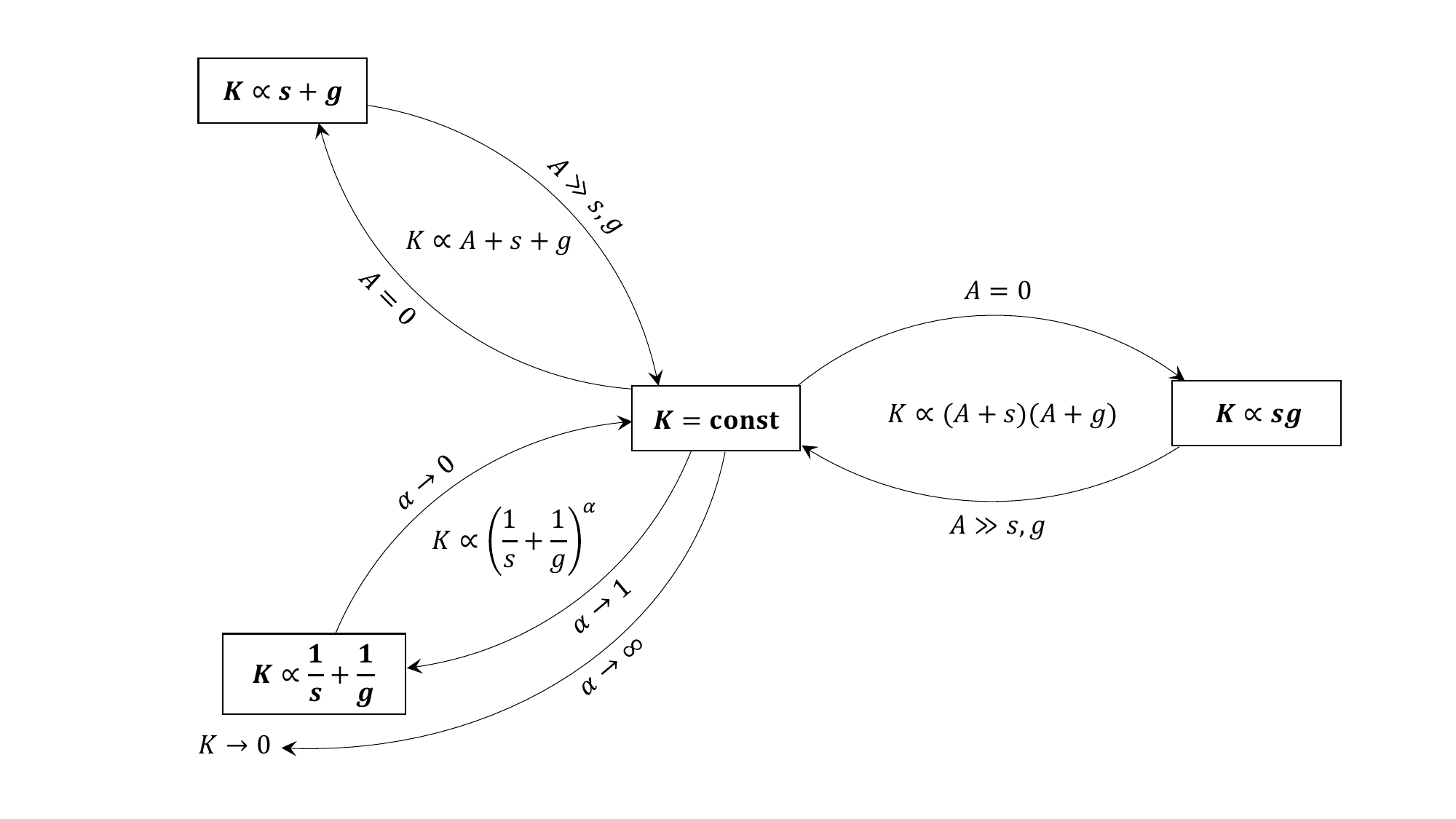}
\caption{Some kernels solved in the combinatorial approach may be transformed with the parameter $A$ to the form corresponding to the constant kernel. These are: linear-chain, $K \propto \left( s^{-1}+g^{-1} \right)^{\alpha}$, condensation, $K \propto (A+s)(A+g)$, and linear combination of constant and additive kernels, $K \propto A+s+g$. In the case of the latter two, by taking $A=0$, we obtain basic kernels, product and additive, respectively. In this way, explicit solutions to all ``intermediate'' processes between the product, additive, linear-chain, and the constant kernel were obtained. { Of course, the above picture is only a schematic representation: exact dynamics throughout the aggregation process depend on the current values of $A$, $s$, $g$, and the system size. For instance, for $K \propto A+s+g$, if $A$ is large, there is an early stage when the system is dominated by clusters of size $s \ll A$, and the dynamics are dominated by $K \propto A$. Later, when the average cluster size is $s \simeq A$, all terms in $K\propto A+s+g$ are equally significant. At later times, when the average cluster size is $s \gg A$, the dynamics are dominated by the term $K \propto s+g$, which may be observed for sufficiently large $N$.} For the linear-chain kernel, the branch ``$\alpha \to \infty$'' symbolically represents processes with $\alpha > 1$, for which increasing $\alpha$ lowers the relative probability of merging for large clusters. {Precision of the theoretical predictions for the above-mentioned kernels varies {from exact to approximate}, see Sec.~\ref{Exactness} for details.}}
 \label{Figure_Podsumowanie}
\end{figure*}

\subsection{Summary of solved kernels}

\begin{table*}
\setlength\extrarowheight{2.5pt}
\setlength{\tabcolsep}{10pt}
\begin{tabular}{  l l l  }
\hline \hline
$K \propto$ & $y_g$ & Equation for $G(z)$ \\[2.5pt]
\hline \hline
$(s+g)^{-1}+A$ & $\frac{x_g}{g!(g-1)!}$ & --- \\[2.5pt]
\hline
$(A+s+g)^{-1}$ & $\frac{x_g}{g!(g-1)!}$ & $z^2G''+AzG'-AG=\frac{1}{2}G^2$ \\[2.5pt]
\hline
$(A+s+g)^\alpha+B$ & $\frac{x_g}{g!(g-1)!}$ & --- \\[2.5pt]
\hline
$(s+A)^\alpha(g+A)^\alpha$ & $\frac{x_g(g+A)^\alpha}{g!(g-1)!}$ & --- \\[2.5pt]
\hline
$(sg)^{-1}$ & $\frac{x_g}{g!g!}$ & $z^2G''-\frac{1}{2}G^2=0$ \\[2.5pt]
\hline
$(A+sg)^{-1}$ & --- & --- \\[2.5pt]
\hline
$(s+g)^\alpha(sg)^{-\beta}$ & $\frac{x_g}{g^\beta g!(g-1)!}$ & --- \\[2.5pt]
\hline
$A+B(s+g)+sg$ & --- & --- \\[2.5pt]
\hline
$(s-g)^2 (s+g)^{-1}$ & --- & --- \\[2.5pt]
\hline
$s^\alpha \pm g^\beta$ & --- & --- \\[2.5pt]
\hline
$(s^\alpha+g^\alpha)(s^{-\beta}+g^{-\beta})$ & --- & --- \\[2.5pt]
\hline
$(|s-g|+1)^{\pm 1}$ & --- & --- \\[2.5pt]
\hline 
\end{tabular}
\caption{Kernels for which an attempt to find an explicit solution in the combinatorial approach was undertaken, but the final solution was not obtained. For some of the kernels, substitutions $y_g$ were found, but it was not possible to obtain solvable differential equations defining the generating function $G(z)$ \cite{Lepek_thesis_2021}. $A,B \ge 0$ stand for constants, $\alpha$ and $\beta$ are real positive numbers. Further description in the text.}
\label{tab_nierozwiazane}
\end{table*}

In the above sections, we have reviewed known solutions (i.e., explicit expressions for $x_g$, thus, for the average number of clusters of a given size) to basic kernels (constant, additive, product) as well as to more complex ones, such as condensation or generalized linear-chain kernels. The summary of those results may be found in Table~\ref{tab_solved_kernels_forms}, where kernels and their corresponding substitutions and omegas, $\omega_g = x_g / (g-1)!$, are listed together with the simplified versions (without Bell polynomials) of the expression for $\langle n_s \rangle$, where applicable. Up to this point, we have shown the methods used to obtain explicit solutions in the combinatorial framework (generating function method, Lagrange inversion). Nevertheless, only a limited set of kernels is susceptible to the analysis presented above.

Several kernels solved in the combinatorial approach may be transformed into the forms of the basic kernels using their respective parameters. These are: linear-chain kernel, $K=\left( s^{-1}+g^{-1} \right)^{\alpha}$ for $\alpha=0$; condensation kernel for $A=0$, $K=(A+s)(A+g)$; and a linear combination of the constant and additive kernels for $A=0$ as well, $K=A+s+g$. Of course, the precision of the theoretical predictions depends on these parameters (at least for the first two cases); however, in this way, a spectrum of intermediate processes was covered (see Fig.~\ref{Figure_Podsumowanie}).

{ For an extensive discussion on the precision of the theoretical solutions, please see Section~\ref{Remarks}.}

\subsection{Problems when solving other forms}

Kernels for which an attempt to find an explicit solution in the combinatorial approach was undertaken, but such a solution was not finally obtained \cite{Lepek_thesis_2021} are listed in Table~\ref{tab_nierozwiazane}. Here, we will comment briefly on some of them.

For $K\propto(A+s+g)^{-1}$ and $K\propto(sg)^{-1}$, although a nonlinear second order differential equation for generating function $G(z)$ was obtained, its only solution was $G=0$, leading to a contradiction. A particular form of the first kernel, $K\propto(1+s+g)^{-1}$, was described in Ref.~\cite{Calin_2006} as analytically intractable (which is not completely true, as we will see in Sec.~\ref{NoExplicitSolution}).

For some kernel families, finding a substitution $y_g$ was possible. However, the next step of transformations---as for now---has failed. Among those kernels, one can highlight: a sum of basic kernels, $K\propto A+B(s+g)+sg$; analytical approximation of Berry's kernel, $K\propto(s-g)^2 (s+g)^{-1}$ \cite{Aldous_1999}; real-power kernel, $K\propto s^\alpha+g^\beta$; general form given by Fournier and Lauren\c{c}ot, $K\propto(s^\alpha+g^\alpha)(s^{-\beta}+g^{-\beta})$ \cite{Fournier_2005}; kernel with modulus, $K\propto(|s-g|+1)^{\pm 1}$.

The last kernel, $K\propto(|s-g|+1)^{\pm 1}$, may be interesting as for the negative power, $K\propto(|s-g|+1)^{-1}$, it takes an assortative form where similar-size particles react with each other with higher probability. In contrast, for the positive power, $K\propto|s-g|+1$, it takes a disassortative form where similar-size particles are less probable to react. Kernels containing functions as the modulus were traditionally difficult for analytical treatment.

However, as we will see in Sec.~\ref{NoExplicitSolution}, we are not entirely defenseless against the above-mentioned arbitrary kernels, even against those including modulus.

\section{When no explicit solution is found}\label{NoExplicitSolution}

\subsection{Using recursive expression for $x_g$}

In the previous section, it was demonstrated that for a number of kernels, including kernels with arbitrary parameters, it was possible to transform related expressions for $x_g$ to the explicit forms without recursion. Such forms allowed us to obtain explicit expressions for $\left\langle n_s \right\rangle$ for these kernels. However, those transformations required substitutions and the use of a generating function method, which cannot be easily applied to any arbitrary kernel form.

In this section, we will take advantage of the fact that any number $x_s$ (or sequence $\left\{ x_g \right\}$) may be calculated using the recursive relation, Eq.~(\ref{xg_recursive_definition_general}). Of course, using the recursive relation decreases susceptibility to further simplification of the final expression or analysis of the final solution. However, analytical possibilities are also limited due to the use of Bell polynomials, which are hard to use for, e.g., asymptotic analysis. Moreover, in practice, values of partial Bell polynomials used to obtain theoretical predictions in Eq.~(\ref{ns_general}) are calculated using a recursive relation (see Appendix A). 

Using the recursive relation, Eq.~(\ref{xg_recursive_definition_general}), requires a starting value, $x_1$, which may be interpreted as the number of possible internal states of a cluster of size $1$. As for all of the previous kernels, we will use $x_1=1$, which is a reasonable guess and has worked well up to this point. A given kernel $K$ used in Eq.~(\ref{xg_recursive_definition_general}) defines further numbers in the sequence, $x_2$, $x_3$, and so on.

In this way, Eqs.~(\ref{ns_general}), (\ref{std_dev_general}), and (\ref{xg_recursive_definition_general}) unambiguously define a theoretical solution {(an approximation in general)} to the average number of clusters of a given size and its corresponding standard deviation for \textit{any} arbitrary kernel. In the following two subsections, to give a better picture of usability, we will apply these equations to obtain theoretical predictions { (approximations)} for ``unconstrained'' aggregation kernels known from the fields of aerosols and cosmology (namely, growth of aerosols and planetesimals).

As before, to describe a point in time of the aggregation process concisely, we will use $\theta = t / N$ throughout this section.

The code implementing Eq.~(\ref{xg_recursive_definition_general}) and used to produce the results below can be found at GitHub \cite{cpp_libraries}.

\subsection{Example 1: Protoplanetary aggregation} \label{example_proto}

The formation of planets is a complex process that still needs to be better understood. In general, planet formation may be regarded as consisting of several stages. First, micrometer-sized cosmic dust grains aggregate into pebbles (mm/cm), then the creation of planetesimals (10--100 km) occurs. Planetesimals are solid objects thought to exist in protoplanetary disks. They can efficiently accrete leftover pebbles and directly form the cores of planets, i.e., protoplanets of sizes greater than 1000 km, which may later evolve into terrestrial planets or gas giants \cite{Levison_2015, Liu_2020}. The current approach to studying cosmic matter aggregation uses $N$-body simulations as in Ref.~\cite{Levison_2015}.

In this example, we will focus on the creation and evolution of planetesimals. Recent studies suggest that our Solar System may have formed from rings of planetesimals created by pressure bumps rather than from a continuous disk \cite{Izidoro_2022}. In Ref. \cite{Ricard_2017}, some aggregation kernels were proposed as rough approximations of the planetesimal growth for subsequent stages of the aggregation. 

The rate at which a small (proto)planetesimal ($R<10$ km) can grow is proportional to its mass to the 2/3 power. This suggests a coagulation kernel of the form
\begin{equation} 
 K(s,g) \propto \left(  s+g \right) ^{2/3} 
\end{equation}

\noindent as a reasonable approximation for the first phase of the coagulation \cite{Wetherill_1990}. Since the kernel has a power of 2/3, which is smaller than 1, this initial phase does not lead to the formation of a runaway planetary embryo.

\begin{figure} 
\includegraphics[scale=0.55]{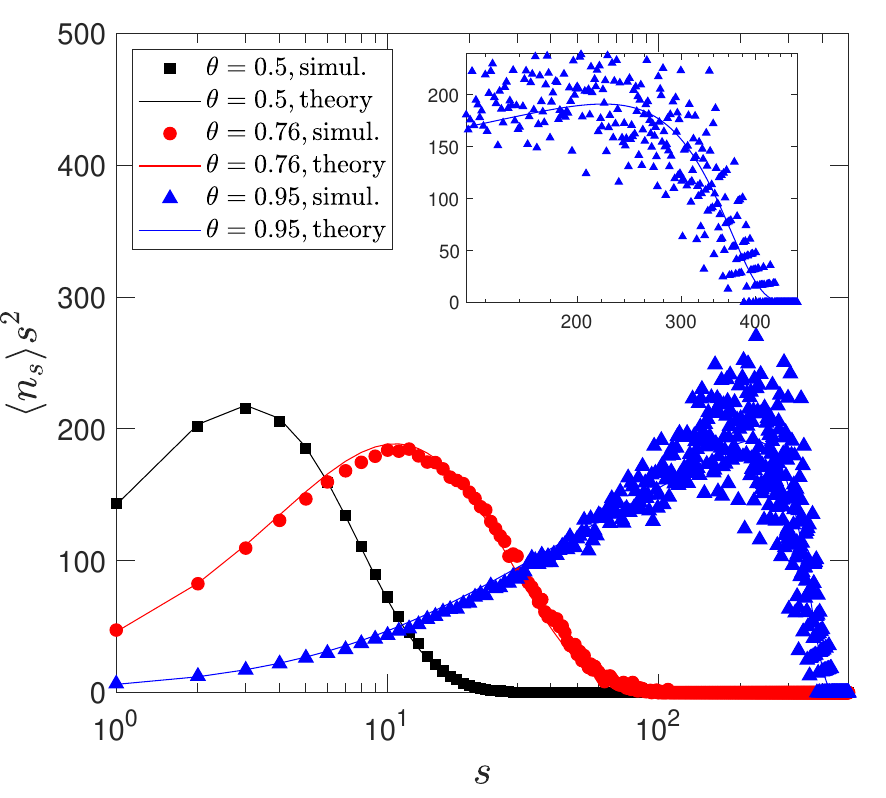}
\caption{Protoplanetary aggregation. An average number of clusters of a given size, $\langle n_{s} \rangle$, multiplied by a squared size for the gravitational aggregation kernel, Eq.~(\ref{icarus_gravitational}) with $\mu=60$, compared to numerical simulations. Squares, circles, and triangles represent numerical simulation results{, averaged over $10^4$ independent runs}; solid lines represent theoretical { approximations}, Eq.~(\ref{ns_general}). Three different stages of the aggregation are presented: half-time ($\theta=0.5$), late stage ($\theta=0.76$), and very late stage ($\theta=0.95$). The system consisted initially of $N=500$ monomers. Inset figure: Zoom to the part of the plot for the highest $s$. Further description in the text.}
\label{Figure_protoplanetary}
\end{figure}

\begin{figure} 
\includegraphics[scale=0.55]{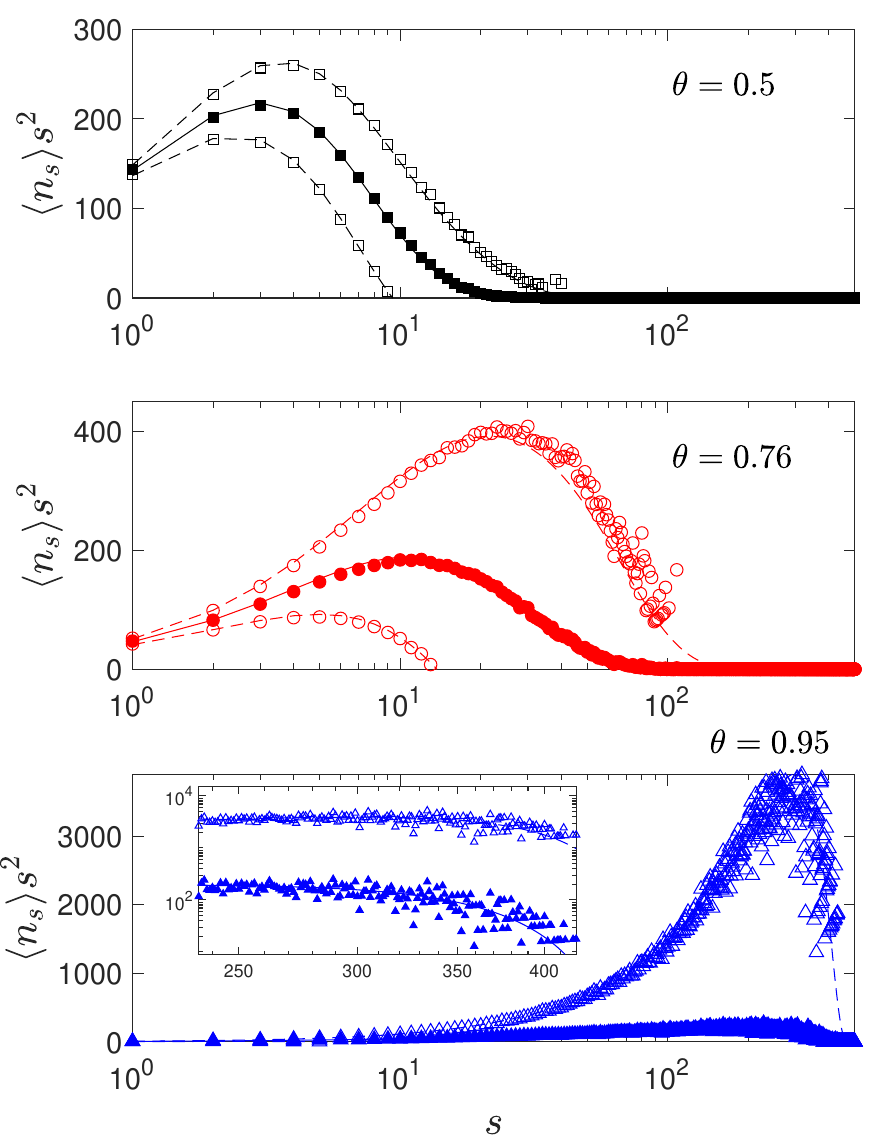}
\caption{Protoplanetary aggregation. Average number of clusters of given size, $\langle n_{s} \rangle$, multiplied by $s^2$, and its standard deviation for the gravitational aggregation kernel, Eq.~(\ref{icarus_gravitational}), compared to the results of numerical simulations for $\theta=0.5$ (top), $\theta=0.76$ (center), and $\theta=0.95$ (bottom). Squares, circles, and triangles represent the results of the numerical simulation {(means)}; lines represent theoretical { approximations}, Eq.~(\ref{ns_general}) and Eq.~(\ref{std_dev_general}). Open symbols and dashed lines represent {$(\langle n_s \rangle \pm \sigma_s)s^2$} obtained numerically and theoretically, respectively. The system consisted initially of $N=500$ monomers, $\mu=60$. Numerical results were averaged over $10^4$ independent runs. Inset in the lower figure: Zoom to the part of the plot for the highest $s$. Theoretical { approximations} follow the shape of the cluster size distribution and its standard deviation, even for the latest stage of the process, and include an extreme rise of the standard deviation. {For $\theta=0.95$, {$\langle n_s \rangle - \sigma_s$} is negative for any $s$; thus, it is not plotted as it has no meaningful sense. The interpretation is that, for any given $s$, {in a single realization,} we can expect to find no clusters of that size at all.}}
\label{Figure_protoplanetary_stddev}
\end{figure}

In the second stage of coagulation, which involves gravitational focusing, a large planetesimal orbits the proto-Sun. It has been shown that a coagulation kernel of the form
\begin{equation} 
 K(s,g) \propto \left(  s+g \right) ^{4/3} 
\end{equation}

\noindent may give a reasonable proxy for this phase of coagulation. This phase leads to a runaway growth of planetesimals, forming a limited number of planetary embryos of lunar or Martian masses \cite{Ricard_2017}.

To describe the runaway formation of planetesimals in which gravitational self-attraction becomes important, the aggregation kernel of the form
\begin{equation} \label{icarus_gravitational}  
 K(s,g) \propto \left(  (s+g)^{2/3} + \frac{(s+g)^{4/3}}{(2\mu)^{2/3}}  \right)
\end{equation}

\noindent was used. Such a formulation accounted for the enhanced cross-sections of the planetesimals of masses larger than $\mu$. The ``runaway'' planetesimals, i.e., the embryos of the protoplanets, were observed for the latest stage of the evolution as large deviations in the size distribution (cf. Fig. 9 in Ref.~\cite{Ricard_2017}). However, the exact magnitude of these deviations could not be determined as significant statistics would be required.

In Fig.~\ref{Figure_protoplanetary}, we present the average number of clusters of given size $s$ (weighted by $s^2$) for the kernel defined by Eq.~(\ref{icarus_gravitational}) with $\mu=60$ for the system consisting of 500 monomers. Theoretical predictions { (approximations)} obtained with Eq.~(\ref{ns_general}) are tested vs. direct numerical simulations averaged over $10^4$ runs. The number of monomers is significantly decreased when compared to the original study, which is due to the computational effort; however, the conclusions remain valid. Three different phases of the evolution are considered: half-time ($\theta=0.5$), a late stage ($\theta=0.76$), and a very late stage of the process ($\theta=0.95$). For the latter case and for $s>180$, the runaway embryos are noticeable in both theoretical and numerical results by increasing deviation from the mean. What is worth noting is the fact that the combinatorial approach follows the mean value of the data almost with perfect precision.

This significant rise in the deviation is further shown in Fig.~\ref{Figure_protoplanetary_stddev} where standard deviation from the mean is plotted for both numerical results and theoretical {approximations}, Eq.~(\ref{std_dev_general}), for the three stages of evolution mentioned before. The combinatorial approach correctly predicts the deviation, which increases with time to explode at the moment when the first embryos are born, suggesting that observing the standard deviation may be critical when predicting the creation of the protoplanetary cores. Such estimates were not achievable with the previous aggregation frameworks (Smoluchowski, Marcus--Lushnikov).

\subsection{Example 2: Growth of aerosols} \label{example_aerosols}

The formation of aerosols, small particles or liquid droplets suspended in a gas, is a complex process influenced by various physical and chemical mechanisms. Aerosols play a pivotal role in the atmosphere, affecting climate, air quality, and weather forecasting. They are used in several industries, including pharmaceutics and food processing \cite{Archer_2020}.

Atmospheric aerosols are the suspensions of particles in the air, ranging in diameter between approximately 1 nm, in the case of molecular particles, to 100 \textmu m, in the case of cloud droplets and dust particles \cite{Friedlander_2000}. Aerosols are subject to complex processes that influence their size distribution over time. Among these, two processes are usually studied within aggregation frameworks: (i) condensation, where nuclei serve as a substrate for the growth by adding further gas molecules, and (ii) coagulation, when aerosol particles undergo collisions, forming larger particles.

\begin{figure*} 
\includegraphics[width=1\textwidth]{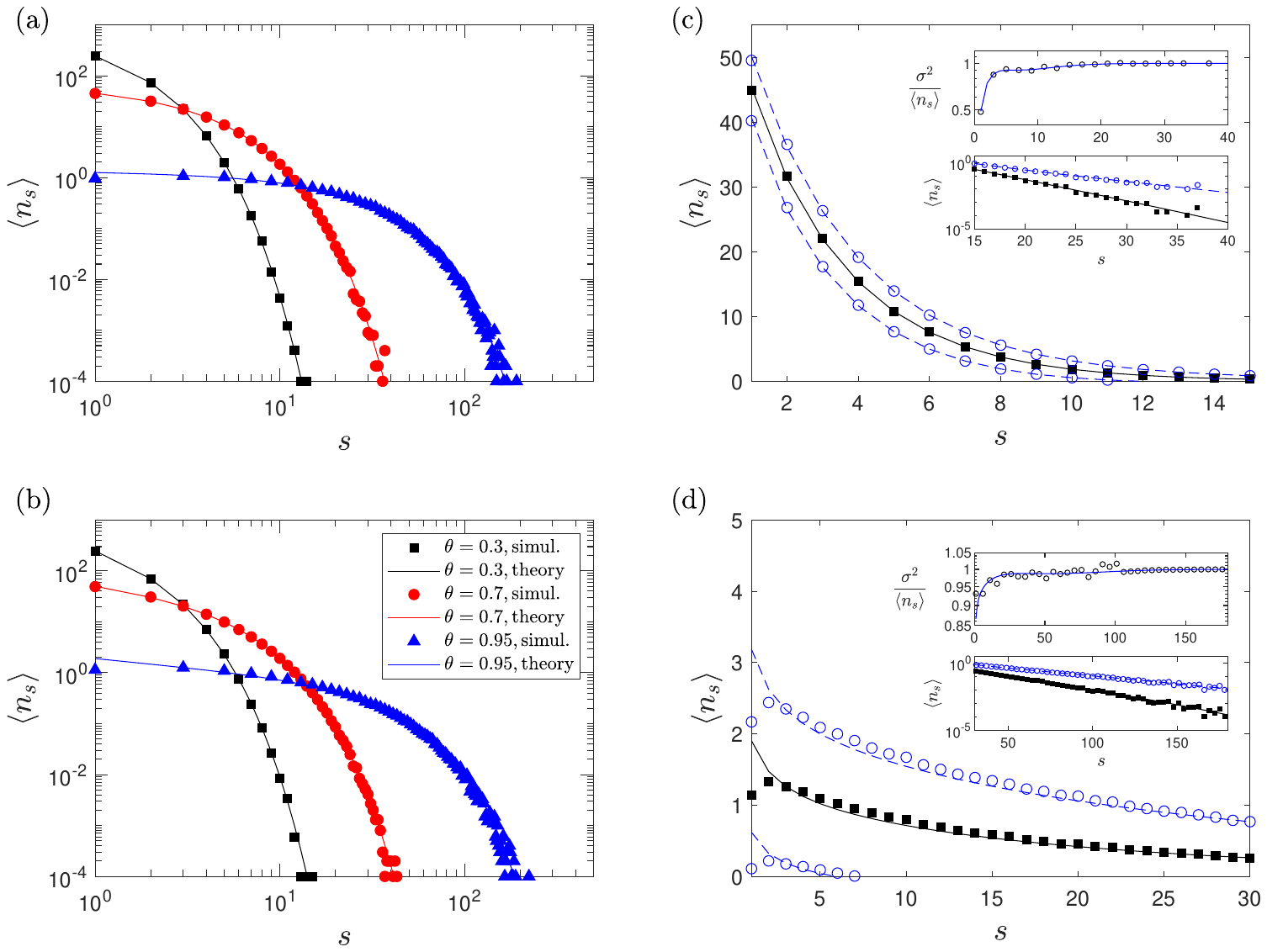}
\caption{Growth of aerosols. Average number of clusters of a given size, $\langle n_{s} \rangle$, and corresponding standard deviation for the continuum regime kernel, Eq.~(\ref{aerosol_CR}), (a) and (c), and for the free--molecular regime kernel, Eq.~(\ref{aerosol_FRM}), (b) and (d). In the left column, three data series are presented: an early stage of the evolution, $t=150$, $(\theta=0.3)$, a later stage of the evolution, $t=350$ ($\theta=0.7$), and a very late stage of the evolution, $t=475$ ($\theta=0.95$; only 25 steps are left to form a single coagulant). Lines represent combinatorial solutions, Eq.~(\ref{ns_general}), and symbols represent data from numerical simulations {(means)}. Right column: (c) $\langle n_{s} \rangle$ and its standard deviation for the continuum kernel for $t=350$ ($\theta=0.7$); (d) $\langle n_{s} \rangle$ and its standard deviation for the free-molecular kernel for $t=475$ ($\theta=0.95$). Black squares represent $\langle n_{s} \rangle$ obtained by numerical simulation; open circles represent corresponding {$\langle n_s \rangle \pm \sigma_s$}. Solid lines and dashed lines represent theoretical $\langle n_{s} \rangle$ and its standard deviation, Eq.~(\ref{std_dev_general}).
Inset figures: Variance divided by $\langle n_{s} \rangle$ (upper inset figures) and $\langle n_{s} \rangle$ for higher $s$ (lower inset figures).
The system consisted initially of 500 monomers. Numerical simulation results were averaged over $10^4$ independent runs. { For the scaling solutions from the usual Smoluchowski theory, see Ref.~\cite{Friedlander_2000}.} }
\label{Figure_brownian_kernels}
\end{figure*}

Several coagulation kernels have been derived for atmospheric aerosols using kinetic theory under some assumptions on the shape and motion of particles \cite{Ferreira_2021}. These assumptions are: (i) uniform distribution in space, (ii) elastic collisions with background air particles where the number of such collisions is much larger than the number of collisions between two coalescing particles, (iii) binary coagulation acts where any collision between coalescing particles yields a coalescing particle, (iv) neglecting condensation phenomena \cite{Ferreira_2021}.

A general form of a coagulation kernel (or collision frequency) for atmospheric aerosols may be expressed as a Brownian motion coagulation kernel, 
\begin{equation} \label{aerosol_general}  
 K(s,g) \propto (D_s+D_g)(r_s + r_g),
\end{equation}
\noindent where $D_s$ and $D_g$ are diffusion coefficient of particles $s$ and $g$, and $r_s$ and $r_g$ are their respective effective collisional radii \cite{Friedlander_2000}. As $D$ and $r$ depend on particle and gas properties (including, e.g., the fractal dimension of particles), their exact formulations depend on the regime being studied. Each regime is defined based on the relation between particle size and the average distance traveled by a particle between two collisions in the air, i.e., mean free path, $\lambda$, and is described by the Knudsen number, $K\!n=\lambda / r$. 

For low $K\!n$, the motion of particles becomes diffusive, and for spherical particles, a limit of Eq.~(\ref{aerosol_general}) may be found as
\begin{equation} \label{aerosol_CR}  
 K(s,g)_{C\!R} \propto (s^{1/3}+g^{1/3})(s^{-1/3}+g^{-1/3})
\end{equation}

\noindent which is called \textit{continuum regime} (diffusive) kernel. 

On the other hand, for large $Kn$, the motion of particles becomes ballistic, and a limit for this regime may be found as 
\begin{equation} \label{aerosol_FRM}  
 K(s,g)_{F\!M\!R} \propto \left(  s^{1/3} + g^{1/3} \right) ^2 \left( s^{-1} + g^{-1} \right)^{1/2}
\end{equation}

\noindent which is called \textit{free-molecular regime} (ballistic) kernel. For derivations of Eqs.~(\ref{aerosol_CR}) and (\ref{aerosol_FRM}), Reader may consult, e.g., Ref.~\cite{Eggersdorfer_2014}.

Both Eqs.~(\ref{aerosol_CR}) and (\ref{aerosol_FRM}) had no explicit solutions previously, neither to the Smoluchowski equation nor in any other approach. Importantly, the above kernels are of use not only in aerosol science as they define more general ballistic and diffusive processes (cf. Tab.~1 in Ref.~\cite{Aldous_1999}). Here, we will use Eqs.~(\ref{ns_general}) and (\ref{std_dev_general}) to obtain theoretical { approximations} for coagulation with kernels (\ref{aerosol_CR}) and (\ref{aerosol_FRM}) and test their performance for different times against numerical simulation.

In Fig.~\ref{Figure_brownian_kernels}, we plotted the average number of clusters of a given size and the corresponding standard deviation for the continuum regime kernel and the free-molecular regime kernel for the system consisting of 500 monomers. We present three data series: an early stage of evolution ($\theta=0.3$), a later stage of evolution ($\theta=0.7$), and a very late stage of  evolution ($\theta=0.95$) when only 25 steps are left to form a single coagulant). For both cases and at any stage of evolution, the combinatorial expressions accurately reproduced the simulation results with excellent precision. As in the case of the planetesimal formation, theoretical {approximations} provided estimates for the standard deviation, which overlapped with the numerical results. Any discrepancy may be found only for the latest stage of evolution and for the smallest sizes $s$. Possible reasons for that will be discussed in Sec.~\ref{Exactness}.

Despite the elegant formulation of the limit kernels, most aerosols in ambient conditions stay in the transition regime \cite{Eggersdorfer_2014}. Various approximations for this regime were proposed \cite{Fuchs_1964, Dahneke_1983, Eggersdorfer_2014, Karsch_2023} but the general scheme may be written as \cite{Zurita_2002}
\begin{equation} \label{aerosol_TR_1}  
 K(s,g)_{trans} = 4\pi (D_s+D_g)(r_s + r_g) \chi (K\!n_D),
\end{equation}

\noindent where $\chi$ is a correction factor that makes the transition from the continuum regime ($K\!n, K\!n_D \ll 1, \chi \to 1$) to the for free-molecule conditions ($K\!n, K\!n_D \gg 1$). The correction factor $\chi$ reads
\begin{equation} \label{aerosol_TR_2}
 \chi (K\!n_D) = \frac{1+ K\!n_D}{1+2K\!n_D(1+K\!n_D)},
\end{equation}

\noindent with $K\!n_D$ being diffusional Knudsen number, $K\!n_D = \lambda_p / r $, with $\lambda_p = 16D / \pi \bar{u}$ being the particle mean free path, $D$ the particle diffusion coefficient, $\bar{u} = \sqrt{8k_B T / \pi m}$ the mean thermal particle velocity and $k_B$ the Boltzmann constant. The diffusional Knudsen number describes dilute conditions more efficiently and is slightly different from the (ordinary) Knudsen number, $K\!n = \lambda / r$, where $\lambda = 2\mu_f / ( p \sqrt{8M/(\pi R T)} )$ is the gas mean free path, $\mu_f$ the gas viscosity, $p$ the pressure, $M$ the molecular weight, $T$ the temperature and $R$ the ideal gas constant \cite{Eggersdorfer_2014}.

The diffusion coefficient $D$ for a particle in Eq.~(\ref{aerosol_TR_1}) may be defined as
\begin{equation} \label{aerosol_TR_3}
 D = \frac{kT~C_c(K\!n)}{3\pi\mu d_m},
\end{equation}

\noindent where $d_m$ is the mobility diameter and $C_c$ is the Cunningham slip correction factor,
\begin{equation} \label{aerosol_TR_4}
 C_c(K\!n) = 1+K\!n \left( A+B~\mathrm{exp} \left( \frac{C}{K\!n} \right) \right),
\end{equation}

\noindent and $A$, $B$ and $C$ are empirically-determined parameters \cite{Rogak_1992}.


The above formulation of the transition regime kernel was presented to comment on the usability of the combinatorial approach to the kernel of a complex form. Despite its involved formulation, such a kernel may be analyzed in the combinatorial approach if only $x_g$ is calculated unambiguously. First, the radii $r_s$, $r_g$, and $d_m$ are, in general, functions of the particle size, including for particles of fractal shape, and these functions were were used to obtain limits (\ref{aerosol_CR}) and (\ref{aerosol_FRM}). Second, several constant coefficients present in Eqs.~(\ref{aerosol_TR_1})--(\ref{aerosol_TR_4}) do not change the final result  $\left\langle n_s\right\rangle $; however, some of them (e.g., $T$) may change during the aggregation. In calculating $x_g$, such a change of the ambient conditions during the aggregation process is not taken into account as the summation in Eq.~(\ref{xg_recursive_definition_general}) assumes that all of the previous sub-clusters were characterized only by their (effective) sizes. This leads to the requirement that all of the important particle-related conditions, which change during the aggregation process, must be expressed as a function of the particle effective size(s), thus, in the form of $f(s,g)$.

Another way of expressing the kernel for the transition regime is to take the harmonic mean of the two limiting regimes \cite{Maricq_2007, Eggersdorfer_2014}, 
\begin{equation}
 K(s,g)_{trans}^{-1} \cong K(s,g)_{C\!R}^{-1} + K(s,g)_{F\!M\!R}^{-1},
\end{equation}

In this case, the use of $K(s,g)_{trans}$ with the combinatorial framework is straightforward.

\subsection{Other fields of potential application}\label{Applications}

{ In Appendix \ref{Appendix_fields}}, we provide examples of processes where the combinatorial approach may find its application to provide theoretical predictions { (approximations in general)}. The examples from previous sections are not included; however, it is worth noting that the use of kernels present in the aerosol example, Eqs.~(\ref{aerosol_CR}) and (\ref{aerosol_FRM}), is not limited to this field only, as they describe general Brownian motion processes. The selection of fields in Appendix \ref{Appendix_fields} should not be considered as a review of new trends in coagulation studies, nor an exhaustive listing. Instead, we focused on studies where coagulation kernels of the form $K(s,g; \dots)$ emerged, {} as such forms can undergo an analysis in the combinatorial approach, similarly to the previous examples. Also, we focused on those processes where real data on cluster statistics are---or supposedly may be---obtainable by experiment.

In this way, we visit dust agglomeration, accretion in Saturn's rings, aggregation of proteins, magnetorheological fluids, polymer growth, coagulation of milk and soot particles, and coalescence in turbulent flows. Short comments on particular topics, { giving only a rough context for the given kernels,} are given in the respective subsections of { Appendix \ref{Appendix_fields}}.



\section{Remarks} \label{Remarks}

\subsection{Validity and precision of solutions} \label{Exactness}

\begin{table*}[ht]
\setlength\extrarowheight{2.5pt}
\setlength{\tabcolsep}{10pt}
\begin{tabular}{ l  l l }
\hline \hline
$K\propto$ & Precision of $\langle n_s \rangle$ & Refs. \\[2.5pt]
\hline \hline
$1$ & Exact{$^\star$} & \cite{Fronczak_2018, Leyvraz_2022} \\[2.5pt]
\hline
$sg$ & Approximate{$^\star$} & \cite{Lepek_2019, Leyvraz_2022} \\[2.5pt]
\hline
$s+g$ & {Formally approximate, suggested as} exact{$^{\star}$} & \cite{Lepek_2019, Leyvraz_2022} \\[2.5pt]
\hline
$A+s+g$  & {Formally approximate, suggested as} exact{$^{\star}$} & \cite{Lepek_2021_ROMP, Leyvraz_2022}
\\[2.5pt]
\hline
$\left( s^{-1}+g^{-1} \right)^\alpha$ &  Approximate, increasing for $\alpha<1$, decreasing for $\alpha>1${$^{\star\star}$} & \cite{Lepek_2021, Leyvraz_2022} \\[2.5pt]
\hline
$(s+A)(g+A)$ & Approximate, increasing with $A${$^{\star\star\star}$} & \cite{Lepek_2021_ROMP} \\[2.5pt]
\hline
$(s+g)^2(sg)^{-1}$ & Approximate & \cite{Lepek_thesis_2021} \\[2.5pt]
\hline 
$sg\left(1+(s+g)^{-1}\right)$ & Approximate & \cite{Lepek_thesis_2021} \\[2.5pt]
\hline 
\end{tabular}
\caption{Kernels solved in the combinatorial approach, i.e., where an explicit expression for $x_g$ was found. In the second column, we summarize the precision of the theoretical predictions given by the expression for $\langle n_s \rangle$, Eq.~(\ref{ns_general}). { The above summary is based on comparisons to the numerical simulations and on theoretical asymptotic analysis performed in Ref.~\cite{Leyvraz_2022}. In the general case, the combinatorial approach yields approximate solutions that, when analyzed in the large system size limit, are incompatible with known results from the scaling theory (see the main text and Appendix C for details). However, the three basic kernels and their combinations ($^\star$), due to their special scaling properties, behave differently from the general case. For the constant and additive kernels, it has been shown that the combinatorial solutions yield exact results both in the large-scale limit of $\langle n_s \rangle$ as well as for the full probability distribution as explicitly evaluated for a small system. Although not explicitly enumerated, the same shall apply for the combination $K=A+s+g$ \cite{Leyvraz_2022}. For the product kernel, the precision of the results drops after the gelling point but preserves the existence of the giant cluster. Asymptotic analysis has shown that the combinatorial solution for this case may be regarded as exact in the large-scale limit for the time before the gelling point, while it yields a known (inaccurate) solution after the gelling point. $(^{\star\star})$ In case of the {linear-chain (LC)} kernel, decreasing $\alpha$ means that the kernel form tends to the one of the constant kernel; thus, behavior and precision also tend to the behavior and precision of the constant kernel. The discrepancies for small-cluster sizes (with $\alpha=1$) were studied in Ref.~\cite{Leyvraz_2022} as a result of the incompatibility with the scaling theory. $(^{\star\star\star})$ Analogously, for the condensation kernel, increasing $A$ means that the contribution of $s$ and $g$ decreases and the kernel behavior becomes more similar to the behavior of the constant kernel, resulting in increasing precision {(of course, exact dynamics throughout the aggregation process depends on the current values of $A$, $s$, $g$, and the system size)}. For $A=0$, the solution is identical to the one for $K=sg$, and all the concerns related to the product kernel apply.}}
\label{tab_solved_kernels_precision}
\end{table*}

{

The precision of the combinatorial solutions tested vs. numerical simulation varies from exact to approximate (see Table~\ref{tab_solved_kernels_precision}).

Lower precision may be observed for the latest stages of evolution and particular forms of kernels, especially for the kernels with behavior significantly different from the behavior of the constant kernel, e.g., gelling kernels. This inaccuracy is manifested in Figs.~\ref{Fig_kernele_zbiorczy} and \ref{Fig_AppE}, where the precision of the prediction is low for the post-gel phase, and in Fig.~\ref{Figure_brownian_kernels}d, where one can observe minor discrepancies between theory and simulation for the smallest cluster sizes. Such discrepancies for small-size aggregates were also studied in detail theoretically and by simulation in Ref.~\cite{Leyvraz_2022} for the case of the LC kernel.

In Ref.~\cite{Leyvraz_2022}, it has been shown analytically that both combinatorial solution to the constant kernel and the solution to the additive kernel are exact solutions in the sense of the large-scale limit ($N,t \rightarrow \infty$, and $t/N=\mathrm{const}$), as compared to the exact solution of the original Marcus--Lushnikov model and the Smoluchowski equation. Moreover, the full probability distribution for these cases was tested against explicitly evaluated probabilities for a system of $N=20$ particles and coincided exactly. However, this was not true in the general case. The solution to the product kernel seems exact in the limit sense only for the pre-gel phase. For this case, after the gelling point, the combinatorial solution coincides with the Stockmayer result (which assumes no interaction between the finite particles and the gel). Those three kernels (the constant, additive, and product) and their combinations have the important property that their small cluster-size and large cluster-size behaviors coincide. However, in general, such a property does not hold, and applying an asymptotic approach to the combinatorial solutions yields the concentration profile that agrees qualitatively in the limit of large cluster sizes but deviates from known scaling behavior in the small cluster sizes. Such a problem is observed clearly for the LC kernel (see Fig.~\ref{Fig_kernele_zbiorczy}e and \ref{Fig_kernele_zbiorczy}f, cf. Fig.~1 in Ref.~\cite{Leyvraz_2022}).

Above, we only reported concisely the most important findings from Ref.~\cite{Leyvraz_2022}. For more detailed elaboration and relation to the scaling theory (the so-called scaling solutions), please refer to Appendix C. A short introduction to the scaling theory itself may be found there.

Initially, the combinatorial expressions were derived for the constant kernel as an exact solution to the aggregation process \cite{Fronczak_2018}. However, its extension to arbitrary processes is, in general, an approximation. Where does this approximation come from? The most restrictive element of formal derivations in Sec.~\ref{Approach} is Eq.~(\ref{probability_general}), which assumes equal probability of all possible time realizations of the system at state $\Omega$. Apparently, this assumption is not valid for an arbitrary case.

How is the (restrictive) idea of equiprobable ways leading to a given system state, included in assumption~(\ref{probability_general}), realized in the formalism? One idea is that the problem lies in the fact that the combinatorial expression for the distribution of merging acts in time, Eq.~(\ref{distribution_time}), is a clear result of Eq.~(\ref{probability_general}). It was written with the assumption that reaction rates do not depend on cluster size (which, for kernels other than constant, is not true). 

For the constant and additive kernel, as well as for their combination, the kernel-dependent factor vanishes from under the sum in Eq.~(\ref{xg_recursive_definition_general}). { This is due to the fact that, for all these kernels, if we consider a single cluster of size $s$ at time $t$ and all its possible preceding states at $t-1$, then all these possible preceding arrangements have equal relative probability of transition to the resulting cluster. For instance, considering a cluster of size $s=6$, the possible preceding states are: (1,5), (2,4), and (3,3). For the constant kernel, all the possible transitions have a probability of $K=1$; for the additive kernel, it is $K(g+h)=6$; and for their combination, it is $K(A+g+h)=6+A$. For a general case, such a feature does not hold. Of course, the above consideration tells nothing about the probabilities of those preceding states but establishes a similarity between these three kernels and distinguishes them from other kernel types.}

The Reader is kindly encouraged to find the above problems in the Open Problems Section.
}


\subsection{Other limitations}\label{Limitations}

The framework presented here was initially developed to solve the ideal process of aggregation only, and when applied to more involved problems (real-world cases), of course, several limitations became noticeable.

One of them may be that real-world aggregation processes are usually accompanied by fragmentation, even if the effective result has the form of an aggregate, as in blood clotting \cite{Guria_2009} or Saturn rings \cite{Brilliantov_2015}. As for now, there has been no attempt to describe fragmentation by analogous combinatorial expressions.

As mentioned in Sec.~\ref{Remarks_on_time}, the discrete-time scale counted as subsequent coagulation acts results in the fact that the kernel in the combinatorial approach is defined up to a multiplicative constant and, therefore, results in difficulties in comparing theoretical predictions to the experimental data. However, such a comparison may be performed if only sufficient information on the coagulating system is available. The solutions in the combinatorial approach, as presented here, do not answer how much time a given process takes in physical time. Thus, they do not provide the reaction rate in seconds (or hours, days) as this is often tested experimentally. However, this limitation may be addressed { by finding the relevant Smoluchowski (physical) time with Eq.~(\ref{time_S_time_comb}), and, in further studies,} by incorporating time delays between subsequent coagulation acts modeled with, e.g., an exponential function as was done in the Marcus--Lushnikov approach.
 
Aggregating particles are subject to local and global conditions. The local conditions affect two coagulating particles in the moment of the merging act and, in the combinatorial approach, should be fully modeled by their effective sizes, which are described by $K$ in the expression for $x_g$. On the other hand, global conditions (as, for instance, $\mu$ from the protoplanetary aggregation example) need deeper examination. If $\mu$ remains constant during the evolution of the system, it may be easily incorporated into the kernel formulation. In turn, if $\mu$ may change in time, affecting reaction rates of the particles, such a situation is not covered by the combinatorial expressions, as the summation in the expression for $x_g$ does not take into account that previous ``possible'' coagulation acts of the sub-clusters may have occurred in different global conditions. Therefore, any influence on the reaction rate must be incorporated into $K$ as a function of $s$ and $g$.

Of course, using Bell polynomials for calculations in the combinatorial approach creates the need for exceptional computational power if larger systems are considered. Further comments may be found in Appendix A. Still, finding respective reliable statistics by numerical simulation for the systems considered in this study consumed time longer by an order of magnitude.

\section{Summary}\label{Summary}

{
As we know, exact solutions to the Smoluchowski equation are not necessarily the exact solutions to the coagulation process itself, particularly in finite systems, which is due to the requirements of the Smoluchowski approach (infinite {number of particles}, continuous cluster concentrations). The combinatorial approach was initially designed to overcome these shortcomings. However, recent developments { presented in this mini-review} revealed that the quality of its results is more involved {(see Sec.~\ref{Exactness})}.

For the constant {kernel and (presumably)} the additive kernel, the {combinatorial solutions given by Eqs.~(\ref{EQ42}) and (\ref{EQ45})} may be regarded as exact for any time point of the process. Comparing theoretical predictions to the results of numerical simulations may be only a suggestion, but, in this case, it has been shown that the combinatorial expressions in the large-scale limit agree with the solutions to the Smoluchowski equation and to the {original continuous-time} Marcus--Lushnikov model \cite{Leyvraz_2022}. For the linear combination of the constant and additive kernels, a numerical suggestion exists that the combinatorial solution may be regarded as exact \cite{Lepek_2021_ROMP}. It stays in agreement with the special scaling behavior of the three basic kernels and their combinations, which may be a further suggestion. For the gelling kernels, where a gel phase appears during the evolution, combinatorial solutions give only approximate predictions reproducing the gel phase only qualitatively as compared to the {numerical} simulation \cite{Lepek_2019, Lepek_2021_ROMP, Leyvraz_2022}. In the general case, the large-scale and large-time limit of the combinatorial solution deviates from the known results from the scaling theory for small clusters \cite{Leyvraz_2022}.
}

In this elaboration, we have guided the Reader through the up-to-date state of the combinatorial approach to coagulating systems, its initial assumptions, building the expressions, and kernels of known solutions. We have presented possibilities of use in cases where no explicit solution can be found.

In general, the set of Eqs.~(\ref{omega_s}), (\ref{ns_general}), (\ref{std_dev_general}), and (\ref{xg_recursive_definition_general}) may be used to obtain theoretical predictions for any given kernel (thus, including, e.g., general forms given by Fournier and Lauren\c{c}ot \cite{Fournier_2005}, or listed in well-known reviews \cite{Aldous_1999, Wattis_2006}) and any given time of the aggregation process (of course, with varying precision, { being an approximation for a general case}). So far, explicit solutions for the average number of clusters of a given size have been obtained for several families of kernels; such a step may be regarded as a novelty in the field, in particular, considering providing a precise estimate for the standard deviation of this average. The importance of such an estimate was presented within the example of the planetesimal formation and may be of use in applied fields. Alternative (Monte Carlo) methods to quantify the probability of rare events are being developed \cite{Dandekar_2023}.

Why should we look for analytical solutions at all? Obtaining analytical solutions (in particular, exact solutions) for at least some classes of kernels helps to assess the accuracy and computational efficiency of different numerical schemes. Numerical solutions are, in principle, limited in time, while (exact) solutions unambiguously illustrate evolution for the whole time interval \cite{Osinsky_2022_JPhysA}.

Nowadays, it is known that several parameters of the aggregating particles depend on their fractal dimension. For instance, the mean mobility diameter, $d_m$, depends on the fractal dimension of the particle, $d_f$. In the case of the linear chains \cite{Lepek_2021}, this dependency was implicitly incorporated into the expression for $K$. However, as shown above in this work, explicit dependency on $d_f$ does not pose a serious problem and theoretical { approximations} may be easily found for such kernel formulation (a vivid example may be protein aggregation with $K \ \propto (sg)^\gamma (s^{1/d_f}+g^{1/d_f})(s^{-1/d_f}+g^{-1/d_f})$, $\gamma=1-1/d_f$ \cite{Zidar_2018}).

Most aggregation studies, such as the one presented here, deal with the systems in a diluted state to prevent analysis of three-body collisions. Recently, other approaches were used to study aggregation under high concentrations, e.g., using Langevin dynamics \cite{Wang_2021}; however, description with kinetic equations was also attempted \cite{Fogelson_2010}.

Further work on the combinatorial approach would cover, for instance, extending the expressions for the initial conditions other than homogeneous ones and improving the accuracy of estimates in the post-gel phase. These issues are listed in the Open Problems section.

\section{Open problems}\label{Problems}

{
In this section, we provide a summary of open problems present in the reviewed combinatorial approach to aggregating systems.}

To this point, finding a ``solution'' of a given kernel in the combinatorial approach has been based on transforming the recursive expression for $x_g$ to the explicit form using substitutions and, when needed, the generating function method. However, this method fails when a proper substitution cannot be found. Can we use other methods to find explicit solutions? 

Also, most of the explicit solutions obtained so far were resistant to simplification to the form that would not contain Bell polynomials. Can we find other relations (other than those presented in Appendix A) to do so? 

{
In the following subsections, please find further (and more fundamental) problems.

\subsection{Precision of solutions}
}

Although asymptotic analysis for Bell polynomials is involved, it was found to be a key tool to investigate the large-scale behavior of combinatorial solutions \cite{Leyvraz_2022}. A general asymptotic expression for the corresponding $c_s(t)$ in the large-size limit was obtained and compared to the known exact results for the three basic kernels and the LC kernel to show complex issues and discrepancies for the general case. What kind of improvements can be applied to the combinatorial formalism to achieve agreement with the scaling theory for any arbitrary case?

A naive observation from Table~\ref{tab_solved_kernels_precision} may suggest that kernel forms that do not contain operations other than addition give exact results when compared to the {numerical simulation of the aggregation process}. For those kernels, the kernel-dependent factor in Eq.~(\ref{xg_recursive_definition_general}) vanishes from under the sum {(see also Sec.~\ref{Exactness})}. Is that the case? Why and how does this impact the final result? A deeper understanding of the mechanism underlying the effective precision of the expressions will be essential to find appropriate limits of their usability. { Also, a formal proof that the combinatorial approach solves the constant and additive kernels ``exactly'' is still needed.}


{ Can we study the size-dependent correctness of the solutions by numerical means?  How would prediction errors scale with $N$? This could be of interest for any applied research studying finite systems far from the large-size limit, as the one in Ref.~\cite{Lepek_2021}.}
\newline

\subsection{Post-gel phase}

As clearly shown in the review, combinatorial expressions have a significant handicap, being low precision of theoretical predictions for the time after the transition point for gelling kernels, i.e., for the post-gel phase of evolution. In general, it may be seen that kernels of lower reaction rates give more precise estimates than gelling kernels, the most precise for the kernels similar to the constant kernel (e.g., linear-chain kernel for low $\alpha$).

Although we have precisely reformulated $x_g$ for an arbitrary kernel, including the product kernel, the reason for that may be that the rest of the combinatorial expressions (e.g., the expression for distributing coagulation acts in time, Eq.~(\ref{distribution_time})) were written with the (implicit) assumption of the constant kernel (size-independent rate). The expression~(\ref{distribution_time}) is particularly dubious as it assumes equiprobability of the two coagulation acts even if these acts originate from the two clusters of significantly different sizes (while the acts related to the greater cluster should be chosen with higher probability).

Any modification leading to better precision after the transition point would be a significant step forward.

\subsection{Non-monodisperse initial conditions}

One of the apparent issues with the combinatorial approach is that its present formulation is limited to the monodisperse initial condition. This is clearly an open problem, as the assumption on the monodisperse initial condition was explicitly used to obtain all of the combinatorial expressions derived up to this point. Even considering a mixture of monomers and dimers as the initial condition (as in, e.g., Ref.~\cite{Fronczak_2019} in Marcus--Lushnikov approach) poses a serious task.

Only one ``non-monodisperse'' case seems to be easily achievable---here, we will modify the combinatorial expressions for $\left\langle n_s\right\rangle$ to cover homogeneous initial conditions other than monodisperse ones, i.e., to cover initial conditions consisting of dimers only, or trimers only, and so on, thus, in general, of $d$-mers only. To do so, we simply observe that the system consisting initially of $d$-mers may be regarded as a re-counted system of monomers. Through the example below, the Reader is invited to contribute to this problem.

To perform such re-counting, let us assume a system of $N$ monomeric units containing initially dimers only (thus, containing $N/2$ clusters). Such a system of ``larger'' initial particles will have rescaled properties: the maximum size of a cluster that \textit{could} grow in the system would be $g/2$, and the total number of clusters in the system will count $k/2$. Of course, in this case, non-zero values of  $\left\langle n_s\right\rangle$ are possible only when $s$ is even. 

Eq.~(\ref{ns_general}) takes the form
\begin{equation} \label{ns_general_initial_cond_dimers}  
 \left\langle n_s\right\rangle =\binom{{N/2}}{{s/2}} \omega_{{s/2}} \frac{B_{{N/2}-{s/2},{k/2}-1}\left(\left\{{\omega}_{{g/2}}\right\}\right)}{B_{{N/2},{k/2}}\left(\left\{{\omega}_{{g/2}}\right\}\right)},
\end{equation}

\noindent for $s~\text{mod}~2 = 0$ and $\left\langle n_s\right\rangle =0$ for other cases. 

Similarly, for arbitrary $d$-mers, we can write
\begin{equation} \label{ns_general_initial_cond}  
 \left\langle n_s\right\rangle =\binom{N/d}{{s/d}} \omega_{{s/d}} \frac{B_{{N/d}-{s/d},{k/d}-1}\left(\left\{{\omega}_{{g/d}}\right\}\right)}{B_{{N/d},{k/d}}\left(\left\{{\omega}_{g/d}\right\}\right)},
\end{equation}

\noindent for $s~\text{mod}~d=0$ and $\left\langle n_s\right\rangle=0$ for other cases. Or, in a more compact form, using a star to denote rescaled parameters,
\begin{equation} \label{ns_general_initial_cond_star}  
 \left\langle n_s\right\rangle =\binom{N^*}{s^*}{\omega }_{s^*}\frac{B_{N^*-s^*,k^*-1}\left(\left\{\omega_{g^*}\right\}\right)}{B_{N^*,k^*}\left(\left\{{\omega }_{g^*}\right\}\right)}.
\end{equation}

Analogously, we modify the related part of the expression for the standard deviation, Eq.~(\ref{std_dev_general_addition}), to the form of
\begin{equation} \label{std_dev_initial_cond_general_addition}  
 \left\langle n_s\left(n_s-1\right)\right\rangle = \binom{N^*}{s^*,s^*}{\left({\omega}_{s^*}\right)}^2\frac{B_{N^*-2s^*,k^*-2}\left(\left\{{\omega}_{g^*}\right\}\right)}{B_{N^*,k^*}\left(\left\{{\omega}_{g^*}\right\}\right)}.
\end{equation}

\begin{figure} 
\includegraphics[scale=0.55]{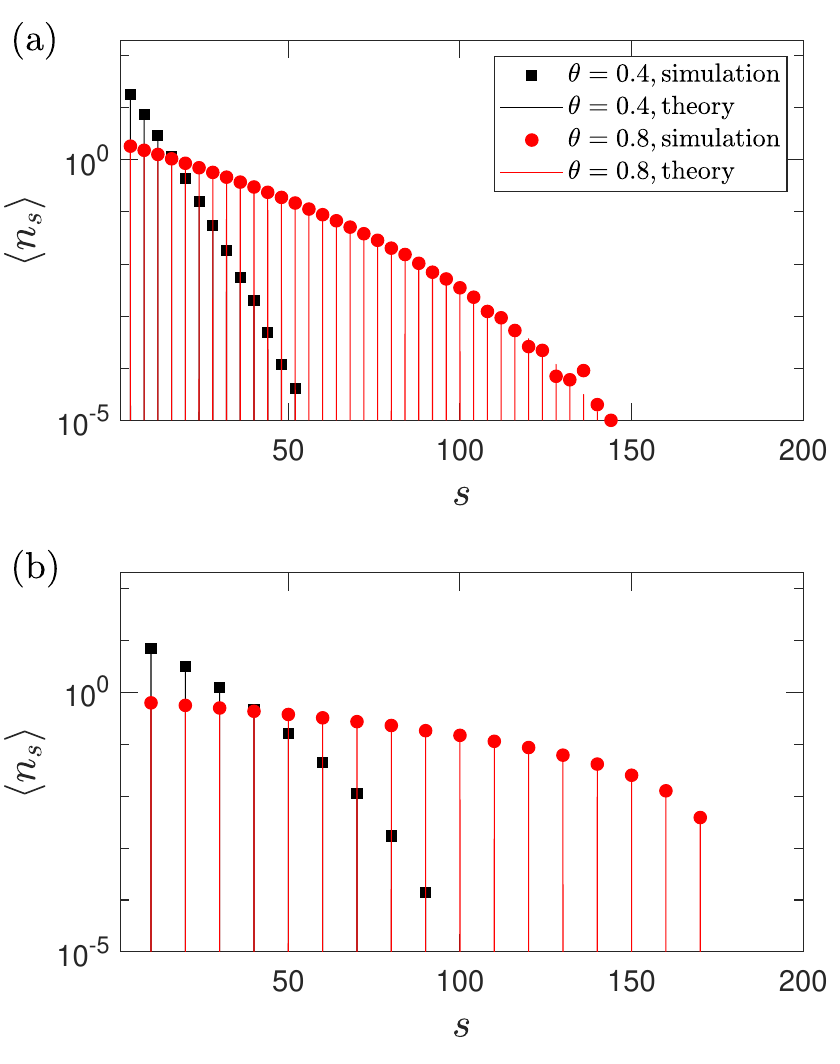}
\caption{Numerical simulations vs. Eq.~(\ref{ns_general_initial_cond_star}). The system consisted of $N=200$ monomers grouped initially into 4-mers (a) and 10-mers (b). Two aggregation moments are presented, $\theta=0.4$ and $\theta=0.8$. Both in numerical results (symbols) and {exact} theoretical solutions (lines), only those cluster sizes that are multiples of the initial cluster sizes are occupied. The simulation results were averaged over $10^5$ independent runs.}
\label{Figure_hom_init_cond}
\end{figure}

A working example of the use of Eq.~(\ref{ns_general_initial_cond_star}) for the constant kernel is presented in Fig.~\ref{Figure_hom_init_cond}.

\section*{Acknowledgments}

{
We want to thank the Reviewers for their significant contribution to the final shape of the article and our colleague Grzegorz Siudem from the Faculty of Physics, Warsaw University of Technology, for his help with the asymptotic analysis of Bell polynomials.
}

The research was funded by POB Cybersecurity and Data Science (AF) and POSTDOC PW programs (PF, M\L{}) of Warsaw University of Technology within the Excellence Initiative: Research University (IDUB).

\section*{Conflict of interest}
The authors declare that they have no conflict of interest.

\appendix

\section{Mathematical tools and derivations}

\subsection{Partial Bell polynomials} \label{AppA_1}

Partial Bell polynomials (also known as incomplete or the second kind of, for short, Bell polynomials) are defined as
\begin{multline} \label{Bell_polynom_def}
 B_{N,k}\left(x_1,x_2,\dots ,x_{N-k+1}\right)=B_{N,k}\left(\left\{x_g\right\}\right) \\ 
 = N!\sum_{\left\{n_g\right\}}{   \frac{x_1^{n_1} x_2^{n_2} \dots}{ n_1!n_2! \dots (1!)^{n_1} (2!)^{n_2} \dots }   }  \\ = N!\sum_{\left\{n_g\right\}}{\prod^{N-k+1}_{g=1}{\frac{1}{n_g!}{\left(\frac{x_g}{s!}\right)}^{n_g}}},
\end{multline} 

\noindent where the summation is taken over all non-negative integers $\left\{n_g\right\}$ that satisfy
\begin{equation}
 \sum^N_{g=1}{n_g=k} \;\;\;\; \textrm{and} \;\;\;\; \sum^N_{g=1}{gn_g=N}.
\end{equation}

By using Bell polynomials, one can obtain complete information on the partition of a given set. For instance, considering $ N = 6 $ monomeric units and $ k = 3 $ clusters, we have: $ B_{6,3} \left(x_1, x_2, x_3, x_4 \right) = 15x^2_1 x_4 + 60 x_1 x_2 x_3 + 15x^3_2 $, telling us that there are 15 ways of partitioning the set of 6 monomers into two clusters of size 1 and one cluster of size 4 (for short, $1 + 1 + 4 $); 60 ways of partitioning to clusters of sizes $ 1 + 2 + 3 $; and 15 ways to obtain partition $2+2+2$. The total number of available partitions equals the Stirling number: $ S \left(6,3 \right) = 15+60+15 = 90 $. Indeed, the Stirling number is equivalent to the Bell polynomial on a set of ones,
\begin{equation} \label{Bell_ones}
 S\left(N,k\right) = B_{N,k}\left(1,1,\dots ,1\right).
\end{equation}

However, how shall we interpret the meaning of the numbers $x_g$ themselves? For instance, let us take a look at the system of $N=12$ monomers grouped into $k=6$ clusters. Coefficients $x_g$, e.g., $x_1$, $x_3$, may be interpreted as relative \textit{preference} of clusters of given size, consisting of, respectively, one or three elements. In the case of $\{x_g\}=\{1,1,1,1,1,1\}$, all available partitions are realized with the same probability. In the case of $\{x_g\}=\{40,1,1,1,1,1\}$, clusters of size 1 are favored. In the case of $\{x_g\}=\{1,40,1,1,1,1\}$, clusters of size 2 (dimers) are strongly favored. In turn, the case $\{x_g\}=\{1,1,40,1,1,1\}$ favors clusters of size 3, but the constraint ($k=6$) forces that the partition for trimers only is not possible, and over-representation of trimers results in over-representation of monomers. {In the above explanation, we used the number of 40 as an example value significantly higher than 1}.

Derivatives of Bell polynomials are defined as \cite{Johnson_2002}
\begin{equation} \label{Bell_pochodne_definicja}
\frac{ \partial B_{N,k}\left(\left\{x_g\right\}\right) }{ \partial x_s } = \begin{cases}
      \binom{N}{s} B_{N-s,k-1}\left(\left\{x_g\right\}\right) & \text{for $s \in A $}\\
      ~0 & \text{for  $s \in S \setminus A$},
    \end{cases}   
\end{equation}

\noindent where $A= \left\{ 1,2, \dots,N-k+1 \right\}$ and $S=\left\{1,2,\dots,N\right\}$. We use the above derivatives throughout the transformations in Sec.~\ref{Approach}.

An extensive study on Bell polynomials may be found in the devoted chapter in Ref.~\cite{Pitman_2006} (\textit{Bell polynomials, composite structures and Gibbs partitions}). Several helpful relations were revealed in Refs.~\cite{Comtet_1974, Wang_2009}. 

Here, we will present the relations used in Sec.~\ref{Solutions} to simplify Bell polynomials.

First relation \cite{Pitman_2006},
\begin{equation} \label{relacja_gg-1}
B_{N,k} \left(\left\{g^{g-1} \right\} \right) = \binom{N}{k}kN^{N-k-1} = {\binom{N-1}{k-1}} N^{N-k}.
\end{equation}

Second relation \cite{Comtet_1974},
\begin{equation} \label{relacja_abz}
\begin{split}
B_{N,k}\left(abx_1,ab^2x_2,\dots,ab^{N-k+1}x_{N-k+1}\right) \\= 
 a^kb^NB_{N,k}\left(x_1,x_2,\dots,x_{N-k+1}\right),
\end{split}
\end{equation}

\noindent for short,
\begin{equation}
B_{N,k} ( \{ab^gx_g \}) = a^kb^NB_{N,k} (\{x_g\}).
\end{equation}

{
With the above relation, Bell polynomial $B_{N,k}(\left\{  g!/2^{g-1} \right\})$ can be transformed in the following way:
\begin{widetext}
\begin{align}  
  \phantom{B_{N,k}\left(\left\{ \frac{g!}{2^{g-1}} \right\}\right)}
  &\begin{aligned} \label{A8}
    \mathllap{B_{N,k} \left(\left\{  \frac{g!}{2^{g-1}} \right\}\right)} & = B_{N,k}\left(\left\{ 2 \cdot 2^{-g} g! \right\}\right)
  \end{aligned}\\
  &\begin{aligned}
    \mathllap{} & = B_{N,k}\left( 2 \cdot 2^{-1} 1!, 2 \cdot2^{-2} 2!, \dots, 2 \cdot 2^{-(N-k+1)} (N-k+1)! \right)
  \end{aligned}\\
    &\begin{aligned} \label{uproszczenie_const_1}
    \mathllap{} & = 2^k 2^{-N} \binom{N-1}{k-1} \frac{N!}{k!}.
  \end{aligned}
\end{align}
\end{widetext}
}

Third relation \cite{Comtet_1974},
\begin{equation} \label{relacja_g!}
\begin{split}
B_{N,k}(\{g!\}) = B_{N,k}(1!,2!,\dots,(N-k+1)!) \\ = \binom{N-1}{k-1}\frac{N!}{k!}.
\end{split}
\end{equation}

Fourth relation \cite{Wang_2009},
\begin{equation} \label{relacja_falling}
 B_{N,k}\left(\left\{  g!g   \right\}\right)  = 
  \frac{1}{k!} \sum^{k}_{j=0} (-1)^{k-j} \binom{k}{j} (j+k+N-1)_N,
\end{equation}

\noindent where the so-called falling factorial $(a)_b$ was defined as
\begin{equation} 
(a)_b = a(a-1) \dots (a-b+1).
\end{equation}

Calculating an incomplete Bell polynomial value from the definition, Eq.~(\ref{Bell_polynom_def}), is NP-hard problem; thus, an efficient way to calculate these polynomials is a recursive relation,
\begin{multline} \label{Bell_recursion}  
 B_{n,k} \left( a_1, a_2, \dots, a_m, \dots, a_{n-k+1} \right) \\ = \sum_{m=1}^{n-k+1} \binom{n-1}{m-1} a_{m} B_{n-m,k-1},
\end{multline}
 
\noindent where $B_{0,0} = 1$, $B_{n,0} = 0$ for $n\ge1$, and $B_{0,k} = 0$ for $k\ge1$.

Although several computational environments implement Bell polynomials (e.g., Wolfram Mathematica) and relevant code can be easily found on the Internet for some others (e.g., MATLAB), a problem in applying theoretical expressions is the explosion of digits of Bell polynomials for $N>70$. As this number immediately exceeds the precision of the standard programming environments, calculations need to be performed with the help of arbitrary precision computation packages. An implementation for a C++ environment using an arbitrary precision package (GNU MPFR for C++) can be found at GitHub \cite{cpp_libraries}.

{
\subsection{Generating function method} \label{Gen_Func_Meth}
}

Below, we will show how to obtain Eq.~(\ref{xg_binomials}) from Eq.~(\ref{xg_recursive_definition_const}). This derivation \cite{Lepek_2019} is worth reviewing because of (at least) two important reasons. First, it proves that Eqs.~(\ref{xg_binomials}) and (\ref{xg_recursive_definition_const}) define the same sequence. The recursive form will be later used to obtain $x_g$ for \textit{any} given kernel. Second, the methodology to obtain Eq.~(\ref{xg_binomials}) from Eq.~(\ref{xg_recursive_definition_const}), i.e., the generating function method, was the main mathematical tool applied to derive explicit solutions to several important kernels (e.g., product, additive, condensation, electrorheological).

Expanding binomials { in Eq.~(\ref{xg_recursive_definition_const})} we have
\begin{equation} \label{EQ12} 
 x_g=\frac{1}{2}\sum^{g-1}_{h=1}{\frac{g!}{h!\left(g-h\right)!}\frac{\left(g-2\right)!}{\left(h-1\right)!\left(g-h-1\right)!}}x_h x_{g-h}. 
\end{equation}

We introduce a new parameter,
\begin{equation} \label{EQ13} 
 y_g=\frac{x_g}{g!\left(g-1\right)!}.
\end{equation}

Elementary transformations of Eq.~(\ref{EQ12}) and using substitution, Eq.~(\ref{EQ13}), result in
\begin{equation} \label{EQ14} 
 {\left(g-1\right)y}_g=\frac{1}{2}\sum^{g-1}_{h=1}{y_h y_{g-h}}. 
\end{equation}

Now, we multiply both sides of~(\ref{EQ14}) by $\sum^{\infty }_{g=1}{z^g}$ to obtain
\begin{equation} \label{EQ15}  
 \sum^{\infty }_{g=1}{{\left(g-1\right)y}_gz^g}=\frac{1}{2}\sum^{\infty }_{g=1}{\sum^{g-1}_{h=1}{\left(y_h z^h\right)\left(y_{g-h}z^{g-h}\right)}}.
\end{equation}

Transforming left-hand side gives
\begin{equation} \label{EQ16}
\begin{split}
 \sum^{\infty }_{g=1}{{\left(g-1\right)y}_gz^g} &= \sum^{\infty }_{g=1}{{gy}_gz^g}-\sum^{\infty }_{g=1}{y_gz^g} \\ &= z\sum^{\infty }_{g=1}{gy_gz^{g-1}}-G\left(z\right) \\ &= z\frac{\partial }{\partial z}\sum^{\infty }_{g=1}{y_gz^g-}G\left(z\right) \\ &= z\frac{\partial }{\partial z}G\left(z\right)-G\left(z\right),
\end{split}
\end{equation}

\noindent where
\begin{equation} \label{EQ17} 
 G(z)\equiv \sum^{\infty }_{g=1}{y_gz^g} 
\end{equation}

\noindent is the generating function for $y_g$. Analogously, after transforming right-hand side of Eq.~(\ref{EQ15}) we obtain
\begin{equation} \label{EQ18} 
\begin{split}
 \frac{1}{2}\sum^{\infty }_{g=1}&{\sum^{g-1}_{h=1}{\left(y_hz^h\right)\left(y_{g-h}z^{g-h}\right)}} \\ & = \frac{1}{2}\left(\sum^{\infty }_{h=1}{y_hz^h}\right)\left(\sum^{\infty }_{g-h=1}{y_{g-h}z^{g-h}}\right) \\ & = \frac{1}{2}G\left(z\right)G\left(z\right).
\end{split}
\end{equation}

The summation boundaries have been changed due to the observation that, in the case of $y_hz^h$, the sum takes all expressions from 1 to infinity and, in the case of $y_{g-h}z^{g-h}$, the sum also takes all the expressions from $g-h=1$ to infinity. 

By the above transformations{, Eqs.~(\ref{EQ16}) and (\ref{EQ18}), from Eq.~(\ref{EQ15})} we obtained differential equation for $G(z)$,
\begin{equation} \label{EQ19} 
 z\frac{\partial }{\partial z}G(z)-G(z)=\frac{1}{2}{\left(G(z)\right)}^2. 
\end{equation}

As Eq.~(\ref{EQ19}) is an { ODE which can be solved by separating variables, namely,
\begin{equation}
 \frac{{d} G(z) }{\frac{1}{2} G^2(z)+G(z)} = \frac{{d} z}{z},
\end{equation}
}

\noindent its solution may be written as
\begin{equation} \label{EQ20}
\begin{split}
 G(z) &= \frac{2Cz}{1-Az} = 2Cz\left(\frac{1}{1-Cz}\right) \\ &= 2Cz\sum^{\infty }_{g=0}{{\left(Cz\right)}^g} = 2\sum^{\infty }_{g=1}{{\left(Cz\right)}^g}
 \end{split}
\end{equation}

\noindent with $C$ being a constant.

Recalling the fact that $G\left(z\right)=\sum^{\infty }_{g=1}{y_gz^g}$ and the substitution $y_g$, we compare the first elements of the series, 
\begin{equation} \label{EQ21} 
 2C^g=\frac{x_g}{g!\left(g-1\right)!}.
\end{equation}

Assuming reasonably that $x_1=1$, we can calculate the constant $C=1/2$ to obtain
\begin{equation} \label{xg_const_appendix} 
 x_g = \frac{g!\left(g-1\right)!}{2^{g-1}}
\end{equation}

\noindent which remains in full compliance with Eq.~(\ref{xg_binomials}). This way, we obtained explicit expression for $x_g$ from a more general (recursive) form. The above reasoning was used to find explicit solutions to several kernels (see Sec.~\ref{Solutions}). 
\newline\newline

{
\subsection{Transformations for standard deviation} \label{trans_for_std_dev}
}

One can calculate standard deviation corresponding to the average number of clusters of a given size as \cite{Fronczak_2018}
\begin{equation} \label{std_dev_general_appendix}  
 {\sigma }_s = \sqrt{\left\langle n_s^2 \right\rangle -{\left\langle n_s\right\rangle }^2} = \sqrt{\left\langle n_s\left(n_s-1\right)\right\rangle +\left\langle n_s\right\rangle -{\left\langle n_s\right\rangle }^2},
\end{equation}

\noindent where (presenting the following transformations may be useful for clarity as they were not presented explicitly before)
\begin{align}
  &\phantom{\left\langle n_s\left(n_s-1\right)\right\rangle}
  \begin{aligned} \label{eq_42}
    \mathllap{\left\langle n_s\left(n_s-1\right)\right\rangle} &= \sum_{\Omega }{n_s( \Omega) (n_s(\Omega)-1) P(\Omega)}
  \end{aligned}\\
  &\begin{aligned}
    \mathllap{} &= \frac{N!}{B_{N,k}\left(\left\{{\omega }_g\right\}\right)} \sum_{\left\{n_g\right\}} {n_s(n_s-1) \prod^{N-k+1}_{g=1}{\frac{1}{n_g!}{\left(\frac{\omega_g}{g!}\right)}^{n_g}} }
  \end{aligned}\\
  &\begin{aligned}
    \mathllap{} &= \frac{N!}{B_{N,k}\left(\left\{{\omega }_g\right\}\right)} \sum_{\left\{n_g\right\}} (n_s-1) { \left( \omega_s \frac{\partial}{\partial \omega_s} \right) \prod^{N-k+1}_{g=1}{\frac{1}{n_g!}{\left(\frac{\omega_g}{g!}\right)}^{n_g}} }
  \end{aligned}\\
  &\begin{aligned}
    \mathllap{} &= \frac{N!}{B_{N,k}\left(\left\{{\omega }_g\right\}\right)} \sum_{\left\{n_g\right\}} { \left( \omega_s \frac{\partial}{\partial \omega_s} \right) \left( \omega_s \frac{\partial}{\partial \omega_s} \right) \prod^{N-k+1}_{g=1}{\frac{1}{n_g!}{\left(\frac{\omega_g}{g!}\right)}^{n_g}} }
  \end{aligned}\\
  &\begin{aligned}
    \mathllap{} &= \frac{1}{B_{N,k}\left(\left\{{\omega }_g\right\}\right)}  \left( \omega_s^2 \frac{\partial^2}{\partial \omega_s^2} \right) B_{N,k}\left(\left\{{\omega }_g\right\}\right)
  \end{aligned}\\
  &\begin{aligned}
   \mathllap{} &=  \binom{N}{s} \omega_s^2 \frac{1}{B_{N,k}\left(\left\{{\omega }_g\right\}\right)}   \frac{\partial}{\partial \omega_s}  B_{N-s,k-1}\left(\left\{{\omega }_g\right\}\right)
  \end{aligned}\\
  &\begin{aligned} \label{eq_49}
   \mathllap{} & = \binom{N}{s}\binom{N-s}{s} {{\omega }_s}^2\frac{B_{N-2s,k-2}\left(\left\{{\omega }_g\right\}\right)}{B_{N,k}\left(\left\{{\omega }_g\right\}\right)}.
   \end{aligned}
\end{align}

Finally,
\begin{equation} \label{std_dev_general_addition_appendix}  
 \left\langle n_s\left(n_s-1\right)\right\rangle =\binom{N}{s,s}{{\omega }_s}^2\frac{B_{N-2s,k-2}\left(\left\{{\omega }_g\right\}\right)}{B_{N,k}\left(\left\{{\omega }_g\right\}\right)}\
\end{equation}

\noindent for $2s \leqslant N$, and $\left\langle n_s\left(n_s-1\right)\right\rangle = 0$ for other cases. For short, $\binom{N}{s,s}=\binom{N}{s}\binom{N-s}{s}$.
\newline\newline

\subsection{Lagrange inversion}

Lagrange inversion method aims to obtain an explicit (series) representation of a function if this function is given in the implicit form. Let us consider
\begin{equation}
    f(G) = F,
\end{equation}

\noindent where $G=G(z)$, $F=F(z)$ and $G$ is a function, for which we need an explicit form, i.e., $G=f^{-1}(F)$.

In the Lagrange inversion method (cf. p.~148 in Ref.~\cite{Krapivsky_2010}), the general form of the inverse function is given as a series,
\begin{equation} \label{lagrange_inversion_general}
G(F) = a + \sum^{\infty}_{n=1}{G_n \frac{ \left( F-f(a) \right)^n }{ n! } },
\end{equation}

\noindent where coefficients $G_n$ are defined as
\begin{equation} \label{lagrange_inversion_Gn}
G_n = \lim_{w \to a} \left[ \frac{d^{n-1}}{dw^{n-1}}  \left( \frac{w-a}{f(w)-f(a)} \right)^n  \right].
\end{equation}

A method to find a sequence $ G_n $ is based on calculating values of Eq.~(\ref{lagrange_inversion_Gn}) for several numbers $ n $, e.g., $ n = 1,2,3,4 $. Next, considering these results, one shall guess an expression for $ G_n $.

Now, we will briefly present the technique of applying the above method to the equations considered in Sec.~\ref{Solutions}.

For the product kernel, we obtained an implicit equation for generating function $G(z)$ in the form of
\begin{equation} \label{G_iloczynowy_dodatek} 
 G(z) e^{-G\left(z\right)}=Cz,
\end{equation}

\noindent where $C$ was a constant (cf. Eq.~(\ref{EQ25})). 

Eq.~(\ref{G_iloczynowy_dodatek}) has form of $ f \left(G \right) = F $, where $ f \left (G \right) $ stands for left-hand side of Eq.~(\ref{G_iloczynowy_dodatek}), and $ F $ stands for right-hand side of Eq.~(\ref{G_iloczynowy_dodatek}). 

Now, we will use Eq.~(\ref{lagrange_inversion_general}). In general, we can choose $a$ arbitrarily. Taking $ a = 0 $ we have $ f (a = 0) = 0 $. We write
\begin{equation} \label{lagrange_inversion_a0_dodatek}
G(F) = \sum^{\infty}_{n=1}{G_n \frac{F^n }{ n! } }
\end{equation}

\noindent and
\begin{equation} \label{lagrange_inversion_Gn_a0}
\begin{split}
G_n &= \lim_{w \to 0} \left[ \frac{d^{n-1}}{dw^{n-1}}  \left( \frac{w}{f(w)} \right)^n  \right] \\ &= \lim_{w \to 0} \left[ \frac{d^{n-1}}{dw^{n-1}}  \left( \frac{w}{we^{-w}} \right)^n  \right] \\ &= \lim_{w \to 0} \left[ \frac{d^{n-1}}{dw^{n-1}}  e^{wn}\right].
\end{split}
\end{equation}

Next, we will determine the values of Eq.~(\ref{lagrange_inversion_Gn_a0}) for the fist integers, $n=1,2,3,4,5$,
\begin{align}
  \phantom{\frac{d^{1}}{dw^{1}} e^{2w}}
  &\begin{aligned}
    \mathllap{\frac{d^{0}}{dw^{0}} e^{w}\ } &= e^w \xrightarrow{w\to0} 1 = 1^0,
  \end{aligned}\\
  &\begin{aligned}
    \mathllap{\frac{d^{1}}{dw^{1}} e^{2w} } & = 2e^{2w} \xrightarrow{w\to0} 2 = 2^1,
  \end{aligned}\\
  &\begin{aligned}
    \mathllap{\frac{d^{2}}{dw^{2}} e^{3w}} = \frac{d^{1}}{dw^{1}} 3e^{3w} = 3^2e^{3w} \xrightarrow{w\to0} 9 = 3^2,
  \end{aligned}\\
  &\begin{aligned} 
    \mathllap{\frac{d^{3}}{dw^{3}} e^{4w}} = \dots = 4^3e^{4w} \xrightarrow{w\to0} 4^3,
  \end{aligned}\\
  &\begin{aligned} 
    \mathllap{\frac{d^{4}}{dw^{4}} e^{5w}} = \dots = 5^4e^{5w} \xrightarrow{w\to0} 5^4.
  \end{aligned}
\end{align}

At this point, we can guess that the sequence of interest is $G_n = n^{n-1}$. We can write down the result of Lagrange inversion for Eq.~(\ref{G_iloczynowy_dodatek}), which is a series representation of function $ G(F)$, namely,
\begin{equation} \label{G(F)_iloczynowy_dodatek}  
 G(F)=\sum_{n\ge 1}{\frac{n^{n-1}}{n!}F^n}.
\end{equation} 

Finally, we find $G(z)$ as
\begin{equation} \label{G(z)_iloczynowy_dodatek}  
 G(z)=\sum_{n\ge 1}{\frac{n^{n-1}}{n!}(Cz)^n}.
\end{equation}

{ The above result complies with the solution to Lambert's $W$-function \cite{NIST}, as used in the main text, Eq.~(\ref{EQ26}). However, for a general case, an explicit representation of $G$ needs to be found using the above method (as for the condensation kernel).}

{
\section{Fields for potential application} \label{Appendix_fields}
}

{ In this {appendix}, we provided selected subdomains of the aggregation studies, where particular kernel forms are known explicitly. As such, they can undergo an analysis in the combinatorial approach with finding $x_g$ recursively to obtain theoretical { approximations}. For each case, we provide a short description of the context for the given kernels. There are two other examples, not present in this {appendix}; these are planetesimal creation and aerosol growth, described in detail and accompanied by figures in Sections~\ref{example_proto} and \ref{example_aerosols}. In the descriptions below, $d_f$ stands for the fractal dimension of particles.}
\newline

\tocless\subsubsection{Dust agglomeration} 

As in the early solar system, collisions between the dust grains were caused by Brownian motion: ballistic for small dust aggregates and dominated by diffusion for large aggregates \cite{Blum_2006}. An evolution of initially monodisperse samples of spherical $\mathrm{SiO}_{2}$ particles was studied experimentally and by simulation \cite{Blum_2000, Blum_2004}, suggesting aggregation with kernels $K_{b}$ and $K_{d}$ for ballistic and diffusive regimes, respectively,
\begin{equation}
K_b \propto \left(  s^{0.9} + g^{0.9} \right) \left( s^{-1} + g^{-1} \right)^{1/2},
\end{equation}
\begin{equation}
K_d \propto \left(  s^{-0.91} + g^{-0.91} \right) \left( s^{1/d_f} + g^{1/d_f} \right).    
\end{equation}

Scaling relations and approximations of kernels for $s \gg g$ are also proposed in Ref.~\cite{Blum_2006}.
\newline

\tocless\subsubsection{Saturn's rings}
 
In Ref.~\cite{Brilliantov_2015}, a set of rate equations is studied with aggregation and decomposition parts, simplified to a generalized effective product aggregation kernel,
\begin{equation}
K \propto (sg)^{\mu},    
\end{equation}

\noindent with $\mu = 1/12$ and $\mu = 1/3$ for which a steady-state solution in the Smoluchowski framework is known.
\newline

\tocless\subsubsection{Aggregation of proteins}

Protein aggregation with a kernel of the form
\begin{equation}
    K \propto (sg)^\gamma (s^{1/d_f}+g^{1/d_f})(s^{-1/d_f}+g^{-1/d_f}),
\end{equation}

\noindent where $\gamma=1-1/d_f$, may be found in relation to, e.g., antibody dynamics \cite{Nicoud_2016} and drug efficiency \cite{Zidar_2018}.  Authors usually study the coagulation model by numerically solving the Smoluchowski equation. They obtain useful experimental data. The impact of the aggregate fractal nature on particle size distribution is emphasized. The topic is developing \cite{Pang_2023}.
\newline

\tocless\subsubsection{Magneto-rheological (MR) aggregation}

Although close to ER processes, MR aggregation is still intensively studied under various conditions, mainly via parametric and microscopic models \cite{Pei_2022}.  Simulation models of the microscopic behaviour of MR fluid \cite{Han_2010} exist. Methods for experimental studies are known \cite{Dominguez_2007}, and they are easily transformable to the timeline of coagulation acts. Linear assemblies under the presence of the external field are also observed in aerosol growth \cite{Biswas_2021}.

A general form of the kernel for linear (single-)chain aggregation is
\begin{equation}
K \propto (s^{\gamma}+g^{\gamma}).   
\end{equation}

\tocless\subsubsection{Polymer growth}

Growth of colloidal polymer aggregates (linear, branched, circular, meta-structures, copolymers) is continuously of high interest \cite{Liu_2010, Chen_2012, Yang_2018, Li_2020, Stuij_2021, BinLiu_2022}. Magnetic colloidal chains are expected to be promising microrobotic systems \cite{Mhanna_2022}. Collecting microscopic cluster statistics is perfectly doable and widely used, e.g., by transmission electron microscopy (TEM; see Figs. in Ref.~\cite{Stuij_2021}, also cf. Fig.~2 in Ref. \cite{Li_2020}). It is common that authors track the average number of x-mers but do not relate to aggregation theory (lack of robust theory?). Another example is fibrin polymerization (blood clotting) \cite{Nelson_2021} where $K \propto (s+1)(g+1)$ was used \cite{Guria_2009}. RNA reveals the characteristics of an ideal polymer \cite{Fang_2011}.

For the polymer growth analysis, kernels for linear chains and a branched-chain polymerization kernel, $K \propto (A+s)(A+g)$, are of interest.
\newline

\tocless\subsubsection{Coagulation of milk}

Coagulation of milk (into, e.g., cheese) is an everyday process. However, it is also a perfect example of aggregation physics (containing fractal models of rennet curds and gel formation) \cite{Horne_2017_chapter}. Despite issues with microscopic observation, some cluster statistics may be observed by laser light scattering \cite{Bansal_2007}. Effects of environmental conditions on the coagulation behavior of casein micelles are known \cite{Li_2019}. As for the kernel formulation, the polyfunctional model of the Flory--Stockmayer type is still in use \cite{Salvador_2022},
\begin{equation}
K \propto [4+2(f-2)(s+g)+(f-2)^2 sg],
\end{equation}

\noindent with $f$ being the number of functional sites.
\newline

\tocless\subsubsection{Soot particle coagulation}

Soot, a particulate byproduct of the incomplete combustion of hydrocarbon fuels, is a major air pollutant and a threat to human health. Coagulation of soot nanoparticles is researched by modeling and experimental techniques (TEM, SEM) \cite{Hou_2022, Henderson_2022}. Here,
\begin{equation}
K_{F\!M\!R} \propto \left(  s^{1/d_f} + g^{1/d_f} \right) ^2 \left( s^{-1} + g^{-1} \right)^{1/2}
\end{equation}
  
\noindent is a generalization of the Brownian free-molecular kernel, Eq.~(\ref{aerosol_FRM}), for fractal-shape aggregates \cite{Maricq_2007}. Formulations for other regimes may be found in Ref.~\cite{Maricq_2007}.
\newline

\tocless\subsubsection{Coalescence in flows}

A {classical} kernel for coalescence in the turbulent flow,
\begin{equation}
K \propto \left(  s^{2/3} + g^{2/3} \right) \left( s^{2/9} + g^{2/9} \right)^{1/2} \mathrm{exp} \! \left(-C \left[ \frac{sg}{s+g} \right]^{4/3} \right),    
\end{equation}

\noindent with $C$ being an empirical constant, was derived for the coalescence of two drops of volumes $s$ and $g$ in turbulent flows \cite{Coulaloglou_1977}, used to model stirred dispersions of drops \cite{Friesen_2003}. { A general form for flow conditions, considering the effects of Brownian motion, flow field and interparticle interactions, was formulated in Ref.~\cite{Lattuada_2016}.} Colloidal dispersions may undergo measurement of the size distribution of droplets using the laser diffraction technique \cite{Low_2020}.
\newline

{
\section{Relation to scaling theory}

\subsection{Introduction to scaling solutions} 
\label{ScalingTheory}

Here, we will briefly recall scaling solutions to the Smoluchowski equation to demonstrate their relation to the combinatorial solutions later.

The usual description of the aggregating systems uses the kinetic Smoluchowski equations of coagulation,
\begin{equation} \label{dc_x_smoluchowski}
\frac{dc_s}{d\tilde{t}}=\frac12\sum_{g,h=1}^\infty K(g,h)c_gc_h\left( \delta_{g+h,s}-\delta_{g,s}-\delta_{h,s} \right)
\end{equation}

\noindent where $c_s(\tilde{t})$ stands for the concentration of aggregates of size $s$ and $K(g,h)$ is the reaction kernel, i.e., the rate at which aggregates of sizes $g$ and $h$ merge.

Basic information about the aggregating system may be recognized from the matrix of reaction rates $K(g,h)$ (p.~153 in Ref.~\cite{Krapivsky_2010}). Only two features of this matrix determine the asymptotic properties of the mass distribution. The first is the homogeneity index, $\lambda$, defined by
\begin{equation}
K(ag,ah) \sim a^{\lambda} K(g,h)
\end{equation}

\noindent that gives the overall mass dependence of the reaction rate. The system is gelling if $\lambda > 1$ and non-gelling otherwise \cite{Leyvraz_2022}.

The second is the index $\nu$, defined by 
\begin{equation}
K(1,h) = K(h,1) \sim h^{\nu} \quad \mathrm{for} \quad h \rightarrow \infty
\end{equation}

\noindent that characterizes the relative importance of reactions between clusters of disparate masses.

According to the behaviour of a system aggregating with the given kernel $K$, the kernel may be classified as one of the three distinct universality classes. These classes are characterized by the following general behavior \cite{Krapivsky_2010}:

Class I. The high reactivity of large clusters causes them to quickly disappear, while small clusters tend to persist {(behaviour qualitatively different to that of Classes II and III)}. Consequently, the cluster mass distribution decays monotonically with mass. The product kernel, $K=gh$ ($\lambda=2$, $\nu=1$), typifies this class of reaction rates. This case corresponds to $\lambda > \nu$.

Class II. If all three types of reactions are of the same order {($\lambda = \nu$)}, the asymptotic behavior is sensitive to details of the reaction rates. The constant kernel, $K=1$ ($\lambda=0$, $\nu=0$), belongs to this class.

Class III. Large–small interactions dominate, and as the reaction develops, small clusters are quickly depleted. Thus, the system has a dearth of small clusters, leading to a peaked mass distribution. This case corresponds to $\lambda < \nu$. An inverse (LC) kernel, $K(g,h)=1/g+1/h$ ($\lambda=-1$, $\nu=0$), may be an example here.


Now, having recalled the indices $\lambda$ and $\nu$ we can proceed to scaling solutions. Scaling is based on the observation that the typical cluster mass grows systematically with time. Thus, a change in time scale corresponds to a change in mass scale, so that the mass distribution becomes invariant when the mass is scaled appropriately with time. We can define the actual scaling solution (or similarity solution) by writing the so-called scaling ansatz \cite{Krapivsky_2010}:
\begin{equation}
c_s(\tilde{t}) = \frac{1}{\hat{s}^2} f \! \left( \frac{s}{\hat{s}} \right) = \frac{1}{\hat{s}^2} f \! \left( u \right),
\end{equation}

\noindent where $c_s(\tilde{t})$ is concentration of clusters of mass (size) $s$ in time $\tilde{t}$, $\hat{s} = \hat{s}(\tilde{t})$ is a typical cluster mass (sometimes referred to as an average cluster mass), $u$ is scaled mass, and $f$ is the scaling function. 

Scaling provides a route to the asymptotic solution when exact solutions are difficult or impossible to obtain. However, it must be clearly noted that the definition by the concentration inherits all the assumptions relevant to the Smoluchowski equation: continuous time, continuous concentrations, { an infinite number of particles, and perfect mixing (spatial invariance) of the system (the latter also applies to Marcus--Lushnikov)}.

The time dependence of the typical mass may be determined as \cite{Krapivsky_2010}:
\begin{equation} \label{typical_mass}
  \hat{s}(t) =
    \begin{cases}
      \tilde{t}^{\frac{1}{1-\lambda}} & \text{for}~ \lambda < 1\\
      e^{\Lambda \tilde{t}} & \text{for}~ \lambda = 1\\
      (\tilde{t}_g - \tilde{t})^{-1} & \text{for}~ 1 < \lambda \le 2,
    \end{cases}       
\end{equation}

\noindent where $\Lambda$ is the separation constant (for details, see p. 154 in Ref.~\cite{Krapivsky_2010}). For non-gelling kernels, the time dependence of the typical mass is determined only by the homogeneity index $\lambda$, with $\nu$ affecting only details. For instance, for the constant kernel ($\lambda=0$), the typical mass is simply $\hat{s}(\tilde{t}) = \tilde{t}$.

The dependence of the scaling function $f$ on the scaled mass $u$ is governed by the complicated nonlinear integro-differential equation, not susceptible to solving for arbitrary kernels. Nevertheless, it has been shown that for large scaled mass, $u \gg 1$, the scaling function has an exponential dependence $f(u) \sim e^{-au}$. On the other hand, for $u \ll 1$, various numerical results indicate that $f(u) \sim u^{-b}$ with $b$ being kernel-dependent (as provided by the classic textbook, Ref.~\cite{Krapivsky_2010}).

In a newer publication, Ref.~\cite{Leyvraz_2022}, we can find additional and relevant information. For instance, if $s,\tilde{t} \to\infty$ in a way that $s/ \hat{s}(\tilde{t})=u$ remains constant, the scaling function may be defined for non-gelling systems by the following limit,
\begin{equation}
    \lim_{\begin{smallmatrix} \tilde{t}\to\infty & \\ s/\hat{s}(\tilde{t})=u \end{smallmatrix}} [s^2c_s(\tilde{t})] = f(u)
\end{equation}

\noindent and for gelling systems as 
\begin{equation}
    \lim_{\begin{smallmatrix} \tilde{t}\to\infty & \\ s/\hat{s}(\tilde{t})=u \end{smallmatrix}} [s^\tau c_s(\tilde{t})] = f(u).
\end{equation}

By additionally defining { $\mu$ \cite{vanDongen_1985, Leyvraz_2022},}
\begin{equation}
    K(g,h) \simeq g^{\mu} h^\nu ~~~~~{(g \ll h; \lambda=\mu+\nu)},
\end{equation}

\noindent one describes the behavior in the strongly asymmetric case for $g \ll h$. Both the large-$u$ ($u \gg 1$) and small-$u$ ($u \ll 1$) behavior of $f(u)$ are determined by the set of exponents $\lambda$, $\mu$, and $\nu$. If the system is non-gelling ($\lambda \le 1$), the small-$u$ behavior of $f(u)$ depends on $\mu$ as below.

For $\mu>0$ {(which corresponds to Class I)}, $f(u) \sim u^{1-\lambda}$.

For $\mu=0$ {(Class II)}, small-$u$ behavior must be determined separately in each case.

For $\mu<0$ {(Class III)}, $f(u)$ goes to zero faster than any power.

On the other hand, for large scaled mass $u$, we always have $f(u) \sim u^{2-\lambda} \mathrm{exp}(-\mathrm{const} \cdot u)$ except when $\nu=1$, in which case the exponent arising before the exponential must be determined for every particular case \cite{Leyvraz_2022}.

} 

{

\subsection{Asymptotic analysis and relation to scaling theory} \label{Asymptotics}

In this section, we will give a brief outline of the asymptotic analysis of the combinatorial solutions, the analysis as performed in Ref.~\cite{Leyvraz_2022}, and describe concisely how its results relate to the scaling theory.

Let $t=N-k$ be the discrete time in the combinatorial solutions, i.e., in the discrete-time Marcus--Lushnikov model (cf. Eq.~(\ref{k})) and let $\theta=t/N$ be its continuous equivalent. Thus, $\tilde{t}$ will stand for the continuous time in the kinetic (Smoluchowski) equations. {With this, we align (to some extent) the notation from Ref.~\cite{Leyvraz_2022} with the main text}.

The large-size limit behavior of the average number of clusters of given size from the discrete Marcus--Lushnikov model (combinatorial solutions) and concentrations $c_s(\tilde{t})$ satisfying the kinetic equations (Eq.~(\ref{dc_x_smoluchowski})) are related by 
\begin{equation} \label{AppD_limit_1}
    \lim_{\begin{smallmatrix} t,N\to\infty & \\ t/N=\theta \end{smallmatrix}}  \frac{\langle n_s(t) \rangle}{N} = c_s(\tilde{t}),
\end{equation}
\begin{equation} \label{AppD_limit_2}
    \sum_{s=1}^{\infty} c_s(\tilde{t}) = 1-\theta.
\end{equation}

The above equations provide correct connection between $\tilde{t}$ and $\theta$ and lead to the correct concentration profile $c_s(\tilde{t})$ \cite{Norris_1999, Leyvraz_2022}.

As $\langle n_s(t) \rangle$ in the combinatorial approach is, in general, defined by Bell polynomials, one needs to find the asymptotic behavior of Bell polynomials for $N\to \infty$. A method used in Ref.~\cite{Leyvraz_2022} may be regarded as a kind of the Darwin--Fowler method. Known from statistical mechanics, the Darwin--Fowler method is used to derive the distribution functions with mean probability. One example of use may be deriving the Maxwell--Boltzmann distribution for the canonical ensemble in a classic textbook \cite{Huang_1987_book}. Although not dedicated originally to Bell polynomials, the Darwin--Fowler method operates with polynomials and, as such, may be adopted to the present need. Nevertheless, we feel that such an adoption is not widely known throughout the works devoted to Bell polynomials. This feeling may be supported by the fact that the author of Ref.~\cite{Leyvraz_2022}, even though he called the transformations elementary, failed to provide any reference for that.

Providing a step-by-step description of the Darwin--Fowler method falls beyond the scope of this mini-review article; thus, we will only report the findings from Ref.~\cite{Leyvraz_2022}. 

Having defined
\begin{equation} \label{xi_definition}
    (s-1) \xi_s = \sum_{l=1}^{s-1} K(l,s-l) \xi_l \xi_{s-l}, ~~~~\xi_1 = 1,
\end{equation}
\begin{equation}
    G(w) = \sum_{s=1}^{\infty} \xi_s w^s,
\end{equation}

\noindent and
\begin{equation}
    w^\star(\theta) = \argminE_{0<w<w_c} [(1-\theta) \ln G(w)-\ln w],
\end{equation}

\noindent where $w_c$ is the convergence radius of $G(w)$, it has been found, via the above-mentioned asymptotic analysis, that
\begin{equation} \label{asymptotic_result}
    \lim_{\begin{smallmatrix} t,N\to\infty & \\ t/N=\theta \end{smallmatrix}}  \frac{\langle n_s(t) \rangle}{N} = \frac{1-\theta}{G[w^\star (\theta)]} \xi_s w^\star(\theta)^s.
\end{equation}

From the considerations provided therein, we learn that a combinatorial solution being an exact solution to the discrete Marcus--Lushnikov model implies that it is also an exact solution to the Smoluchowski equations. This is a key conclusion that supports the need to validate the combinatorial solutions against the scaling theory.

For the infinite time limit, $\theta \to 1$, Eq.~(\ref{asymptotic_result}) leads to the following scaling form,
\begin{equation} \label{c_j_t_limit}
    c_s(\tilde{t}) \simeq s^{-\lambda} { \left( \frac{w_c}{w^\star(\theta)} \right)^s } \left[ \sum_{l=1}^{\infty} c_l(\tilde{t}) \right].
\end{equation}

{Due to the large-size limit in Eq.~(\ref{asymptotic_result}), Eq.~(\ref{c_j_t_limit}) holds only for infinite systems and does not provide a reliable concentration profile for finite systems where a single cluster appears at the end of the system's evolution.} With some manipulations, Eq.~(\ref{c_j_t_limit}) may be expressed in the language of the scaling function as
\begin{equation}
    f(u) = \mathrm{const} \cdot u^{2-\lambda} e^{-u}. 
\end{equation}

The above result has been derived for a general case of the combinatorial Bell polynomial-based solution. At this point, the key conclusion is that it stays in good agreement with the scaling theory with the large-cluster size behavior, while it is in strong disagreement with the small-cluster size behavior (cf. last paragraphs in Appendix C 1).

As an example confirming the discrepancies, an inverse (LC) kernel, $K=1/g+1/h$, is recalled, for which exact large-time behavior may be found \cite{Leyvraz_2021, Leyvraz_2022} as
\begin{equation}
    \lim_{\tilde{t} \to \infty} \left[ \sum_{l=1}^\infty c_l(\tilde{t}) \right]^{-1} \exp \left[ \int_0^{\tilde{t}}{ \sum_{l=1}^\infty c_l(\tilde{t}')d\tilde{t}' } \right] c_1(\tilde{t}) = 1,
\end{equation}

\noindent which is incompatible with Eq.~(\ref{c_j_t_limit}). Indeed, true concentration of small clusters goes to zero faster than the concentration resulting from the combinatorial solution, as can be evidently seen in Fig.~\ref{Fig_kernele_zbiorczy}f, Fig.~1 in Ref.~\cite{Leyvraz_2022}, and Fig.~2 in Ref.~\cite{Lepek_2021}. 

On the other hand, for the so-called classical kernels, i.e., the constant, additive, and multiplicative (product) kernels, as well as for their combinations, it is characteristic that their small-$s$ and large-$s$ behaviors are identical. Apparently, this is the reason why the precision of the combinatorial expressions differs from the general case.

An explicit comparison between the exact distribution of the discrete Marcus–Lushnikov model and the probability distribution from the combinatorial approach, Eq.~(\ref{P_omega}), was performed for a small system of $N=20$ particles. For the constant and additive kernels, the distributions were found to coincide exactly, yet to differ in the case of the product kernel.

Indeed, further theoretical analysis performed in Ref.~\cite{Leyvraz_2022} confirms and explains the above observations. By means of the standard approach to solving the classical kernels, i.e., by finding the generating function $G(w)$ for $\xi_j$ \cite{Leyvraz_2003}, one obtains $G$ and $w$ which may be used to substitute into the right-hand side of Eq.~(\ref{c_j_t_limit}).

{
In this way, for the product kernel ($\xi_s = s^{s-2}/j!$), we effectively obtain (see Ref.~\cite{Leyvraz_2022} for details)
\begin{equation}
    c_s(\tilde{t}) = \xi_s \tilde{t}^{s-1} e^{-s\tilde{t}},
\end{equation}

\noindent which is a correct solution to the Smoluchowski equation (cf., e.g., p.~148 in Ref.~\cite{Krapivsky_2010}), the so-called Stockmayer solution,} { expected to hold when there is no interaction between the finite particles and the gel (thus, leading to inaccuracies when compared to numerical simulations as there is no such restriction there).

For the other two basic kernels, the constant and additive, the relations for their respective $G(w)$ are known and may be easily used to show that the combinatorial solutions correspond to the exact solutions of the kinetic equations.

Another puzzling fact is that Eq.~(\ref{xi_definition}), which play an important role in the short-time behavior of the solutions to the Smoluchowski equation, is similar to the one of the fundamental equations in the combinatorial formalism, Eq.~(\ref{EQ14}), which was used to obtain exact results for any arbitrary time for the constant kernel.   
}
} 

{

\section{Smoluchowski solutions to basic kernels} \label{SmolSols}

Besides the formally correct asymptotic analysis given in Appendix C, in practice, there exists a need to evaluate combinatorial solutions against Smoluchowski theory for finite systems. In this appendix, we provide standard Smoluchowski solutions for the three basic kernels, which are solved exactly in the Smoluchowski theory and yield explicit (closed) forms.

The discrete-size Smoluchowski solutions to Eq.~(\ref{dc_x_smoluchowski}) for those three kernels under the monodisperse initial conditions ($c_s(0)=\delta_{s,1}$) are as follows.

For the constant kernel, we have ($0\le \tilde{t}<\infty$) \cite{Aldous_1999}
\begin{equation}
    c_s(\tilde{t}) = \left(1 + \frac{\tilde{t}}{2} \right)^{-2} \left( \frac{\tilde{t}}{2+\tilde{t}} \right)^{s-1}.
\end{equation}

For the additive (sum) kernel, we have ($0\le \tilde{t}<\infty$) \cite{Aldous_1999}
\begin{equation}
    c_s(\tilde{t}) = \frac{e^{-\tilde{t}}}{s!} ( (1-e^{-\tilde{t}})s )^{s-1}  e^{ -(1-e^{-\tilde{t}})s }.
\end{equation}

For the gelling multiplicative (product) kernel, for the pre-gel phase ($0 \le \tilde{t}\le 1$), we have the so-called Stockmayer solution \cite{Aldous_1999,Leyvraz_2022},
\begin{equation}
c_s(\tilde{t}) = \frac{s^{s-2}}{s!} \tilde{t}^{s-1} e^{-s\tilde{t}},
\end{equation}

\noindent and for the post-gel phase ($\tilde{t}>1$), we have the so-called Flory solution \cite{Leyvraz_2022},
\begin{equation}
c_s(\tilde{t}) = \frac{s^{s-2}}{s!} \frac{1}{\tilde{t}} e^{-s}.
\end{equation}

To connect time $\tilde{t}$ in the above Smoluchowski results with the discrete time $t$ of the combinatorial approach, we used Eq.~(\ref{time_S_time_comb}) (see Sec.~\ref{Remarks_on_time} for detailed comments). Given the combinatorial time $t$, for each case, we found $\tilde{t}$ so that both sides of the expression are equal. Concentrations $c_s(\tilde{t})$ from the Smoluchowski theory were translated to the numbers of clusters as $n_s=c_sN$.}

{
The above Smoluchowski solutions were plotted against the respective results of the numerical simulations and theoretical predictions given by the combinatorial solutions in Figs.~\ref{Fig_kernel_const}, \ref{Fig_kernele_zbiorczy}, and \ref{Fig_AppE}. Please see their respective captions for the detailed comments.
}


\begin{thebibliography}{199}


\bibitem{wiki_commons_2005}
J.~Carr: \textit{Blood clot in scanning electron microscopy}, https://commons.wikimedia.org/wiki/ File:Blood\_clot\_in\_scanning\_electron\_microscopy.jpg (last access: 25 September 2025).

\bibitem{Frohlich_2023}
J.A.~Fr\"{o}hlich, N.A.~Ruprecht, R.~Kohlus: {\em Drying Technol.\/} {\bf 41}, 1907--1923 (2023), Nozzle zone agglomeration in spray dryers: Determination of the agglomeration efficiency in the fines return by means of agglomerate properties and residence time distribution. 

\bibitem{Smoluchowski_1917} M.~Smoluchowski: {\em Phys. Z.\/} {\bf 17}, 557 (1916), Drei Vortrage uber Diffusion Bewegung und Koagulation von Kolloidteilchen. 

{
\bibitem{daCosta_1998}  
F.P.~da Costa: {\em J. Nonlinear Sci.\/} {\bf 8}, 619--653 (1998), A finite-dimensional dynamical model for gelation in coagulation processes.
}

\bibitem{Aldous_1999}
D.J.~Aldous: {\em Bernoulli\/} {\bf 5}, 3--48 (1999), Deterministic and stochastic models for coalescence (aggregation and coagulation): a review of the mean field theory for probabilists.

\bibitem{Leyvraz_2003}
A.~Leyvraz:
{\em Phys. Rep.\/}
{\bf 383}, 95--212 (2003),
Scaling theory and exactly solved models in the kinetics of irreversible aggregation. 

\bibitem{Blum_2006}
J.~Blum:
{\em Adv. Phys.\/}
{\bf 55}, 881--947 (2006),
Dust agglomeration. 


\bibitem{Krapivsky_2010}
   P.L.~Krapivsky, S.~Redner and E.~Ben-Naim: {\em A Kinetic View of Statistical Physics\/} (Chapter 5), Cambridge University Press,
   New York 2010.

\bibitem{Banasiak_2019}
   J.~Banasiak, W.~Lamb and P.~Lauren\c{c}ot: {\em Analytic Methods for Coagulation--Fragmentation Models\/}, CRC Press,
   Boca Raton 2019.

{
\bibitem{Rouwhorst_2020_Nature}
J.~Rouwhorst, C.~Ness, S.~Stoyanov, A.~Zaccone, P.~Schall: {\em Nat. Commun.\/} {\bf 11}, 3558 (2020), Nonequilibrium continuous phase transition in colloidal gelation with short-range attraction.

}



\bibitem{Marcus_1968}
A.H.~Marcus: {\em Technometrics\/} {\bf 10}, 133--143 (1968), Stochastic coalescence.

{
\bibitem{Lushnikov_1978}
A.A.~Lushnikov: {\em J. Colloid Interface Sci.\/} {\bf 65}, 276--285 (1978), Coagulation in finite systems.}

\bibitem{Lushnikov_2004}
A.A.~Lushnikov: {\em Phys. Rev. Lett.\/} {\bf 93}, 198302 (2004), From sol to gel exactly.

\bibitem{Lushnikov_2005}
A.A.~Lushnikov: {\em Phys. Rev. E\/} {\bf 71}, 046129 (2005), Exact kinetics of the sol-gel transition.

\bibitem{Lushnikov_2011}
A.A.~Lushnikov:
{\em J. Phys. A\/}
{\bf 44}, 335001 (2011),
Exact kinetics of a coagulating system with the kernel K = 1.

{
\bibitem{vanDongen_1987_fluct}
P.G.J.~Van Dongen, M.H.~Ernst: {\em J. Stat. Phys. \/} {\bf 49}, 879--926 (1987), Fluctuations in coagulating systems.

\bibitem{vanDongen_1987_fluct_II}
P.G.J.~Van Dongen: {\em  J. Stat. Phys. \/} {\bf 49}, 927--975 (1987), Fluctuations in coagulating systems. II.
}

\bibitem{Leyvraz_2022}
F.~Leyvraz: {\em Phys. Rev. E\/} {\bf 106}, 024133 (2022), Rate equation limit for a combinatorial solution of a stochastic aggregation model.

\bibitem{Fronczak_2019} A.~Fronczak, M.~Łepek, P.~Kukliński, P.~Fronczak:
{\em Phys. Rev. E\/}
{\bf 99}, 012104 (2019),
Coagulation with product kernel and arbitrary initial conditions: Exact kinetics within the Marcus--Lushnikov framework.


\bibitem{2018_Matsoukas}
T.~Matsoukas: {\em Kinetic Gelation\/}, in {\em Generalized Statistical Thermodynamics. Understanding Complex Systems\/}, Springer, Cham 2018.



\bibitem{grassberger1}
S.-W.~Son, C.~Christensen, G.~Bizhani, P.~Grassberger, M.~Paczuski:
{\em Europhys. Lett.\/}
{\bf 95}, 58007 (2011),
Irreversible aggregation and network renormalization.

\bibitem{grassberger2}
S.-W.~Son, C.~Christensen, G.~Bizhani, P.~Grassberger, M.~Paczuski:
{\em Phys. Rev. E\/}
{\bf 84}, 040102 (2011),
Exact solutions for mass-dependent irreversible aggregations.




\bibitem{Fronczak_2018}
A.~Fronczak, A.~Chmiel and P.~Fronczak:
{\em Phys. Rev. E\/}
{\bf 97}, 022126 (2018),
Exact combinatorial approach to finite coagulating systems. 

\bibitem{Lepek_2019}
M.~\L{}epek, P.~Kukli\'{n}ski, A.~Fronczak, P.~Fronczak: {\em Rep. Math. Phys.\/} {\bf 84}, 117--130 (2019), Exact combinatorial approach to finite coagulating systems through recursive equations. 

\bibitem{Lepek_2021}
M.~\L{}epek, A.~Fronczak, P.~Fronczak: {\em Physica D\/} {\bf 415}, 132756 (2021), Combinatorial solutions to coagulation kernel for linear chains.

\bibitem{Lepek_2021_ROMP}
M.~\L{}epek, A.~Fronczak, P.~Fronczak: {\em Rep. Math. Phys.\/} {\bf 88}, 89--113 (2021), Coalescence with arbitrary-parameter kernels and monodisperse initial conditions: A study within combinatorial framework.



\bibitem{Osinsky_2022}
A.I.~Osinsky, N.V.~Brilliantov: {\em Phys. Rev. E\/} {\bf 105}, 034119 (2022), Anomalous aggregation regimes of temperature-dependent Smoluchowski equations.

\bibitem{Kaushik_2023}
S.~Kaushik, S.~Kumar: {\em J. Comput. Appl. Math.\/} {\bf 419}, 114710 (2023), A novel optimized decomposition method for Smoluchowski’s aggregation equation.

{
\bibitem{Kang_1986}
K.~Kang, S.~Redner, P.~Meakin, F.~Leyvraz:
{\em Phys. Rev. A\/}
{\bf 33}, 1171--1182 (1986),
Long-time crossover phenomena in coagulation kinetics.
}




\bibitem{Eibeck_2000}
A.~Eibeck, W.~Wagner:
{\em SIAM J. Sci. Comput.}
{\bf 22}(3), 802--821 (2000),
An Efficient Stochastic Algorithm for Studying Coagulation Dynamics and Gelation Phenomena.

{
\bibitem{Lushnikov_2006_review}
A.A.~Lushnikov:
{\em Phys. D}
{\bf 222}, 37–53 (2006),
Gelation in coagulating systems.}



{
\bibitem{Ernst_1984}
M.H.~Ernst, E.M.~Hendriks, F.~Leyvraz: {\em J. Phys. A: Gen. Phys\/} {\bf 17}, 2137--2144 (1984), Smoluchowski's equation and the $\theta$-exponent for branched polymers.
}

\bibitem{vanDongen_1987}
P.G.J.~Van Dongen: {\em Physica A \/} {\bf 145}, 15 (1987), Solutions of Smoluchowski’s coagulation equation at large cluster sizes.

{
\bibitem{NIST}
F.W.J.~Olver et al.: NIST Digital Library of Mathematical Functions (Release 1.2.4 of 15 March 2025), https://dlmf.nist.gov (last access: 8 August 2025).
}

\bibitem{Wattis_2009}
Z.~Mimouni, J.A.D.~Wattis:
{\em Physica A}
{\bf 388}, 1067--1073 (2009),
Similarity solution of coagulation equation with an inverse kernel.

{
\bibitem{Leyvraz_2021}
F.~Leyvraz: {\em Phys. Rev. E\/} {\bf 103}, 022123 (2021), Exact asymptotic solution of an aggregation model with a bell-shaped distribution.
}

\bibitem{film_semenova}
N.~Semenov: \textit{Electrorheology}, a movie at YouTube, www.youtube.com/watch?v=ybyeMw1b0L4 (last access: 8 August 2025).



\bibitem{Semenov_2021}
A.~Danilin, K.~Kydralieva, N.~Semenov, E.~Kelbysheva: {\em Mater. Today Proc.\/} {\bf 34} (1), 239--242 (2021), Electrorheological properties of polyimide nanoparticles suspensions.

\bibitem{Dominguez_2007}
P.~Dom\'{i}nguez-Garc\'{i}a, S.~Melle, J.M.~Pastor, M.A.~Rubio: {\em Phys. Rev. E\/} {\bf 76}, 051403 (2007), Scaling in the aggregation dynamics of a magnetorheological fluid.

\bibitem{Han_2010}
K.~Han, Y.T.~Feng, D.R.J.~Owen: {\em Int. J. Numer. Meth. Engng\/} {\bf 84}, 1273–1302 (2010), Three-dimensional modelling and simulation of magnetorheological fluids.

\bibitem{Bossis_2013}
G.~Bossis, P.~Lan\c{c}on, A.~Maunier et al.: {\em Physica A\/} {\bf 392} (7), 1567--1576 (2013), Presence: Kinetics of internal structures growth in magnetic suspensions.

\bibitem{Reynolds_2016}
C.~Reynolds: {\em Field induced assembly of paramagnetic colloidal particles}, PhD Thesis, University of Oxford, 2016.

\bibitem{Kumar_2022}
M.~Kumar, A.~Kumar, R.K,~Bharti, H.N.S.~Yadav, M.~Das: {\em Mater. Today Proc.} {\bf 56}, A6-A12 (2022), A review on rheological properties of magnetorheological fluid for engineering components polishing.

\bibitem{Miyazima_1987}
S.~Miyazima, P.~Meakin, F.~Family:
{\em Phys. Rev. A}
{\bf 36} (3), 1421--1427 (1987),
Aggregation of oriented anisotropic particles.

\bibitem{Fraden_1989}
S.~Fraden, A.J.~Hurd, R.B.~Meyer:
{\em Phys. Rev. Lett.}
{\bf 63}, 2373--2376 (1989),
Electric-field-induced association of colloidal particles.

\bibitem{Melle_2001}
S.~Melle, M.A.~Rubio, G.G.~Fuller:
{\em Phys. Rev. Lett.}
{\bf 87}, 115501 (2001),
Time scaling regimes in aggregation of magnetic dipolar particles.

\bibitem{Mimouni_2007}
Z.~Mimouni:
{\em C. R. Physique}
{\bf 8}, 115--120 (2007),
Cinetique d'agregation en chaines dans une suspension colloidale soumise a un champ electrique alternatif.





\bibitem{Walter_1973}
H.~Walter: {\em J. Aerosol Sci.\/} {\bf 4} (1), 1-15 (1973), Coagulation and size distribution of condensation aerosols.

\bibitem{Wagner_1982}
P.E.~Wagner: {\em Aerosol growth by condensation\/}, in: W.H.~Marlow (ed.): {\em Aerosol microphysics II. Topics in current physics\/} {\bf 29}, Springer, Berlin, Heidelberg 1982.

\bibitem{Long_1974}
A.B.~Long: {\em J. Atmos. Sci.\/} {\bf 31}, 1040--1052 (1974), Solutions to the droplet collection equation for polynomial kernels.

\bibitem{Gillespie_1975}
D.T.~Gillespie: {\em J. Atmos. Sci.\/} {\bf 32}, 600--607 (1975), Three models for the coalescence growth of cloud drops.

\bibitem{Dziekan_2017}
P.~Dziekan, H.~Pawlowska: {\em Atmos. Chem. Phys.\/} {\bf 17}, 13509--13520 (2017), Stochastic coalescence in Lagrangian cloud microphysics.

\bibitem{Liu_2022}
H.~Liu, J.~Shao, W.~Jiang, X.~Liu: {\em Atmosphere\/} {\bf 13}, 326 (2022), Numerical modeling of droplet aerosol coagulation, condensation/evaporation and deposition processes.

\bibitem{Wegener_1954}
P.P.~Wegener: {\em J. Appl. Phys.\/} {\bf 25}, 1485 (1954), Water vapor condensation process in supersonic nozzles.

\bibitem{Yang_2017}
Y.~Yang, J.H.~Walther, Y.~Yan, C.~Wen: {\em Appl. Therm. Eng.\/} {\bf 115}, 1357--1362 (2017), CFD modeling of condensation process of water vapor in supersonic flows.

\bibitem{Spouge_1983}
J.L.~Spouge: {\em J. Stat. Phys.\/} {\bf 31}, 363--378 (1983), The size distribution for the $A_gRB_{f-g}$ model of polymerization.

\bibitem{Baskakov_2007}
I.V.~Baskakov: {\em FEBS J.\/} {\bf 274} (15), 3756--3765 (2007), Branched chain mechanism of polymerization and ultrastructure of prion protein amyloid fibrils.

\bibitem{Fang_2011}
L.T.~Fang, W.M.~Gelbart, A.~Ben-Shaul: {\em J. Chem. Phys.\/} {\bf 135}, 155105 (2011), The size of RNA as an ideal branched polymer.

\bibitem{Gray_2009}
A.~Gray‐Weale, R.G.~Gilbert: {\em J. Polym. Sci. A Polym. Chem.\/} {\bf 47} (15), 3914--3930 (2009), General description of the structure of branched polymers.

\bibitem{Vilaplana_2010}
F.~Vilaplana, R.G.~Gilbert: {\em Macromolecules\/} {\bf 43} (17), 7321--7329 (2010), Two-dimensional size/branch length distributions of a branched polymer.

\bibitem{Marshall_2013}
B.D.~Marshall, W.G.~Chapman: {\em J. Chem. Phys.\/} {\bf 138}, 174109 (2013), Three new branched chain equations of state based on Wertheim's perturbation theory.

\bibitem{Spouge_1983b}
J.L.~Spouge: {\em J. Phys. A Math. Gen.\/} {\bf 16} (4), 767--774 (1983), Solutions and critical times for the monodisperse coagulation equation when $a_{ij}=A + B(i + j) + Cij$.



\bibitem{Lepek_thesis_2021}
M.~\L{}epek: {\em Coagulation phenomena in selected complex systems} (in Polish), PhD Thesis, Warsaw University of Technology, Warsaw 2021.

\bibitem{Calin_2006}
C.D.~Calin, M.~Shirvani, H.J.~van Roessel: {\em WIT Trans. Eng. Sci.\/} {\bf 52} (2006), Numerical results for coagulation equation with bounded kernels, particle source and removal.

\bibitem{Fournier_2005}
N.~Fournier, Ph.~Lauren\c{c}ot: {\em Commun. Math. Phys.\/} {\bf 256}, 589–609 (2005), Existence of self--similar solutions to Smoluchowski’s coagulation equation.


\bibitem{cpp_libraries} M.~\L{}epek, \textit{C++ code for simulation of aggregation processes with arbitrary kernels and for theoretical predictions using combinatorial approach and incomplete Bell polynomials}, https://github.com/mlepek/aggregation (last access: 8 August 2025).



\bibitem{Levison_2015}
H.F.~Levison, K.A.~Kretke, M.J.~Duncan: {\em Nature\/} {\bf 524}, 322--324 (2015), Growing the gas-giant planets by the gradual accumulation of pebbles.

\bibitem{Liu_2020}
B.~Liu, J.~Ji: {\em Res. Astron. Astrophys.\/} {\bf 20}, 164 (2020), A tale of planet formation: from dust to planets.

\bibitem{Izidoro_2022}
A.~Izidoro, R.~Dasgupta, S.N.~Raymond, R.~Deienno, B.~Bitsch, A.~Isella: {\em Nat. Astron.\/} {\bf 6}, 357–366 (2022), Planetesimal rings as the cause of the Solar System’s planetary architecture.







\bibitem{Ricard_2017}
Y.~Ricard, D.~Bercovici, F.~Albar\`{e}de:
{\em Icarus\/} {\bf 285}, 103--117 (2017), Thermal evolution of planetesimals during accretion.

\bibitem{Wetherill_1990}
G.W.~Wetherill:
{\em Icarus\/} {\bf 88}, 336--354 (1990), Comparison of analytical and physical modeling of planetesimal accumulation.




\bibitem{Archer_2020}
J.~Archer, J.S.~Walker, F.K.A.~Gregson, D.A.~Hardy, J.P.~Reid: {\em Langmuir\/}
{\bf 26}, 12481--12493 (2020), Drying kinetics and particle formation from dilute colloidal suspensions in aerosol droplets.

\bibitem{Friedlander_2000}
 S.K.~Friedlander: {\em Smoke, dust, and haze. Fundamentals of Aerosol Dynamics\/} (Chapter 7), Oxford University Press, New York, Oxford 2000.

\bibitem{Ferreira_2021}
M.A.~Ferreira: {\em Coagulation equations for aerosol dynamics\/}, in G.~Albi, S.~Merino-Aceituno, A.~Nota, M.Zanella (eds.): {\em Trails in Kinetic Theory\/}, SEMA SIMAI Springer Series, vol 25, Springer, Cham 2021.

\bibitem{Eggersdorfer_2014}
M.L.~Eggersdorfer, S.E.~Pratsinis: {\em Adv. Powder Technol.\/}
{\bf 25}, 71--90 (2014), Agglomerates and aggregates of nanoparticles made in the gas phase.


\bibitem{Fuchs_1964}
N.A.~Fuchs: {\em The mechanics of aerosols\/}, Pergamon, New York 1964.

\bibitem{Dahneke_1983}
B.~Dahneke: {\em Simple kinetic theory of Brownian diffusion in vapors and aerosols\/}, in R.E. Meyer (ed.): {\em Theory of Dispersed Multiphase Flow\/}, Academic Press, New York, 1983, pp. 97–133.

\bibitem{Karsch_2023}
M.~Karsch, A.~Kronenburg: {\em J. Aerosol Sci.\/} {\bf 173}, 106228 (2023), Modelling nanoparticle agglomeration in the transition regime: A comparison between detailed Langevin Dynamics and population balance calculations.

\bibitem{Zurita_2002}
M.~Zurita-Gotor, D.E.~Rosner: {\em J. Colloid Interface Sci.\/} {\bf 255}, 10--26 (2002), Effective Diameters for Collisions of Fractal-like Aggregates: Recommendations for Improved Aerosol Coagulation Frequency Predictions.

\bibitem{Rogak_1992}
S.N.~Rogak, R.C.~Flagan: {\em J. Colloid Interface Sci.\/} {\bf 151}, 203--224 (1992), Coagulation of aerosol agglomerates in the transition regime.

\bibitem{Maricq_2007}  
M.M.~Maricq: {\em J. Aerosol Sci.\/} {\bf 38}, 141--156 (2007), Coagulation dynamics of fractal-like soot aggregates.

\bibitem{Guria_2009} 
G.Th.~Guria, M.A.~Herrero, Ks.E.~Zlobina: {\em Discrete Contin. Dyn. Syst.\/} {\bf 25}, 175--194 (2009), A mathematical model of blood coagulation induced by activation sources.

\bibitem{Brilliantov_2015}
N.~Brilliantov, P.L.~Krapivsky, A.~Bodrova, F.~Spahn, H.~Hayakawa, V.~Stadnichuk, J.~Schmidt: {\em PNAS\/} {\bf 112}, 9536–9541 (2015), Size distribution of particles in Saturn’s rings from aggregation and fragmentation.

\bibitem{Wattis_2006}
J.A.D.~Wattis: {\em Physica D\/} {\bf 222}, 1--20 (2006), An introduction to mathematical models of coagulation-–fragmentation processes: A discrete deterministic mean-field approach.


\bibitem{Dandekar_2023}
R.~Dandekar, R.~Rajesh, V.~Subashri, O.~Zaboronski: {\em Comput. Phys. Commun.\/} {\bf 288}, 108727 (2023), A Monte Carlo algorithm to measure probabilities of rare events in cluster-cluster aggregation.

\bibitem{Osinsky_2022_JPhysA}
A.I.~Osinsky, N.~Brilliantov: {\em J. Phys. A: Math. Theor.\/} {\bf 55}, 425003 (2022), Exact solutions of temperature-dependent Smoluchowski equations.


\bibitem{Zidar_2018}
M.~Zidar, D.~Kuzman, M.~Ravnik: {\em Soft Matter\/} {\bf 14}, 6001 (2018), Characterisation of protein aggregation with the Smoluchowski coagulation approach for use in biopharmaceuticals.


\bibitem{Wang_2021}
X.~Wang, Y.~Liu, T.~Wu, M.~Yu: {\em Appl. Sci.\/} {\bf 11}, 6815 (2021), Polymerization and collision in high concentrations for Brownian coagulation.

\bibitem{Fogelson_2010}  
A.L.~Fogelson, J.P.~Keener: {\em Phys. Rev. E\/} {\bf 81}, 051922 (2010), Toward an understanding of fibrin branching structure.

\bibitem{Johnson_2002}
W.P.~Johnson: {\em Am. Math. Mon.\/} {\bf 109}, 217234 (2002), The curious history of the Fa\`a di Bruno's formula.

\bibitem{Pitman_2006}
J.~Pitman: {\em Combinatorial Stochastic Processes\/}, Springer-Verlag, Berlin 2006.

\bibitem{Comtet_1974}
L.~Comtet:
{\em Advanced Combinatorics: The Art of Finite and Infinite Expansions\/}, Reidel Publishing Company, Dordrecht, Holland / Boston, U.S., 1974.

\bibitem{Wang_2009}
W.~Wang, T.~Wang:
{\em Comput. Math. Appl.\/}
{\bf 58}, 104--118 (2009),
General identities on Bell polynomials.


\bibitem{Blum_2000}
J.~Blum, G.~Wurm, S.~Kempf et al.: {\em Phys. Rev. Lett.\/} {\bf 85}, 2426 (2000), Growth and form of planetary seedlings: results from a microgravity aggregation experiment.

\bibitem{Blum_2004}
J.~Blum, R.~Schr\"{a}pler: {\em Phys. Rev. Lett.\/} {\bf 93}, 115503 (2004), Structure and mechanical properties of high-porosity macroscopic agglomerates formed by random ballistic deposition, https://doi.org/10.1103/PhysRevLett.93.115503.








\bibitem{Nicoud_2016}
L.~Nicoud, J.~Jagielski, D.~Pfister,S.~Lazzari,J.~Massant, M.~Lattuada, M.~Morbidelli: {\em J. Phys. Chem. B\/} {\bf 120}, 3267--3280 (2016), Kinetics of Monoclonal Antibody Aggregation from Dilute toward Concentrated Conditions.



\bibitem{Pang_2023}
K.T.~Pang, Y.S.~Yang, W.~Zhang, Y.S.~Ho, P.~Sormanni,
T.C.T.~Michaels, I.~Walsh, S.~Chia: {\em Biotechnol. Adv.\/} {\bf 67}, 108192 (2023), Understanding and controlling the molecular mechanisms of protein aggregation in mAb therapeutics.

\bibitem{Pei_2022}
P.~Pei, Y.~Peng: {\em J. Magn. Magn. Mater.\/} {\bf 550}, 169076 (2022), Constitutive modeling of magnetorheological fluids: A review.




\bibitem{Biswas_2021}
P.~Biswas, P.~Ghildiyal, G.W.~Mulholland, M.R.~Zachariah: {\em J. Colloid Interface Sci.\/} {\bf 592}, 195–204 (2021), Modelling and simulation of field directed linear assembly
of aerosol particles.



\bibitem{Liu_2010}
K.~Liu, Z.~Nie, N.~Zhao, W.~Li, M.~Rubinstein, E.~Kumacheva: {\em Science\/} {\bf 329}, 197 (2010), Step-growth polymerization of inorganic nanoparticles.

\bibitem{Chen_2012}  
Q.~Chen, S.C.~Bae, S.~Granick: {\em J. Am. Chem. Soc.\/} {\bf 134}, 11080--11083 (2012), Staged self-assembly of colloidal metastructures.



\bibitem{Yang_2018}  
T.~Yang, D.W.M.~Marr, N.~Wu: {\em Colloids Surf. A\/} {\bf 540}, 23--28 (2018), Superparamagnetic colloidal chains prepared via Michael-addition.

\bibitem{Li_2020}  
W.~Li, B.~Liu, C.~Hubert, A.~Perro, E.~Duguet, S.~Ravaine: {\em Nano Res.\/} {\bf 13}, 3371–3376 (2020), Self-assembly of colloidal polymers from two-patch silica nanoparticles.

\bibitem{Stuij_2021}  
S.~Stuij, J.~Rouwhorst, H.J.~Jonas, N.~Ruffino, Z.~Gong, S.~Sacanna, P.G.~Bolhuis, P.~Schall: {\em Phys. Rev. Lett.\/} {\bf 127}, 108001 (2021), Revealing polymerization kinetics with colloidal dipatch particles.

\bibitem{BinLiu_2022}  
B.~Liu, W.~Li, E.~Duguet, S.~Ravaine: {\em ACS Macro Lett.\/} {\bf 11}, 156--160 (2022), Linear assembly of two-patch silica nanoparticles and control of chain length by coassembly with colloidal chain stoppers.

\bibitem{Mhanna_2022}  
R.~Mhanna, Y.~Gao, I.~Van Tol, E.~Springer, N.~Wu, D.W.M.~Marr: {\em Langmuir\/} {\bf 38}, 5730--5737 (2022), Chain assembly kinetics from magnetic colloidal spheres.

\bibitem{Nelson_2021} 
A.C.~Nelson, M.A.~Kelley, L.M.~Haynes, K.~Leiderman: {\em Curr. Opin. Biomed. Eng.\/} {\bf 20}, 100350 (2021), Mathematical models of fibrin polymerization: past, present, and future.



\bibitem{Horne_2017_chapter}  
D.S.~Horne, J.A.~Lucey: {\em Rennet-Induced Coagulation of Milk\/}, in: P.L.H.~McSweeney, P.F.~Fox, P.D.~Cotter, D.W.~Everett (Eds.): {\em Cheese: Chemistry, Physics and Microbiology\/}, Academic Press, 2017.

\bibitem{Bansal_2007}  
N.~Bansal, P.F.~Fox, P.L.H.~McSweeney: {\em J. Agric. Food Chem.\/} {\bf 55}, 3120--3126 (2007), Aggregation of rennet-altered casein micelles at low temperatures.

\bibitem{Li_2019}  
Q.~Li, Z.~Zhao: {\em Food Chem.\/} {\bf 291}, 231--238 (2019), Acid and rennet-induced coagulation behavior of casein micelles with modified structure.

\bibitem{Salvador_2022}  
D.~Salvador, Y.~Acosta, A.~Zamora, M.~Castillo: {\em Foods\/} {\bf 11}, 1243 (2022), Rennet-induced casein micelle aggregation models: a review.

\bibitem{Hou_2022}  
D.~Hou, L.~Pascazio, J.~Martin, Y.~Zhou, M.~Kraft, X.~You: {\em J. Aerosol Sci.\/} {\bf 159}, 105866 (2022), On the reactive coagulation of incipient soot nanoparticles.

\bibitem{Henderson_2022} 
L.~Henderson, P.~Shukla, V.~Rudolph, S.K.~Bhatia: {\em Combust. Flame\/} {\bf 245}, 112303 (2022), Modelling the formation, growth and coagulation of soot in a combustion system using a 2-D population balance model.




\bibitem{Coulaloglou_1977}  
C.A.~Coulaloglou, L.L.~Tavlarides: {\em Chem. Eng. Sci.\/} {\bf 32}, 1289--1297 (1977), Description of interaction processes in agitated liquid-liquid dispersions.

\bibitem{Friesen_2003}  
W.I.~Friesen, T.~Dabros: {\em J. Chem. Phys.\/} {\bf 119}, 2825–2839 (2003), Constant-number Monte Carlo simulation of aggregating and fragmenting particles.

{
\bibitem{Lattuada_2016}  
M.~Lattuada, A.~Zaccone, H.~Wuc, M.~Morbidelli: {\em Soft Matter\/} {\bf 12}, 5313 (2016), Population-balance description of shear-induced clustering, gelation and suspension viscosity in sheared DLVO colloids.
}

\bibitem{Low_2020}  
L.E.~Low, S.P.~Siva, Y.K.~Ho, E.S.~Chan, B.T.~Tey: {\em Adv. Colloid Interface Sci.\/} {\bf 277}, 102177 (2020), Recent advances of characterization techniques for the formation, physical properties and stability of Pickering emulsion.









{
\bibitem{vanDongen_1985}
P.G.J.~van Dongen, M.H.~Ernst: {\em Phys. Rev. Let. \/} {\bf 54}, 1396--1399 (1985), Dynamic scaling in the kinetics of clustering.
}


{
\bibitem{Norris_1999}
J.R.~Norris:
{\em Ann. Appl. Probab. \/}
{\bf 9},  78--109 (1999),
Smoluchowski's coagulation equation: uniqueness, nonuniqueness and a hydrodynamic limit for the stochastic coalescent.


\bibitem{Huang_1987_book}
K.~Huang: {\em Statistical Mechanics\/} (Chapter 9), John Wiley \& Sons, New York 1987.
}






\end{thebibliography}
\end{document}